\documentclass[useAMS,usenatbib,babel]{mn2e}
\usepackage{amsmath}
\usepackage[breaklinks=true]{hyperref}
\usepackage[usenames]{color}
\usepackage{graphicx}
\usepackage{gensymb}
\usepackage{times}

\begin{document} 
\title[Galaxy orientation with the cosmic web]{Galaxy orientation with the cosmic web across cosmic time 
}

\author[S. Codis, A. Jindal et al.]{\parbox{\textwidth}{
S.~Codis$^{1,2}$\thanks{E-mail: codis@iap.fr}\!\!, A.~Jindal$^{3,2}$\!, 
N.~E.~Chisari$^{4}$\!, 
D.~Vibert$^{5}$\!, 
Y.~Dubois$^{1}$\!,  
C.~Pichon$^{1,6,7}$\!, 
J.~Devriendt$^{4}$
}
\vspace*{6pt}\\
\noindent
$^{1}$CNRS and Sorbonne Universit\'e, UMR 7095, Institut d'Astrophysique de Paris, 98 bis Boulevard Arago, F-75014 Paris, France\\
$^{2}$ Canadian Institute for Theoretical Astrophysics, University of Toronto, 60 St. George Street, Toronto, ON M5S 3H8, Canada\\
$^{3}$ Department of Astronomy \& Astrophysics, University of Toronto, 60 St. George Street, Toronto, ON M5S 3H8, Canada\\
$^{4}$ Department of Physics, University of Oxford, Keble Road, Oxford, OX1 3RH,UK\\
$^{5}$ Aix Marseille Universit\'e, CNRS, Laboratoire d'Astrophysique de Marseille, UMR 7326, 38 rue F. Joliot-Curie, 13388, Marseille, France\\
$^6$ School of Physics, Korea Institute for Advanced Study (KIAS), 85 Hoegiro, Dongdaemun-gu, Seoul, 02455, Republic of Korea\\
$^7$ Institute for Astronomy, University of Edinburgh, Royal Observatory, Blackford Hill, Edinburgh, EH9 3HJ, United Kingdom\\
}

\pubyear{2018}
\maketitle

\begin{abstract}
This work investigates the alignment of galactic spins with the cosmic web across cosmic time using  the cosmological hydrodynamical simulation Horizon-AGN. The cosmic web structure is extracted via the persistent skeleton as implemented in the {\sc DisPerSE} algorithm. It is found that the spin of low-mass galaxies is more likely to be aligned with the filaments of the cosmic web and to lie within the plane of the walls while more massive galaxies tend to have a spin perpendicular to the axis of the filaments and to the walls. The mass transition is detected with a significance of 9 sigmas.
 This galactic alignment is consistent with the alignment of the spin of dark haloes found in pure dark matter simulations and with predictions from (anisotropic) tidal torque theory. However, unlike haloes, the alignment of low-mass galaxies is weak and disappears at low redshifts while the orthogonal spin orientation  of massive galaxies is strong and increases with time, probably as a result of mergers. At fixed mass, alignments  are correlated with galaxy morphology: the high-redshift  alignment is dominated by spiral galaxies while  elliptical centrals {  are mainly responsible for the perpendicular signal}. 
 These predictions for spin alignments with respect to cosmic filaments and unprecendently walls are successfully compared with existing observations.
The alignment of the shape of galaxies with the different components of the cosmic web is also investigated. A coherent and stronger signal is found  in terms of shape at high mass. 
 The two regimes probed in this work  induce competing  galactic alignment signals  for weak lensing, with opposite redshift and luminosity evolution. Understanding the details of these intrinsic alignments will be key to exploit future major cosmic shear surveys like Euclid or LSST.
\end{abstract}


\section{Introduction}
\label{sec:intro}

In the concordance model of cosmology, the large-scale structure originates from primordial Gaussian fluctuations which grow under the laws of gravity in the expanding Universe, 
and hierarchically form galaxies, clusters and super-clusters of galaxies. Those galaxies are not islands randomly distributed in the Universe but form a complex network -- the so-called cosmic web -- made of large voids surrounded by walls and filaments which intersect at nodes where large clusters of galaxies reside \citep{klypin&shandarin83,bkp96,pogosyanetal1998}. 
The origin of the filaments and nodes lies in the local asymmetries of the initial density field: the high-density peaks define the nodes of the cosmic web and completely determine the filamentary bridges in between. 
This structure is later amplified by the gravitational instability. Matter escapes from the voids towards the walls then flows towards the filaments before accreting onto the over-dense nodes \citep{1982GApFD..20..111A,1984AZh....61..837S}. 

Galaxy formation and evolution naturally take place within these large-scale cosmic dynamics, raising the question of the role, if any, of the cosmic web in determining the properties of galaxies.
It is now well-established from observations and simulations that galaxy properties depend on their environment. For instance, massive ellipticals tend to reside in dense regions while low-mass, disk-like galaxies tend to live in the field \citep{1980ApJ...236..351D}. 
There is increasing evidence for galaxy properties being correlated to their large-scale environment \citep{2015MNRAS.451.3249A,2016MNRAS.457.2287A,2017MNRAS.465.3817M,Kraljic2018}. 
This behaviour is explained by the theory of biased clustering \citep{1984ApJ...284L...9K,1988ApJ...333L..45W,2015MNRAS.447.2683A} in which high mass objects preferentially form in over-dense environments. 
The effect is therefore mainly isotropic and due to the local density although an additional dependence on mass assembly is found, the so-called assembly bias \citep{2008ApJ...687...12D,2017JCAP...03..059L,2017MNRAS.468.2984P,2017ApJ...848L...2M}, probably due to the anisotropic nature of the filamentary cosmic web \citep{Musso2018}. 
To unveil the effect of the large-scale structure on galaxy formation beyond its scalar or isotropic part, one strategy is to rely on observables that are not scalar but have a directionality such as their rotation axis or their shape. 

In the standard paradigm of galaxy formation, the intrinsic angular momentum -- or spin -- is thought to arise from tidal torquing \citep[see][for a review]{schaefer09}. Because galaxies do not form everywhere but in  filaments and nodes, \cite{ATTT} showed that the net effect of the cosmic web was to set preferred directions for the orientation of the spins, in agreement with numerical simulations  which showed that massive haloes and galaxies tend to have a spin perpendicular to the filaments while lower mass objects are more likely to have a spin aligned with their host filament {  \citep[e.g][]{2005ApJ...627..647B,calvoetal07,codisetal12,2013ApJ...762...72T,2014MNRAS.440L..46A} \footnote{  Several works have also measured part of these spin-filament alignments in simulations of dark matter only e.g \cite{aubertetal04,sousbie08,2009ApJ...706..747Z} or including baryons \citep{hahn10,dubois14}.}}. Qualitatively, this transition occurring at a (redshift-dependent) mass of about $5\times 10^{12} M_\odot$ at redshift zero for haloes \citep[][i.e. about 1/8$^{\rm th}$ of the mass of non linearity at that redshift]{codisetal12} can be understood from the large-scale dynamics of matter. When the walls collapse to form a filament at their intersection, they create a 
vorticity field aligned with the axis of the filament with typically eight (point reflection symmetric with regard to the central saddle point of the filament) octants of opposite orientation \citep{pichon99,libeskind13,2013arXiv1301.0348L,2014ApJ...793...58W,laigle2014,2017ApJ...838...21Z} so that the first generation of haloes and galaxies also form with a spin aligned with the filament's direction. The region of interest with a given polarity for the vorticity is therefore by symmetry 1/8$^{\rm th}$ of the 
typical size of the filament (in terms of spherical volumes) as pointed out by \cite{ATTT}. 
Later, filaments collapse and galaxies and haloes therefore flow towards the nodes. During this process, some of them merge and convert their orbital angular momentum into spin perpendicular to the filament's axis as they become more massive \citep{codisetal12}. The specific role played by mergers in flipping the spin of massive objects perpendicular to filaments was later confirmed by \cite{2014MNRAS.445L..46W} who followed explicitly flip as a function of merger. 
{  This spin flip can therefore be explained by the difference of mass accretion history depending on halo mass \citep{2015ApJ...813....6K}, environment \citep{2018MNRAS.473.1562W} and migration time to reach in turn the walls, filaments and nodes of the cosmic \citep{2017MNRAS.468L.123W}}.
Understanding the interplay between this purely gravitational process and baryonic effects (feedback from active galactic nuclei (AGN) and supernovae, cooling, etc)  requires the use of full physics hydrodynamical simulation. Using the Horizon-AGN at high redshift, \cite{dubois14} carried out a first  study and found a tentative spin flip at a stellar mass about $10^{10.5} M_\odot$. 
This alignment between spin and filamentary axis was also detected in observations \citep{tempel&libeskind13,Tempel13,2013ApJ...779..160Z,2015ApJ...798...17Z}.

\begin{figure}
\includegraphics[width=\columnwidth]{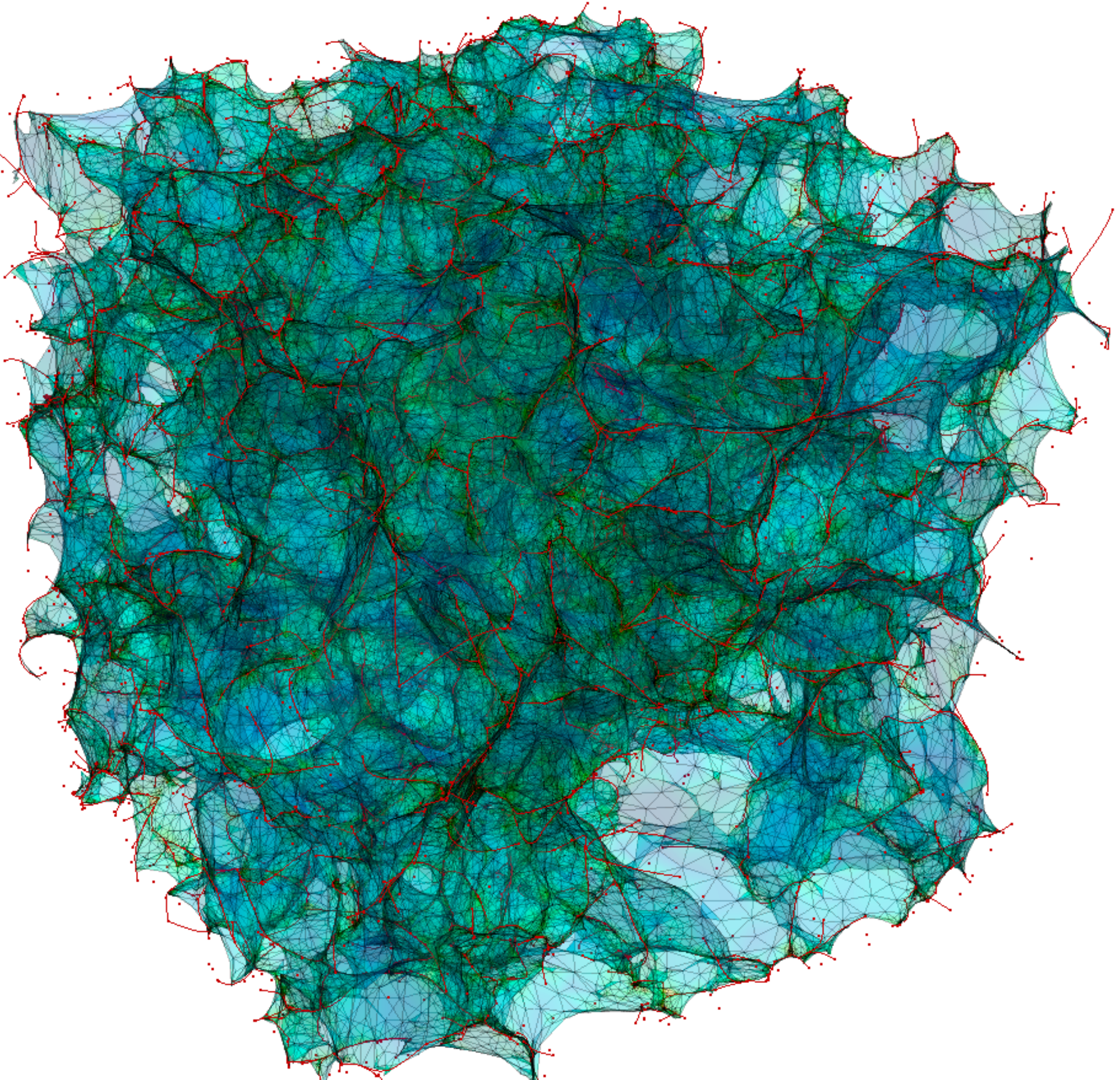}
\caption{
The 3D structure of the walls identified by {\sc DisPerSE}(together with the corresponding skeleton in red) from galaxies in the redshift zero snapshot of the Horizon-AGN simulation. The colour  encodes log  density on the wall from light blue to green-yellow-red.
The triangulation of the walls is clearly visible and is used throughout to define orientation and distance to the walls. The dots represent minima of the discrete morse-smale complex. The filaments (red solid lines) are as expected at the boundary of the walls.
\label{fig:walls3D}}
\end{figure}

Angular momentum naturally shapes galaxies. Together with  mass, its radial stratification  determines the morphology of a galaxy. The more orderly, the thinner the galactic disk  \citep{1970ApJ...160..831S,2014ApJ...784...26O}. The direction of the spin also affects the orientation of the galaxy's shape. For spirals, discs are perpendicular to the spin axis but the shape of elliptical galaxies also correlate with spin axis depending on their triaxiality: prolate ellipticals tend to have their spin aligned with the minor axis of their shape while oblate galaxies are more likely to spin along their major axis. If the spin of galaxies is correlated to their large-scale environment, their shapes should also be coherent on cosmological scales. For ellipticals, this alignment between shape and host filament might even be stronger than for spin, since on the one hand ellipticals are dominated by the random motion of their stars rather than their spin and on the other hand, tidal stretching which affects their distribution of stars \citep{Catelan01} is first order in the fields (shape are expected to respond linearly to a tidal field at first order) while tidal torquing is second order.
 Numerous works have indeed shown that massive haloes tend to align their shape along filaments and walls in dark matter simulations \citep{2006ApJ...652L..75P,2007MNRAS.375..184B}, an effect which is also observed in real data. In particular, the two-point radial alignment between massive clusters is now well-established \citep{2012MNRAS.423..856S,2017MNRAS.468.4502V}.
 
Understanding the effect of the large-scale cosmic environment on galaxy shapes is 
of interest to explain the so-called Hubble sequence \citep{1994ARA&A..32..115R}, i.e. the distribution and redshift evolution of the fraction of spirals and ellipticals. It is also 
paramount for weak lensing as intrinsic alignments \citep[see e.g][for recent reviews]{Joachimi15,2015SSRv..193..139K,Kiessling15,troxel15} can significantly contaminate weak lensing observables and eventually bias 
the cosmological constraints obtained by cosmic shear experiments \citep{Kirk10,Krause15}. Recent improvements in our ability to simulate galaxy formation on cosmological scales now allow us to measure this effect directly in hydrodynamical simulations which can account for all sources of non-linearity and probe the impact of baryonic physics \citep{codis14,Tenneti15a,Chisari15,Chisari16,Velliscig15b,Hilbert16}. 

Indeed, intrinsic alignment `halo models' have been constructed to connect the galaxy shape and orientation to the properties of the host dark matter halo \citep{Schneider10,joachimi13a,Joachimi13b}. These approaches, which often adopt different models for disc and elliptical galaxies, are beginning to be tested with hydrodynamical simulations, for instance with Horizon-AGN \citep{2017MNRAS.472.1163C}. This entails understanding the relation between galaxy and halo shape and orientation. One of the questions about the validity of this type of model is whether it is necessary to incorporate the effects of the larger scale anisotropic environment when modelling intrinsic alignments.
Alignments between galaxies and filaments have been studied extensively  theoretically and numerically  to understand the role played by the environment in shaping galaxies, mostly relying on linear theory \citep[e.g][]{lee&pen00,Catelan01,porcianietal02,ATTT} and dark matter simulations. With the advent of full-physics hydrodynamical simulations, it is now possible to also investigate this issue with virtual galaxies, e.g.  extending  \cite{dubois14} to low redshift.
Yet, very little attention has been paid to the simulated walls of the cosmic web despite being more  
 easily detected (the major axis of the tidal tensor ie the normal to the wall can be more accurately extracted than the other two as argued for instance by \citealt{navarro04}), not only in 3D but also in projection,
and despite being possibly a better testbed for linear tidal alignment effects as they are presumably less dense than filaments and therefore less non-linear \citep{2016ApJ...825...49C}.
Quantifying the alignment of galaxy spin and orientation with respect to cosmic walls 
 is therefore one of the goals of this work.

In observations, several investigations have reported the detection of alignments between galaxy spins and walls defined as the boundaries of voids. \cite{flin86,flin90} discovered that galaxies' spins tend to be aligned with the Local Supercluster plane, which was recently confirmed by \cite{navarro04}.
\cite{trujillo06} used the SDSS and 2dFGRS to show that the rotation axis of spiral galaxies lies preferentially in the plane of these voids.
More recently, \cite{2018ApJ...860..127L} measured a strong tendency for the spins of galaxies, located relatively close to the W-M sheet in the vicinity of Virgo and the Local Void, to align their spin with the plane of the sheet  \footnote{The words `sheet' and `wall' will be used interchangeably in this paper.}.
In contrast, \cite{slosar09} found no detection of alignment between galaxy spins and the voids in the SDSS. \cite{varela11} also used the SDSS as well as morphological classification from the Galaxy Zoo Project and reported that the spin of galaxy discs is more likely to point towards the centre of voids, in contradiction with e.g \cite{trujillo06}.
These somewhat contradictory results are in part due to the difficulty to measure spin vectors, which often involve relating observed shapes and spins using only axis ratios and position angles. In particular, this leads to two or four solutions depending on the galaxy's inclination and therefore increases the noise, which makes any detection challenging. In addition, the signal seems to significantly depend on the population of galaxies (luminosity, morphology, colours, etc) and the type of walls (their size for instance) selected, which makes any comparison challenging.

This work studies the alignment of galaxy spins with the filaments and walls of the cosmic web across cosmic time in Horizon-AGN \citep{dubois14}\footnote{\href{http://www.horizon-simulation.org}{http://www.horizon-simulation.org}}, 
in order to provide theoretical predictions to be compared with observations. It  quantifies how the large-scale alignment of dark matter haloes pervades for galaxies, a key ingredient of halo models. 
For this purpose, this work extends the first attempts of \cite{dubois14} at detecting the spin-filament alignments at $z=1.8$ to all redshifts, to, not only filaments but also cosmic sheets, and with an improved extraction of the cosmic web components. It also investigates how those spin alignments propagate to galaxy shapes and therefore galaxy intrinsic alignments.  
Specifically, this paper focuses on understanding the origin of these alignments, for instance by studying the spin flip as a function of morphology at a fixed mass.

Section~\ref{sec:simu} briefly presents the Horizon-AGN simulation and the {\sc DisPerSE} algorithm used  to identify filaments and walls. 
Section~\ref{sec:spin-fil}  is devoted to the study of the alignment of spins and filaments while Section~\ref{sec:spin-wall} investigates the orientation of the spins with respect to the walls.
Section~\ref{sec:shape-fil} also considers how the shapes of galaxies align with the cosmic web.
Finally, Section~\ref{sec:conclusion} concludes.
Appendix~\ref{sec:halo-gal} investigates how galaxy's and halo's spins correlate. 
Appendix~\ref{app:skel} presents a comparison between skeletons generated by the distribution of galaxies and haloes while
Appendix~\ref{sec:persistence} studies how the skeleton of the cosmic web and galaxy alignments depend on the persistence level chosen in the {\sc DisPerSE} algorithm.
The redshift-evolution of the spin flips and their dependence on galaxy morphology are investigated in Appendix~\ref{sec:Mtr} and ~\ref{sec:vsig}  respectively.
Appendix~\ref{app:ATTT} presents a model for halo spin near walls in Gaussian random fields. 
{  A resolution study is performed in Appendix~\ref{sec:res}.}

\section{Simulating galaxy evolution 
}
\label{sec:simu}
\begin{figure*}
\includegraphics[width=0.98\columnwidth]{fig/skel-sig-5-out-761}\hskip .5 cm
\includegraphics[width=0.98\columnwidth]{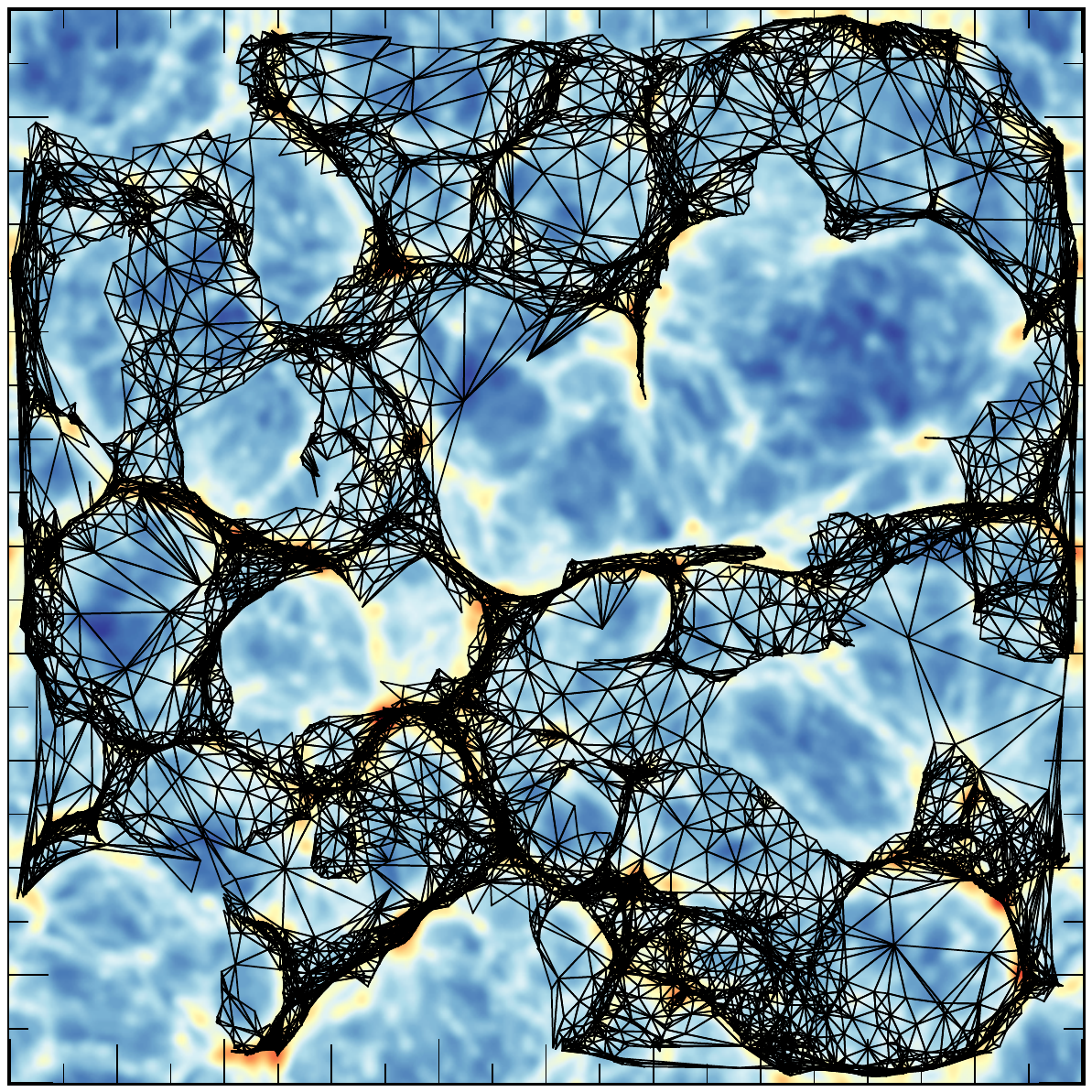}
\caption{Left-hand panel: A two dimensional projection of a slice of the simulation with a width of about 10\% of the simulation box at $z=0$. The black lines represent the filaments extracted by {\sc DisPerSE} for a persistence level of $5\sigma$. The total matter density field computed on a $512^{3}$ grid is shown from blue (low density) to red (high density). As expected, the persistent skeleton follows closely the filaments as guessed by visual inspection. Right-hand panel: same as the left-hand panel but when the tessellation of the walls is shown in black.
\label{fig:Filaments}}
\end{figure*}

\subsection{The Horizon-AGN simulation}

 This work is based on the Horizon-AGN simulation \citep{dubois14}. This ``full physics'' simulation uses
a standard $\Lambda$CDM cosmology compatible with WMAP-7 ~\citep{komatsuetal11}. The total matter density is set to $\Omega_{\rm m}=0.272$, the dark energy density to $\Omega_\Lambda=0.728$, the amplitude of the matter power spectrum is $\sigma_8=0.81$, the baryon density $\Omega_{\rm b}=0.045$, the Hubble constant $H_0=70.4 \, \rm km\,s^{-1}\,Mpc^{-1}$, and the spectral index $n_s=0.967$. The Horizon-AGN simulation follows $1024^3$ dark matter (DM) particles
in a $L_{\rm box}=100 \, h^{-1}\rm\,Mpc$ periodic box. The resulting DM mass resolution is then $M_{\rm DM, res}=8\times 10^7 \, \rm M_\odot$. The simulation was run with
the Adaptive Mesh Refinement code {\sc ramses}~\citep{teyssier02} with
an initial mesh refinement of up to $\Delta x=1\, \rm kpc$ (7 levels of refinement).
All the details of this simulation are given in \cite{dubois14}; only  the most important features are repeated here.

Gas cooling is allowed by means of H and He cooling down to $10^4\, \rm K$ with a contribution from metals, following \cite{sutherland&dopita93}. 
Heating from a uniform UV background takes place after redshift $z_{\rm reion} = 10$ according to the model of~\cite{haardt&madau96}.
Metallicity is modelled as a passive variable for the gas that varies according to the injection of gas ejecta during supernovae explosions and stellar winds.
Star formation is modelled by a Schmidt law:
$\dot \rho_*= \epsilon_* {\rho / t_{\rm ff}}\, ,$ where $\dot \rho_*$ is the star formation rate density, $\epsilon_*=0.02$~\citep{kennicutt98, krumholz&tan07} the constant star formation efficiency, and $t_{\rm ff}$ the local free-fall time of the gas.
Star formation is allowed where
the gas Hydrogen number density exceeds $n_0=0.1\, \rm H\, cm^{-3}$ following a Poisson random process~\citep{rasera&teyssier06, dubois&teyssier08winds} with a stellar mass resolution of $M_{*, \rm res}=\rho_0 \Delta x^3\simeq 2\times 10^6 \, \rm M_\odot$.

Stellar feedback is modelled using a \citet{salpeter55} initial mass function with a low-mass (high-mass) cut-off of $0.1\, \rm M_{\odot}$ ($100 \, \rm M_{\odot}$). 
The mechanical energy from supernovae type II and stellar winds follows the prescription of {\sc starburst99}~\citep{leithereretal99, leithereretal10}, and the frequency of type Ia supernovae explosions is taken from~\cite{greggio&renzini83}. 
 Active Galactic Nuclei (AGN) feedback is modeled according to the model of \cite{duboisetal12agnmodel}.

\subsection{Defining galaxies'  physical properties}
\label{sec:defgal}

Galaxies are identified using the {\sc AdaptaHOP}  finder~\citep{aubertetal04}. This method relies directly on the distribution of star particles to construct the catalogue of galaxies, with
20 neighbours to compute the local density of each particle.
A local threshold of $\rho_{\rm t}=178$ times the average total matter density is applied to select relevant densities.
Only galactic structures identified with more than 50 star particles are included in the mock galaxy catalogue. 
Similarly to galaxies, haloes are identified by applying the {\sc AdaptaHOP} algorithm. The centre of the halo is found using a shrinking sphere method \citep{Power03} and  only  haloes with more than $100$ particles and with more than $80$ times the average density of the box  are kept in the catalogue.  

This paper considers  outputs of the simulation with redshift ranging from approximately $z=2$ to $z=0$. The catalogue of main galaxies (all galaxies that are not considered as a substructure of a more massive galaxy) contains at redshift $z=2$  $\sim 150\,533$ objects with masses between $1.76\times10^{8}$ and $5.20\times10^{11}\, \rm M_{\odot}$, and the halo catalogue contains $\sim 269\,443$ objects with masses between $8.35\times10^{9}$ and $3.78\times10^{13}\, \rm M_{\odot}$.  At redshift $z=0$ the galaxy catalogue contains $\sim 114\,292$ objects with masses between $1.66\times10^{8}$ and $5.28\times10^{12}\, \rm M_{\odot}$, and the halo catalogue contains $\sim 229\,464$ objects with masses between $8.35\times10^{9}$ and $7.42\times10^{14}\, \rm M_{\odot}$. 
The stellar masses quoted in this paper should be understood as the sum over all star particles that belong to a galaxy structure or dark matter particles that belong to a halo structure as identified by AdaptaHOP. 
{  Note that the galactic size-mass relation in Horizon-AGN was shown to be in fairly good agreement with observations in particular thanks to AGN feedback \citep{2016MNRAS.463.3948D}.}

  The spin of galaxies is assigned   the total angular momentum of the star particles which make up a given galactic structure relative to the centre of mass. 
 The intrinsic angular momentum vector $\mathbf{L}$ or spin of a galaxy  can therefore be written as
\begin{equation}
\label{eq:spindef}
  \mathbf{L} = \sum_{ \alpha=1}^{N} m^{( \alpha)} \mathbf{x}^{(\alpha)} \times \mathbf{v}^{(\alpha)} \, ,
\end{equation}
where the $\alpha$ superscript denotes the $\alpha$-th stellar particle
of mass $m^{ (\alpha)}$, 
position
$\mathbf{x}^{ (\alpha)}$ and velocity
  $\mathbf{v}^{(\alpha)}$ relative to the center of mass of that galaxy. 
A similar procedure is used to compute the spin of haloes. 

Section~\ref{sec:shape-fil}   also considers the orientation of the shape of galaxies, by means of the minor axis of their standard inertia tensor
\begin{equation}
  I_{ij} = \frac{1}{M} \sum_{\alpha=1}^{N} m^{(\alpha)} x^{(\alpha)}_i x^{(\alpha)}_j,
  \label{eq:sit}
\end{equation}
where $i,j\in\{1,2,3\}$ correspond to the axes of the simulation box and $M=\sum_{\alpha=1}^{N} m^{(\alpha)}$ is the total stellar mass. The minor axis corresponds to the shortest axis of the ellipsoid and tends to align with the spin, especially when galaxies are rotation dominated \citep{2017MNRAS.472.1163C}.

 This paper focuses on four properties of the galaxies: their spin, shape, mass and morphology characterized by the kinematic ratio between stellar rotation and velocity dispersion $V/\sigma$ defined in \cite{dubois14}. For this purpose,
the spin of the stellar population is first computed in order to define a set of cylindrical coordinates ($r$, $\theta$, $z$), with the $z$-axis oriented along the galaxy spin. We then decompose the velocity of each star into cylindrical components $v_{r}$, $v_{\theta}$, $v_z$, and we define the rotational velocity of a galaxy as the mean $v_{\theta}$ i.e  $V=\bar v_{\theta}$ while the average velocity dispersion of the galaxy is computed using $\sigma^2=(\sigma_{r}^2+\sigma_{\theta}^2+\sigma_z^2)/3$.
 High $V/\sigma$ corresponds to disk-like galaxies dominated by their rotation while low $V/\sigma$ corresponds to pressure-supported ellipticals; the galaxy population  will also be split  into centrals and satellites. For this purpose, galaxies  are matched to their closest halo. A central galaxy is then defined as the most massive galaxy belonging to a given main host halo 
 and satellite galaxies as all other galaxies matched to the same halo \footnote{Note that adding an additional criterion on the distance of the centrals from the halo's center of mass only affects one percent of our sample of central galaxies and therefore does not change the results shown in this paper.}.

\subsection{Extracting the persistent cosmic web}

\begin{figure*}
\includegraphics[width=0.95\columnwidth]{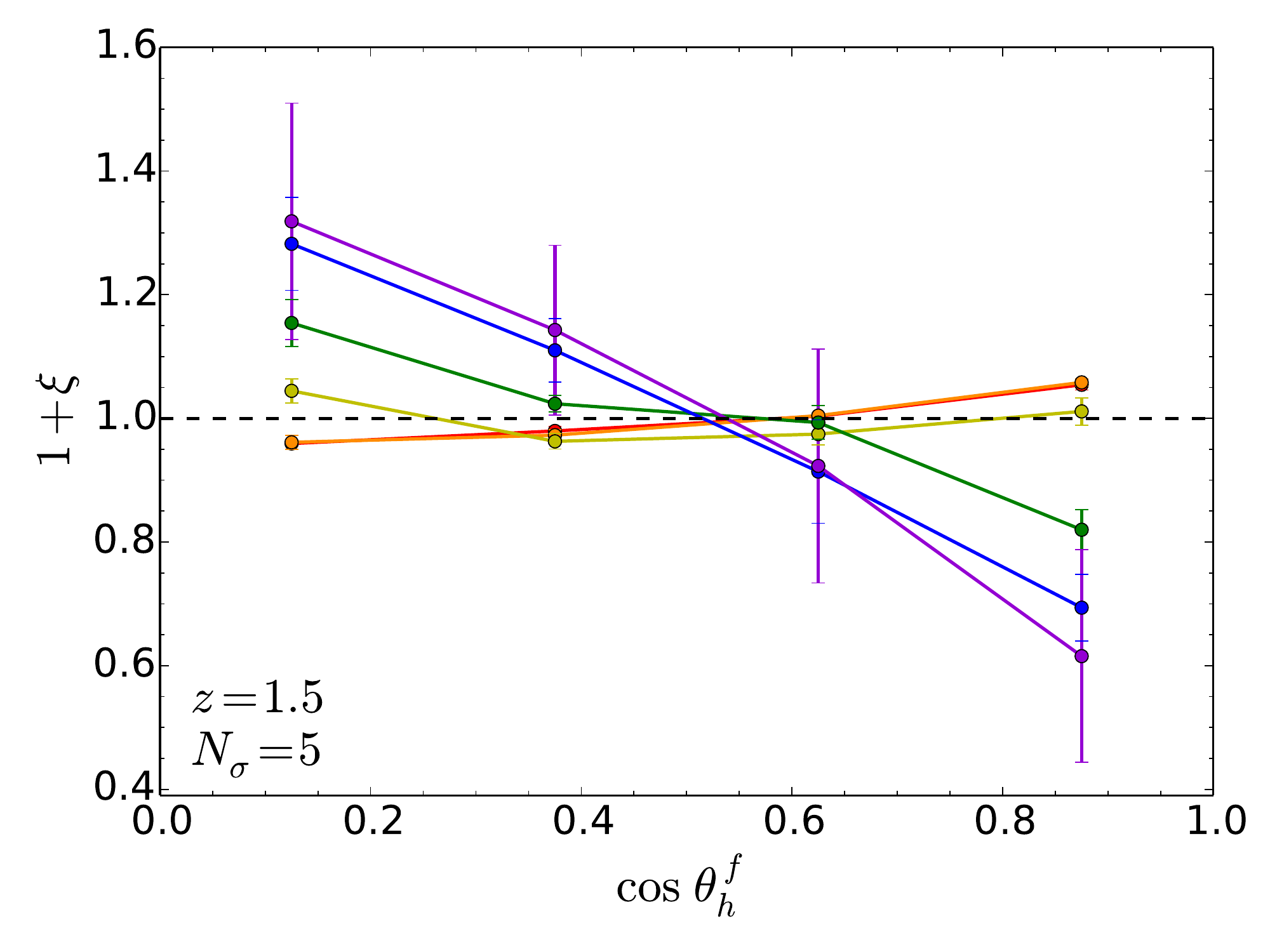}
\includegraphics[width=0.95\columnwidth]{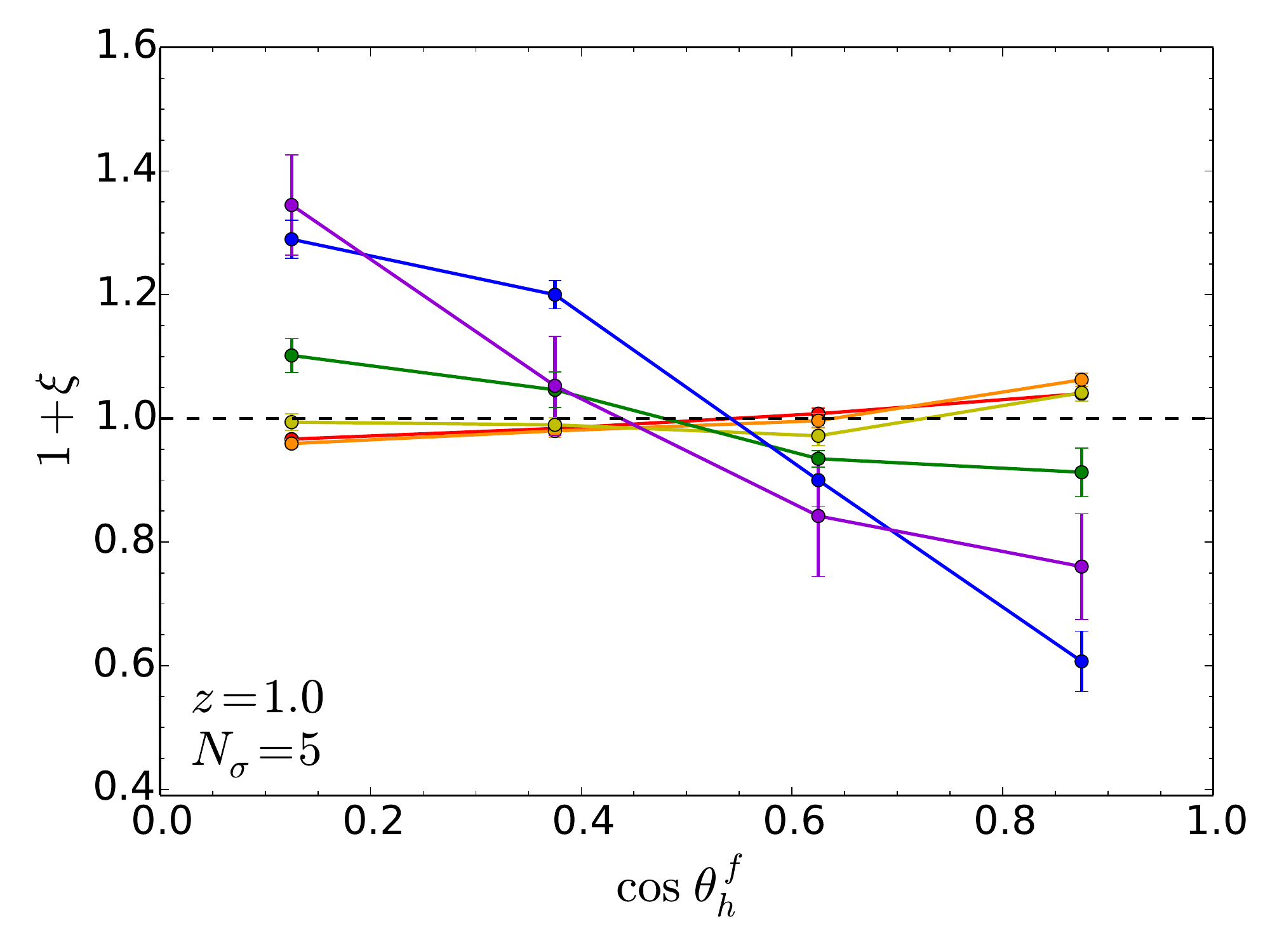}
\includegraphics[width=0.95\columnwidth]{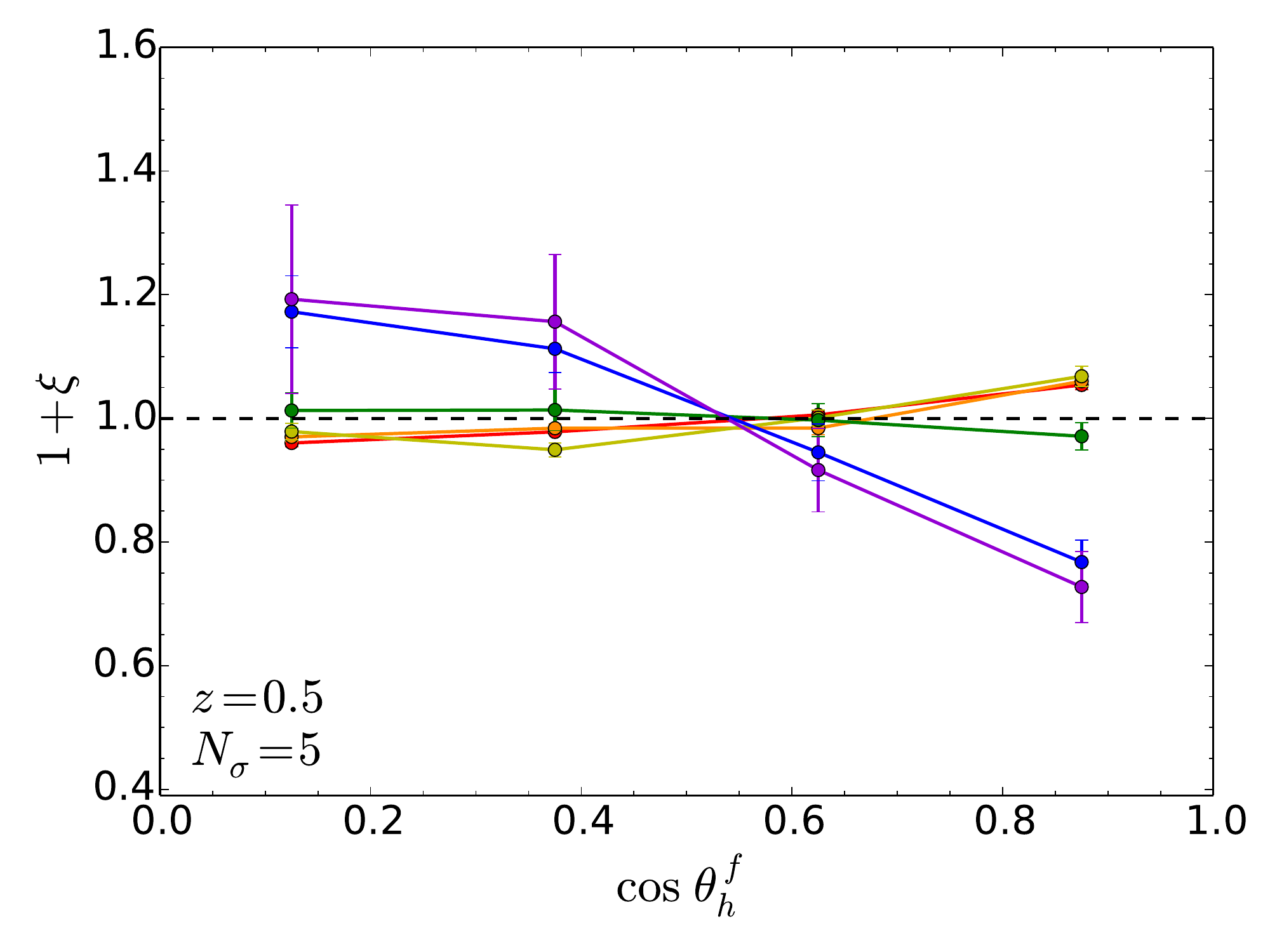}
\includegraphics[width=0.95\columnwidth]{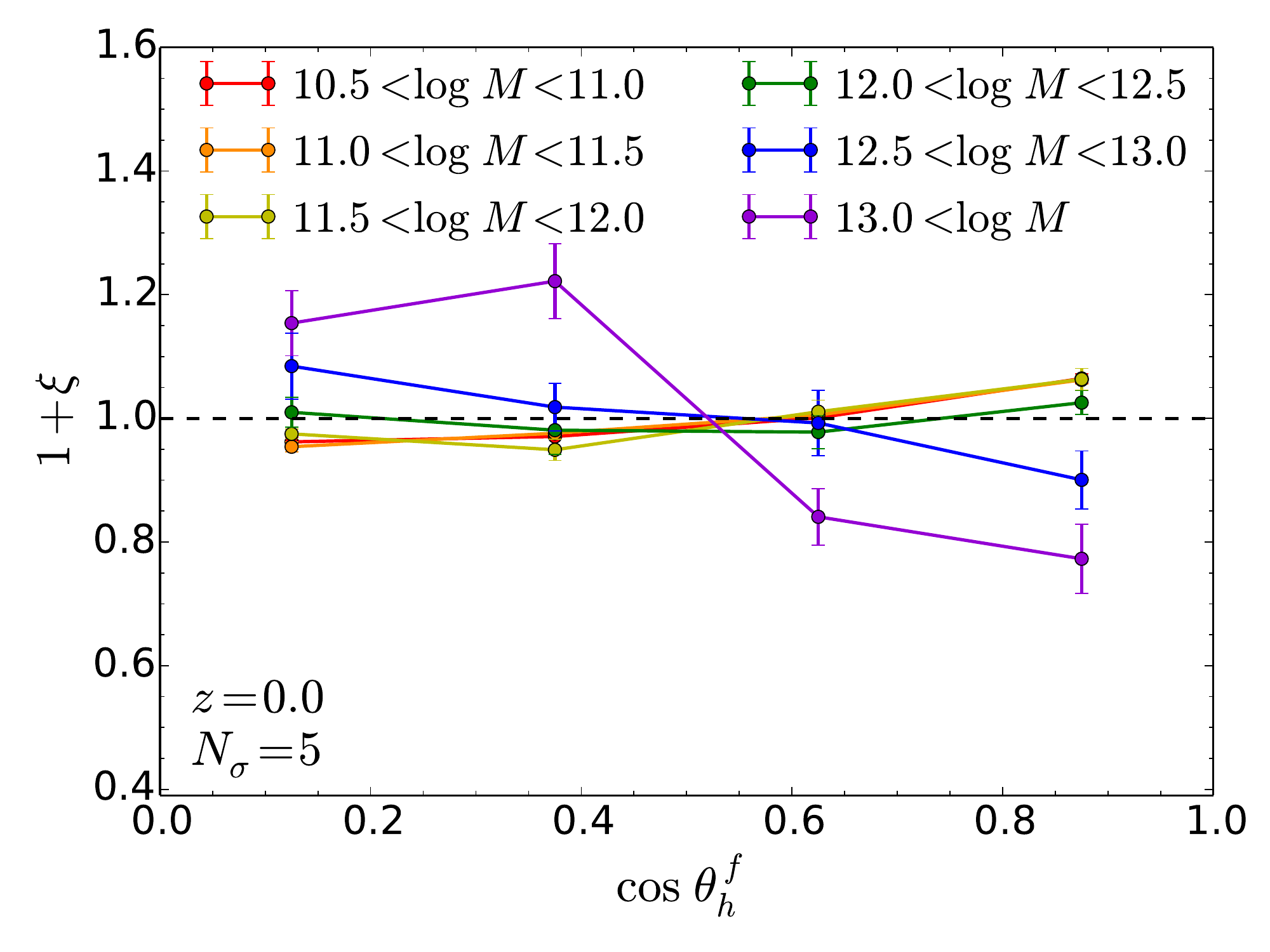}
\caption{Alignment between the spin of dark matter haloes and the closest cosmic filament for a persistence level of $N_{\sigma}$ = 5, various halo masses from red (low mass) to purple (high mass) as labelled and different redshifts from $z=1.5$ (top left-hand panel) to $z=0$ (bottom right-hand panel). Error bars represent the error on the mean among 8 subcubes of the simulation. High mass haloes tend to have a spin perpendicular to filaments while the spin of lower mass haloes is more likely to be aligned with them.  
}
\label{fig:HaloFilamentAlign}
\end{figure*}

In order to investigate the effect of the large-scale environment on galaxies' spins and shapes,  the filaments and walls of the cosmic web are extracted using a ridge extractor topological algorithm, {\sc DisPerSE} \citep{2013arXiv1302.6221S}, which is publicly available \footnote{\href{{http://www.iap.fr/users/sousbie/disperse.html}}{http://www.iap.fr/users/sousbie/disperse.html}}. 
The persistent skeleton  presented in \cite{2011MNRAS.414..350S} 
 is a generalisation of the skeleton picture, in which
Morse theory was put forward by \cite{novikov,sousbie08,pogo09,sousbie09} to define the skeleton as the set of critical lines joining the maxima of the (density) field through saddle points following the gradient,
 but now for discrete tracers. For this purpose,  {\sc DisPerSE}  uses discrete homology  to build the so-called discrete Morse-Smale complex on the point process. It then identifies ascending manifolds of dimension one and two as the filaments and walls of the cosmic web. The notion of smoothing scale is replaced by topological persistence which allows it to assign a level of significance  to each topologically connected  pair of critical points. This effectively  mimics an adaptive smoothing depending on the local level of noise. This multiple-scale analysis is an asset  when attempting to extract robustly 
a weak alignment signal from noisy data, especially in the context of the large scale structures, which are intrinsically multi-scale.
The caveat arises when comparing to theory which is more easily implemented at fixed scale \citep[see e.g.][and Appendix~\ref{app:ATTT}]{pogo09}.

In this paper,
 the persistent cosmic web is extracted from a Delaunay tessellation of the galaxy or halo distribution for different thresholds of persistence from $N_\sigma=1$ to $8$. This procedure removes any persistence pair with probability less than $N_\sigma$ times the dispersion to appear in a Gaussian random field and therefore filters out noisy structures, keeping only the most prominent structures as $N_{\sigma}$ increases.
As an illustration, Figure~\ref{fig:walls3D} shows the walls and filaments  as extracted by {\sc DisPerSE} for $N_\sigma=5$  in the redshift $z=0$ snapshot,
 while Figure~\ref{fig:Filaments} displays a slice of the simulation with the corresponding filaments and walls.

\begin{figure*}
\includegraphics[width=0.95\columnwidth]{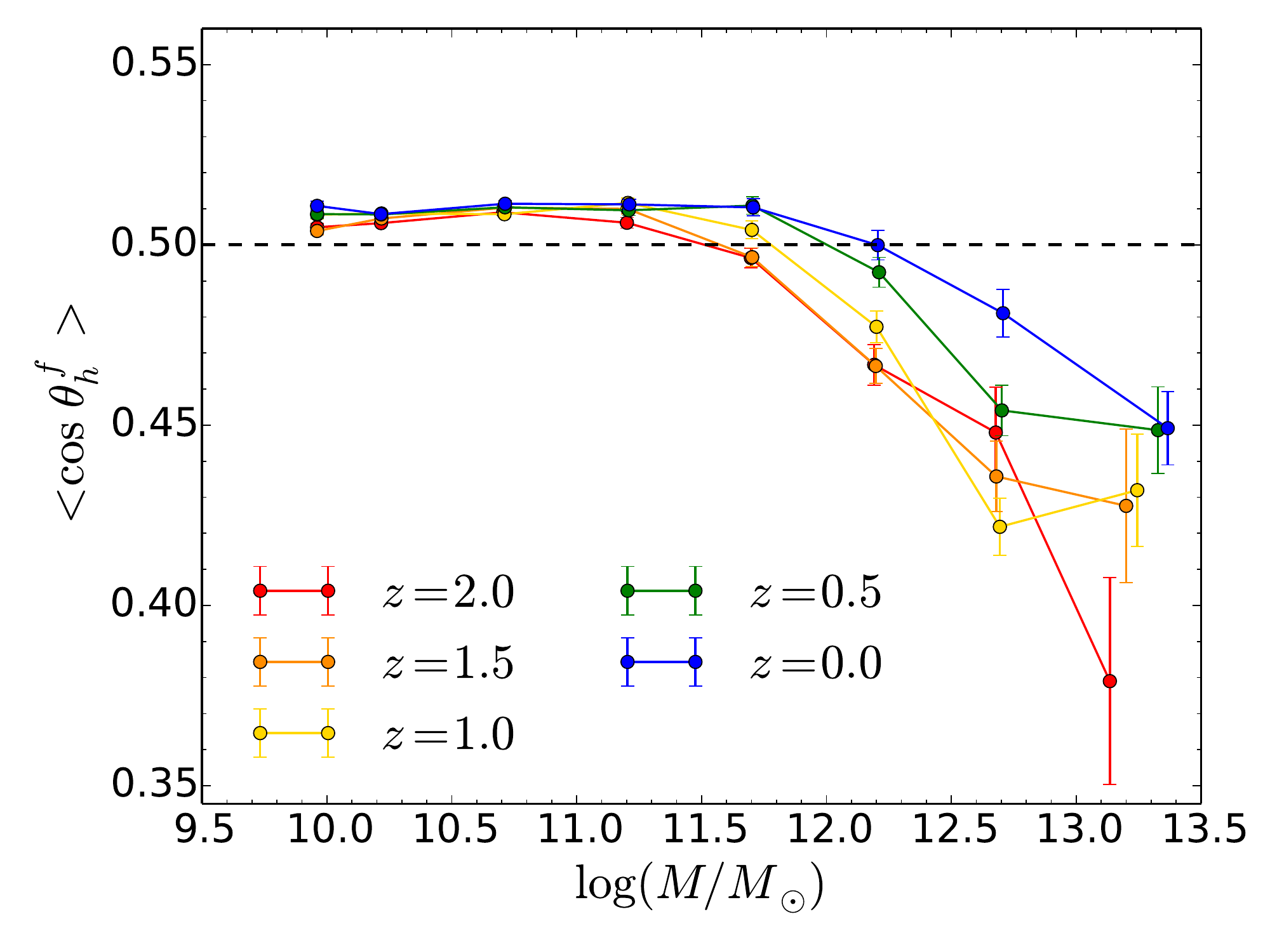}
\includegraphics[width=0.95\columnwidth]{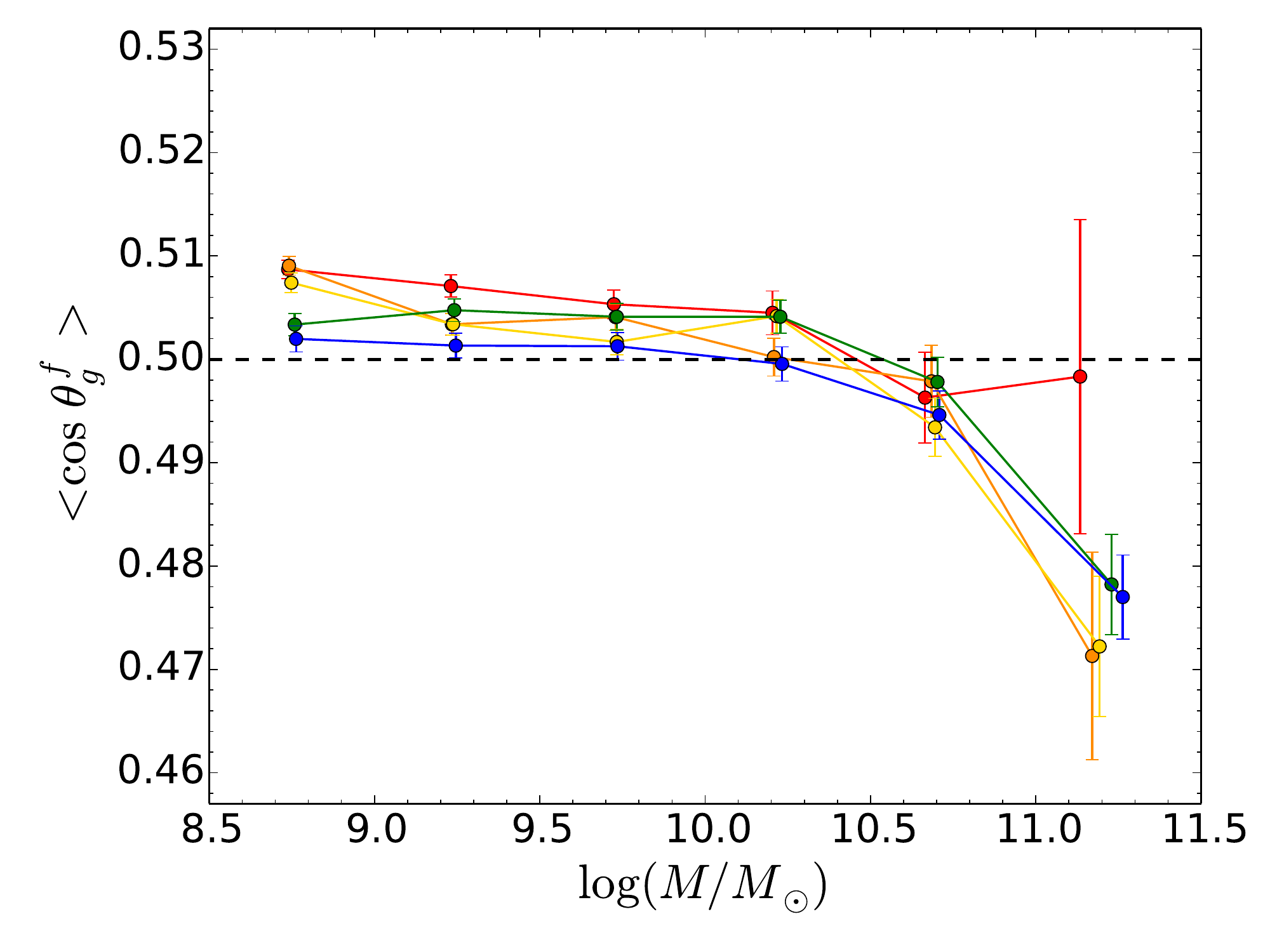}
\includegraphics[width=0.95\columnwidth]{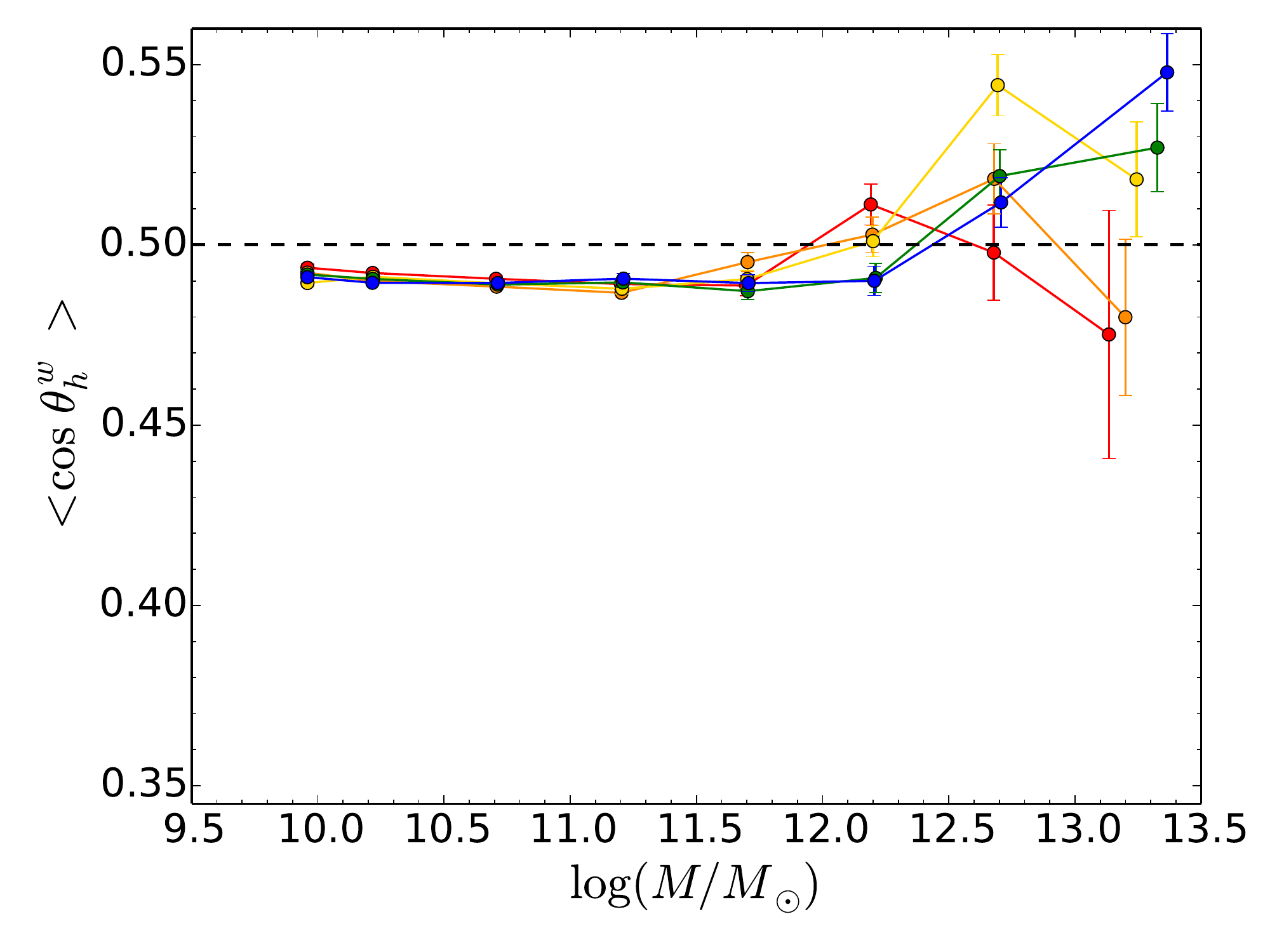}
\includegraphics[width=0.95\columnwidth]{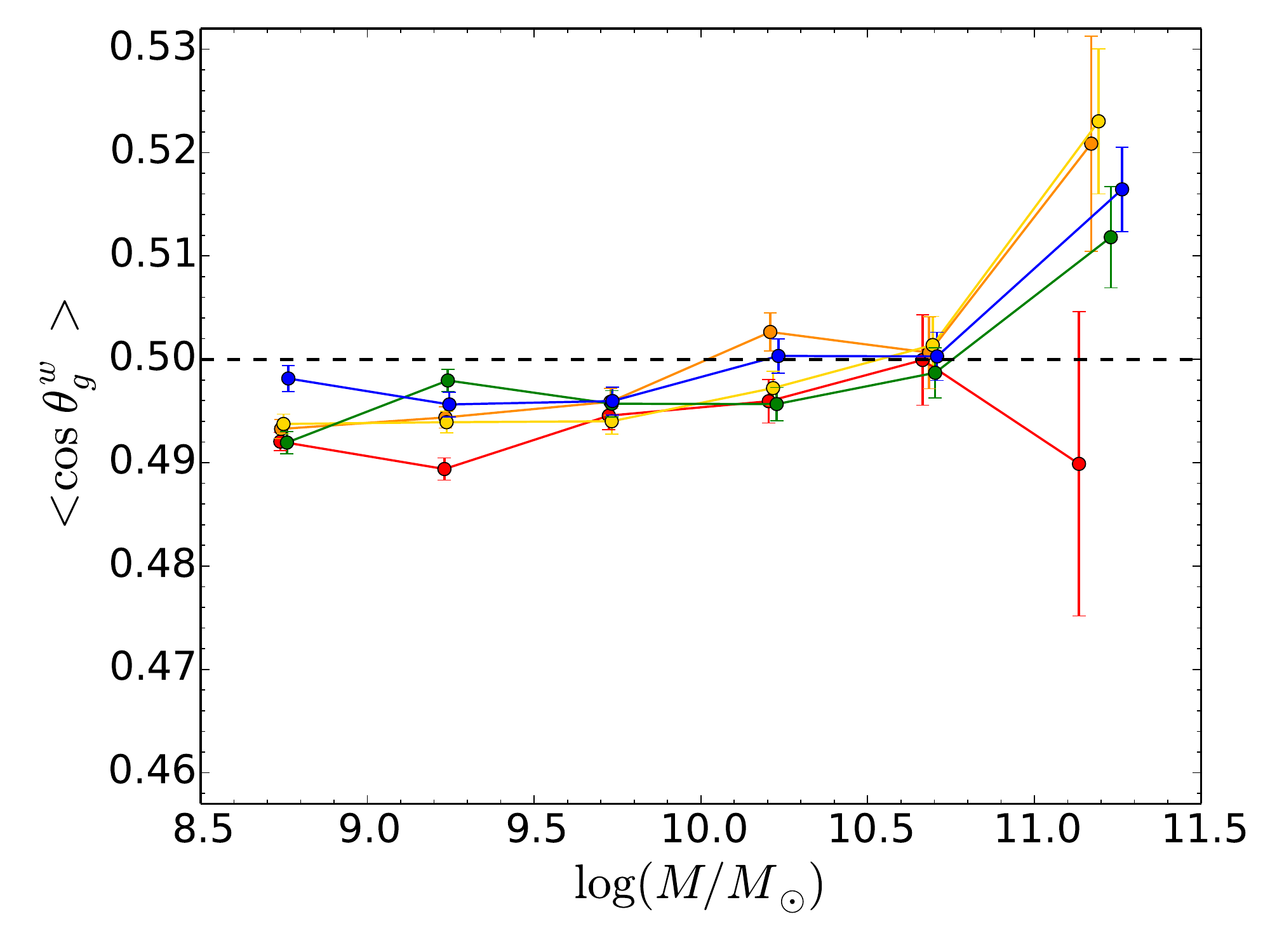}
\caption{Alignment between the spins of objects and the environment, with haloes on the left-hand panels and galaxies on the right-hand panels, and alignment with filament axes for the top panels and with normal vectors to walls for the bottom panels. A persistence level of $N_{\sigma}= 5$ is used to extract the skeleton from the distribution of galaxies. The x-axis values are defined as the mean mass in each mass bin defined in Figure~\ref{fig:HaloFilamentAlign}. Note the difference in y-axis range for haloes and galaxies. }
\label{fig:CosThetaVsMass}
\end{figure*}

\begin{figure*}
\includegraphics[width=0.95\columnwidth]{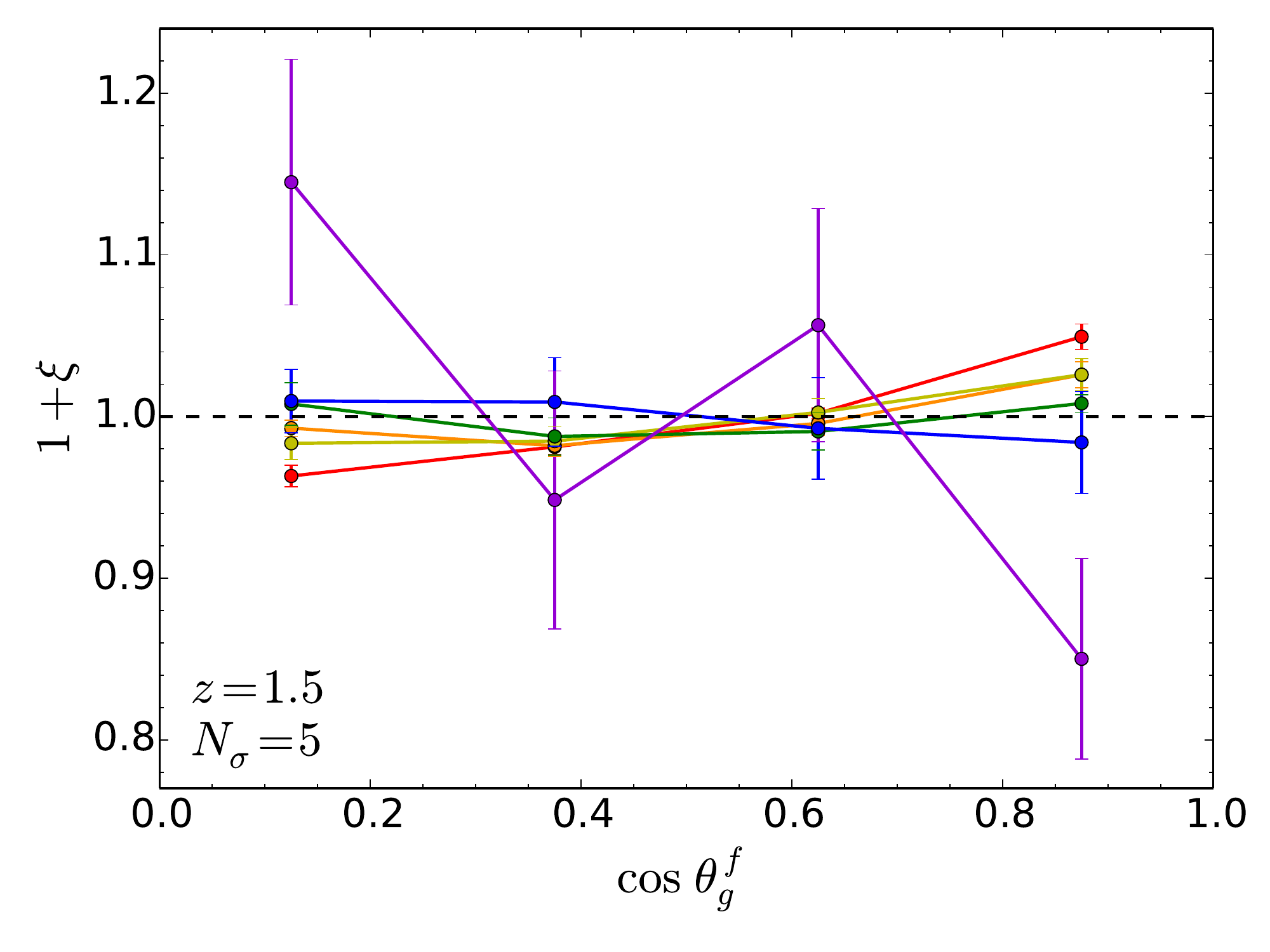}
\includegraphics[width=0.95\columnwidth]{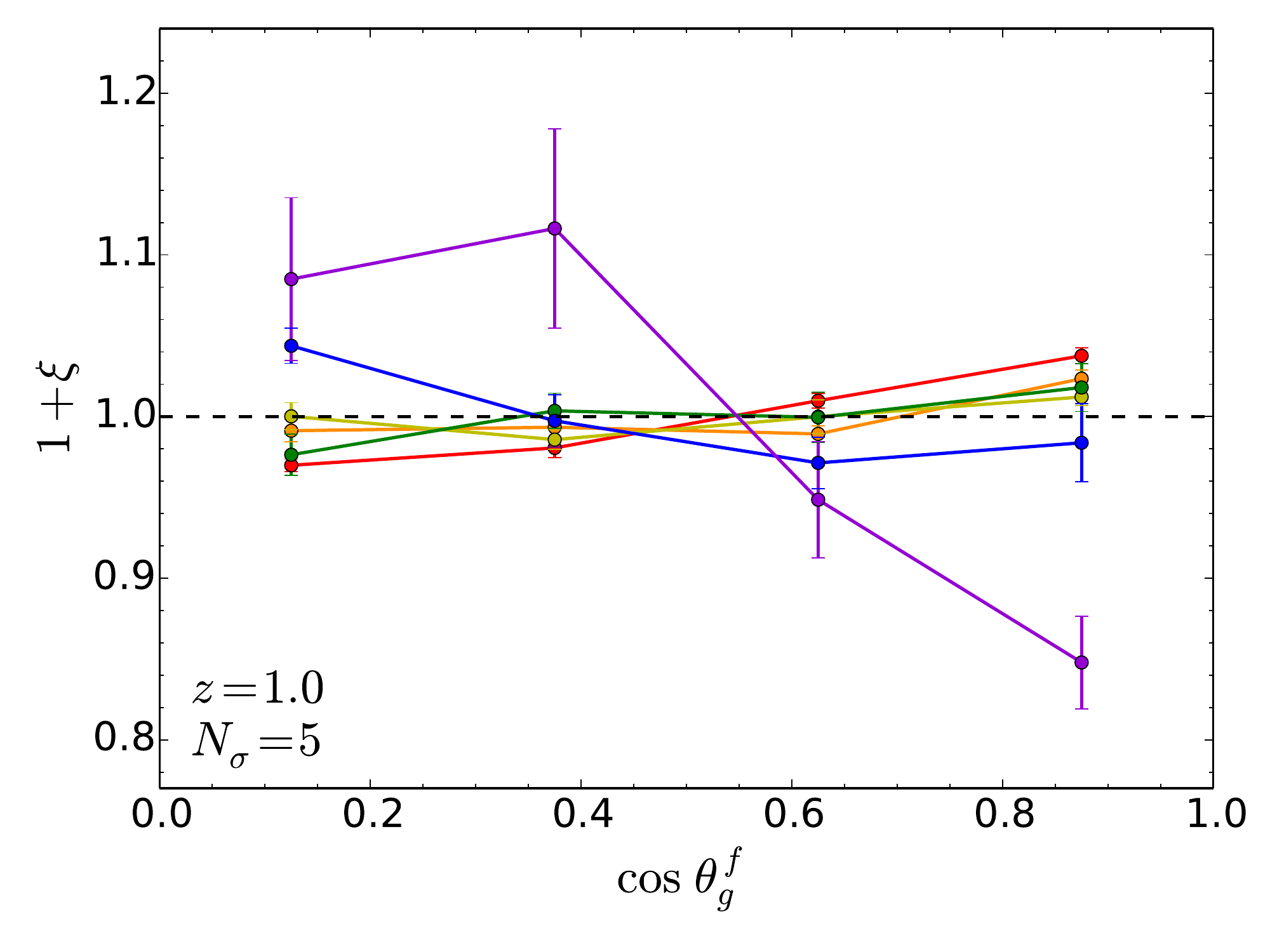}
\includegraphics[width=0.95\columnwidth]{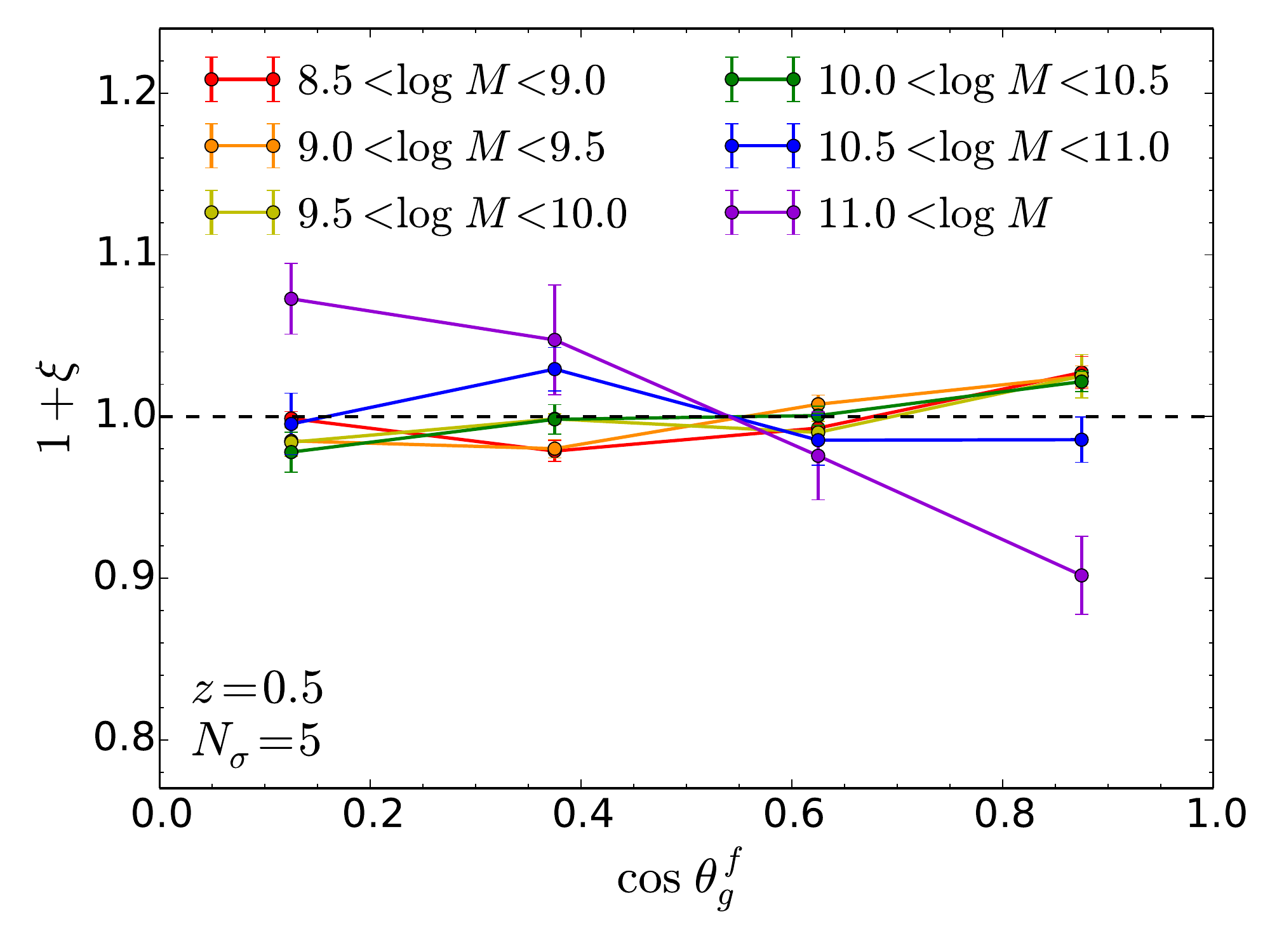}
\includegraphics[width=0.95\columnwidth]{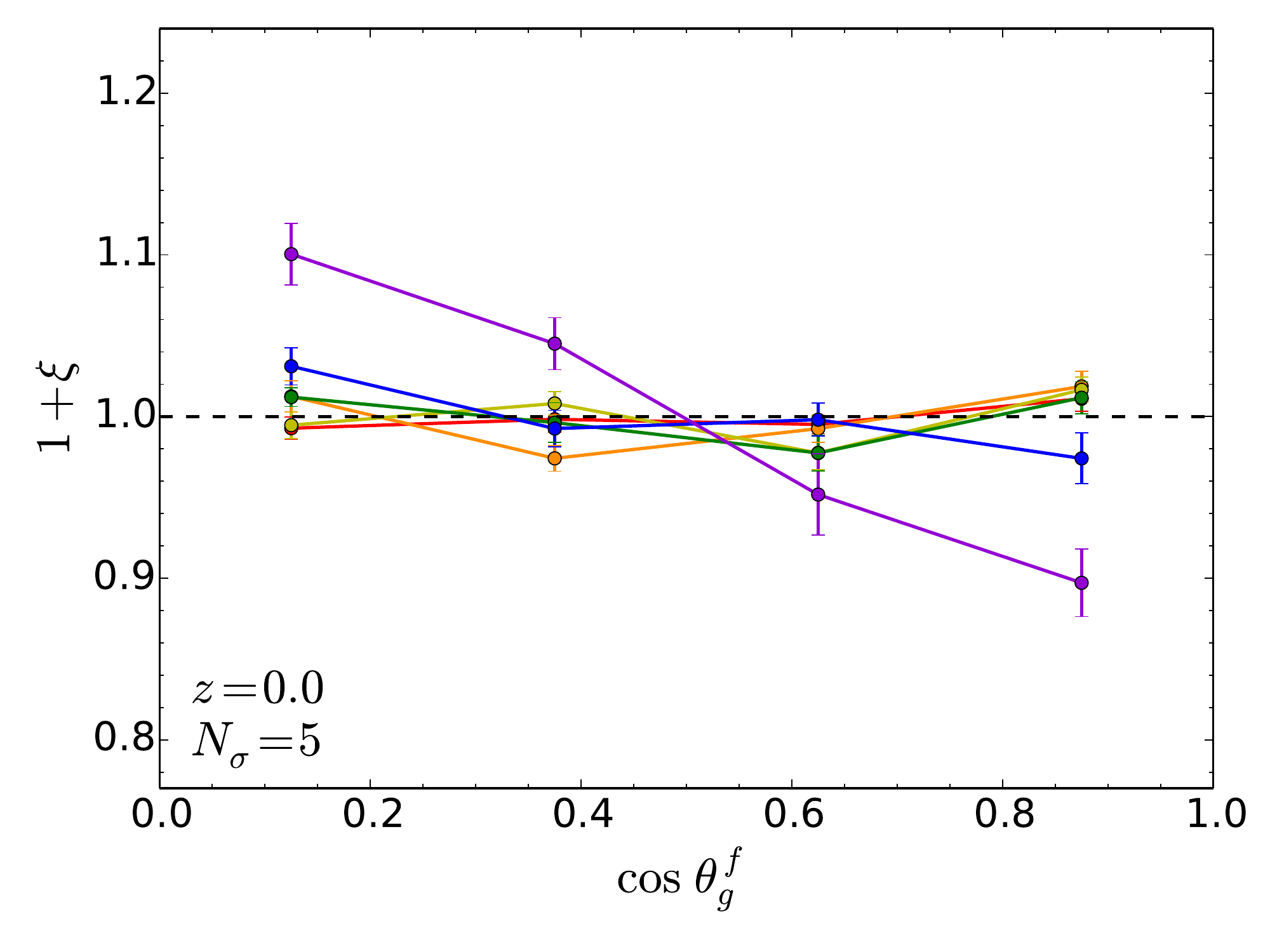}
\caption{Alignment between the spin of galaxies and the closest cosmic filament structure for a persistence level of $N_{\sigma}$ = 5 and for different redshifts from $z=1.5$ (top left-hand panel) to $z=0$ (bottom right-hand panel) and different stellar masses from red to purple as labelled. Massive galaxies are more likely to align their spin perpendicular to the filament axis while the spin of less massive ones tend to be parallel to filaments in particular at high redshifts.
}
\label{fig:GalFilamentAlign}
\end{figure*}

\section{Spin flips along filaments}
\label{sec:spin-fil}

Let us quantify in this section the redshift evolution of spin alignment with respect to the cosmic web, first for dark matter haloes, then  galaxies.  These alignments will be quantified in particular as a function of mass and morphology.

\subsection{Extracting filaments}
First, cosmic filaments are extracted from the simulation. For this purpose,
the positions of the galaxies are fed into the {\sc DisPerSE} algorithm which extracts the cosmic web,
walls and filaments included. Eventually, filaments are given as a set of connected points, and segments are pairs of consecutive points whose direction will be computed from the separation vector of those two consecutive points.
Let us emphasize that only filaments based on the galaxy catalogue are considered as this is closer to what could be done in observational datasets. Appendix~\ref{app:skel} presents a comparison between skeletons extracted from galaxy and halo calalogues: the use of one or the other does not make any significant difference for this work.
The topological features  are computed for various persistence levels to control the level of robustness desired but only results obtained for a persistence of 5$\sigma$ are shown in the main text. 
 The dependence of the result on the persistence threshold is discussed in Appendix~\ref{sec:persistence} and shown to be rather small, in agreement with results found from pure dark matter simulations \citep{codisetal12}.

For each galaxy or halo, the closest segments of the persistent skeleton are found and   the cosine $\cos\theta_{h,g}^{f}$ of the angle between the filament axis and the spin of the object  ($h$ for haloes, $g$ for galaxies)  is measured. This cosine is defined to be always positive (we do not attribute an orientation to the filaments, although this could be done depending on the relative masses of its endpoints).  The histogram of the measured values of $\cos\theta_{h,g}^{f}$ is computed for different bins of mass and rescaled by the total number of objects times the size of the bin to get a PDF. The results were shown to be insensitive to the number of closest segments (1, 2 or 3) chosen to define the filament's orientation. The remainder of the paper  always considers the two closest segments to define the orientation of the closest branch of skeleton. 
Error bars are always estimated as the error on the mean obtained from 8 subcubes of the simulation.

\subsection{Dark matter haloes}
\label{sec:filhalo}
Let us first focus on the orientation of the spin of dark matter haloes with respect to their closest filament.

Figure~\ref{fig:HaloFilamentAlign} shows the PDF of the alignment angle, $\cos\theta_{h}^{f}$, between halo spins and filament directions for different redshifts and masses. At all redshifts,   
a transition from alignment to orthogonality is detected: low-mass haloes tend to have a spin aligned with the axis of their host filament while more massive ones are more likely to have a spin perpendicular to the closest filamentary structure. The spin flip occurs roughly around a transition mass ${M}^{h}_{\rm tr}\approx10^{12}M_{\odot}$. 
From the three highest mass bins probed in this work (violet, blue and green curves), it is clear that the transition mass increases towards low redshift. These findings are in very good agreement with previous studies based on much larger  pure dark matter runs like the Horizon $4\pi$ simulation \citep{teyssieretal09}, where the redshift evolution of this transition mass was found to be
$\hat M^h_0 \approx 5(\pm 1)\times 10^{12} M_\odot$ at redshift 0 and decrease with $1+z$ following a power law \citep{codisetal12}. 
This paper also reported a (small) dependence of the transition mass with the scale at which the skeleton was computed.  Appendix~\ref{sec:persistence} investigates how our result evolves with scale by means of the persistence level.
 No significant  transition mass  is  detected here against the persistence level of the skeleton, probably due to the smaller volume hence  poorer statistics of the full physics Horizon-AGN simulation compared to that of the pure DM Horizon $4\pi$ simulation.

Interestingly, the amplitude of the correlation  is of the order $\sim 5-40\%$ at high redshift and $\sim 5-20\%$ at low redshift, which is consistent with results obtained from pure dark matter simulations. 
This is somewhat surprising, as one would naively expect some level of de-correlation on smaller scales; it
indicates that even if baryonic physics was shown to affect the orientation of halo spins \citep[see e.g][]{2017MNRAS.472.1163C}, it does not significantly change the mean alignment between halo spins and filaments,
{  which is mainly sensitive to the outer regions of haloes that are less affected by the presence of baryons}.

\begin{figure}
\includegraphics[width=0.95\columnwidth]{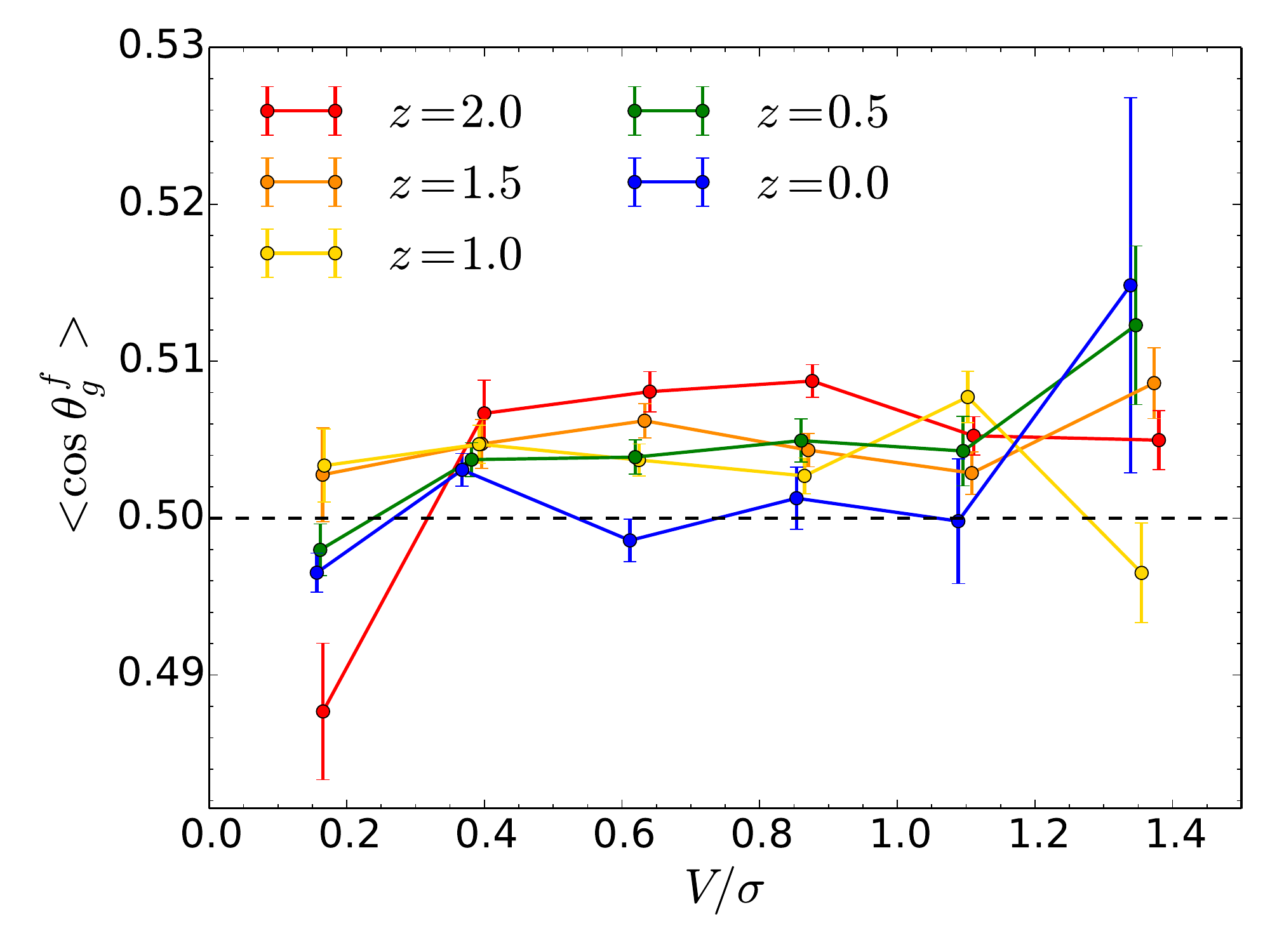}
\includegraphics[width=0.95\columnwidth]{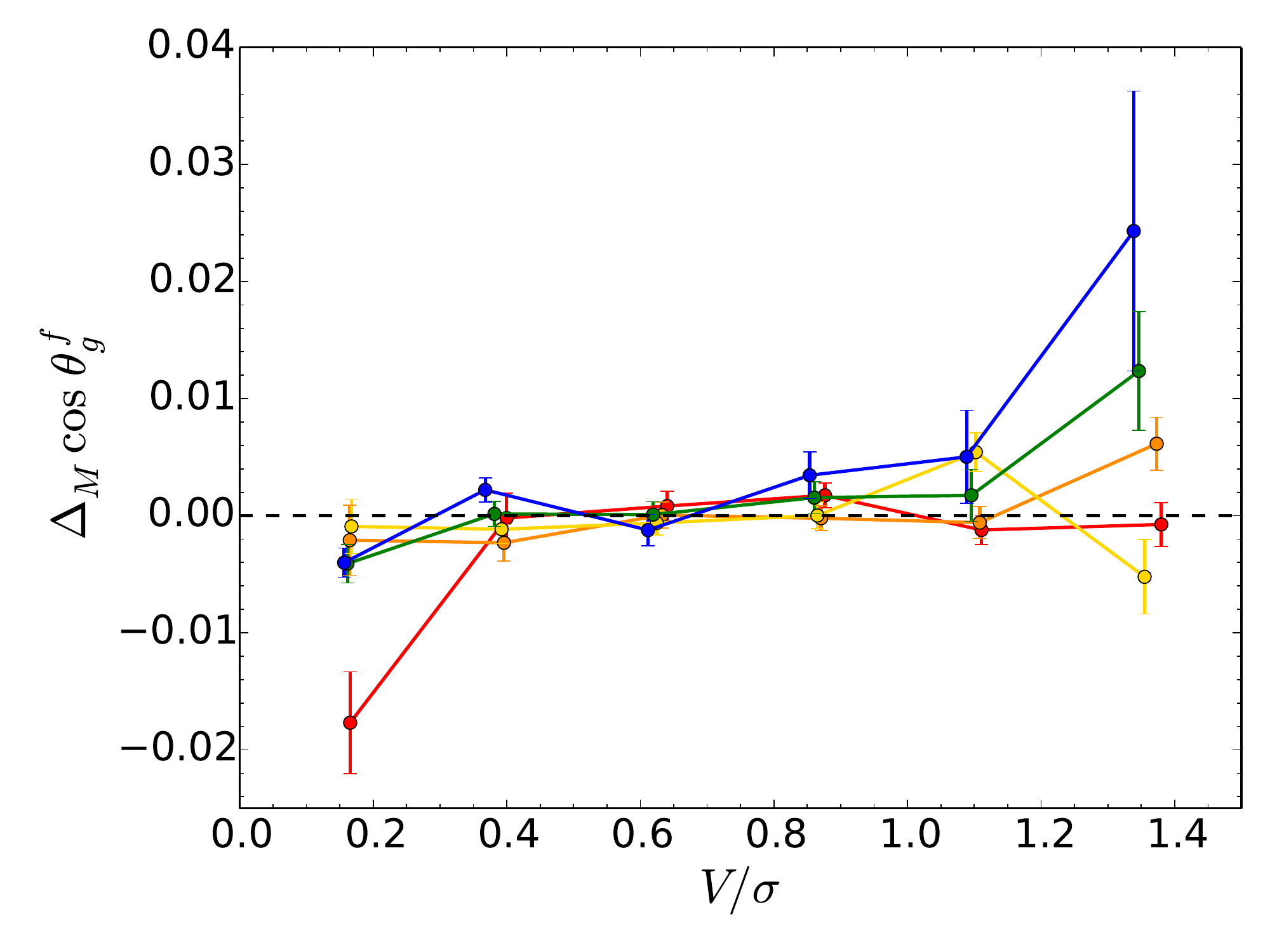}
\caption{Top panel: Average alignment between galactic spins and filaments as a function of $V/\sigma$, for a skeleton with persistence level $N_{\sigma}=5$ and different redshifts as labelled. Bottom panel: Residuals compared to the mean mass-alignment relationship as described in Section~\ref{sec:GalFil}. At fixed mass, elliptical galaxies tend to have a spin more perpendicular to the filaments and spirals parallel.
}
\label{fig:VoSResid}
\end{figure}

Let us also investigate the mass and redshift dependence of the alignment by focusing on the mean alignment angle. This is shown in the top left panel of Figure~\ref{fig:CosThetaVsMass}. As expected, there is again a very clear downward trend, indicating that the spins of more massive haloes tend to be perpendicular to the filament directions while the less massive haloes have a spin more parallel to the axis of the filament. The redshift evolution clearly indicates that the transition mass (when the cosine equals 0.5) increases toward lower redshift (in blue). This transition mass can be compared to the typical mass collapsing at a given redshift. Appendix~\ref{sec:Mtr} 
shows that the transition mass is close to the non-linear mass at redshift 0 but becomes larger for increasingly large redshift, in agreement with the findings of \cite{codisetal12} using a pure dark matter simulation. The redshift evolution is shown to follow closely a power-law behaviour
  \begin{equation}
  \label{eq:fit}
 M_{\rm tr}^{h}(z)\approx M_{0}^{h}(1+z)^{-\gamma_{h}},
 \end{equation}
 with $\log M_{0}^{h}=12.2\pm0.1$ and $\gamma_{h}=1.6\pm0.3$. The value of the $\gamma_h$ index cannot directly be compared with the Horizon-4$\pi$ simulation as the scales probed and the strategy for extracting the skeleton differ from our work. 
 
Note that a maximum mass of alignment around $10^{11}M_{\odot}$ is also marginally detected   (below which the alignment decreases) which was shown to correspond to the typical size of coherent flows of vorticity in filaments \citep{laigle2014}.

\subsection{Galaxies}
\label{sec:GalFil}

As haloes are not observable directly and given that there is some misalignment between galaxy and halo as discussed in Appendix~\ref{sec:halo-gal}  \citep[see also][]{Ten++14,2015MNRAS.453..721V,2017MNRAS.472.1163C}, let us  now investigate whether galaxies also retain some memory of the anisotropic environment in which they form.
Similarly to the procedure followed for dark haloes, for each galaxy  the angle between the spin of its stellar population and the two closest segments of filament is measured.

\begin{figure*}
\includegraphics[width=0.95\columnwidth]{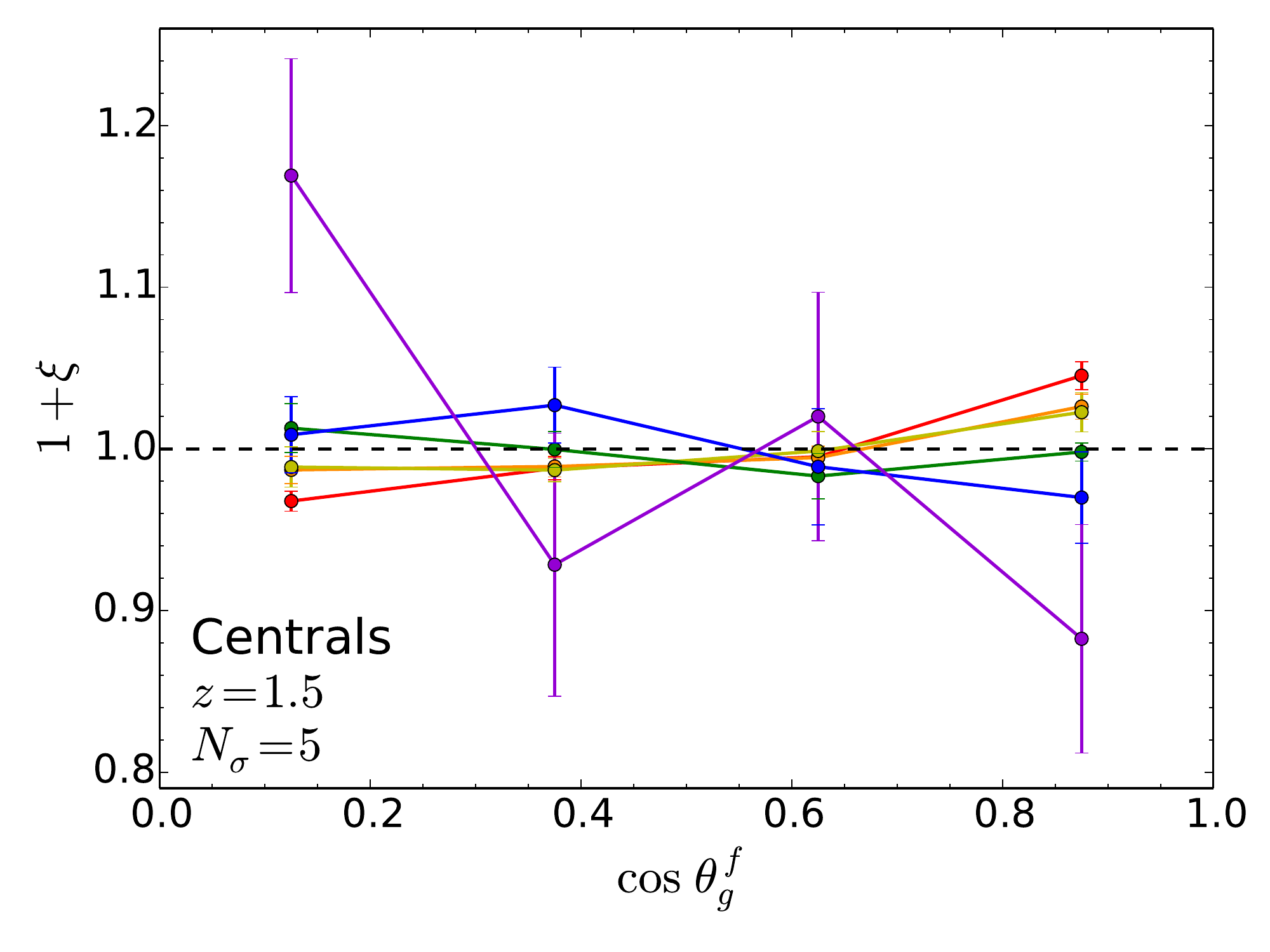}
\includegraphics[width=0.95\columnwidth]{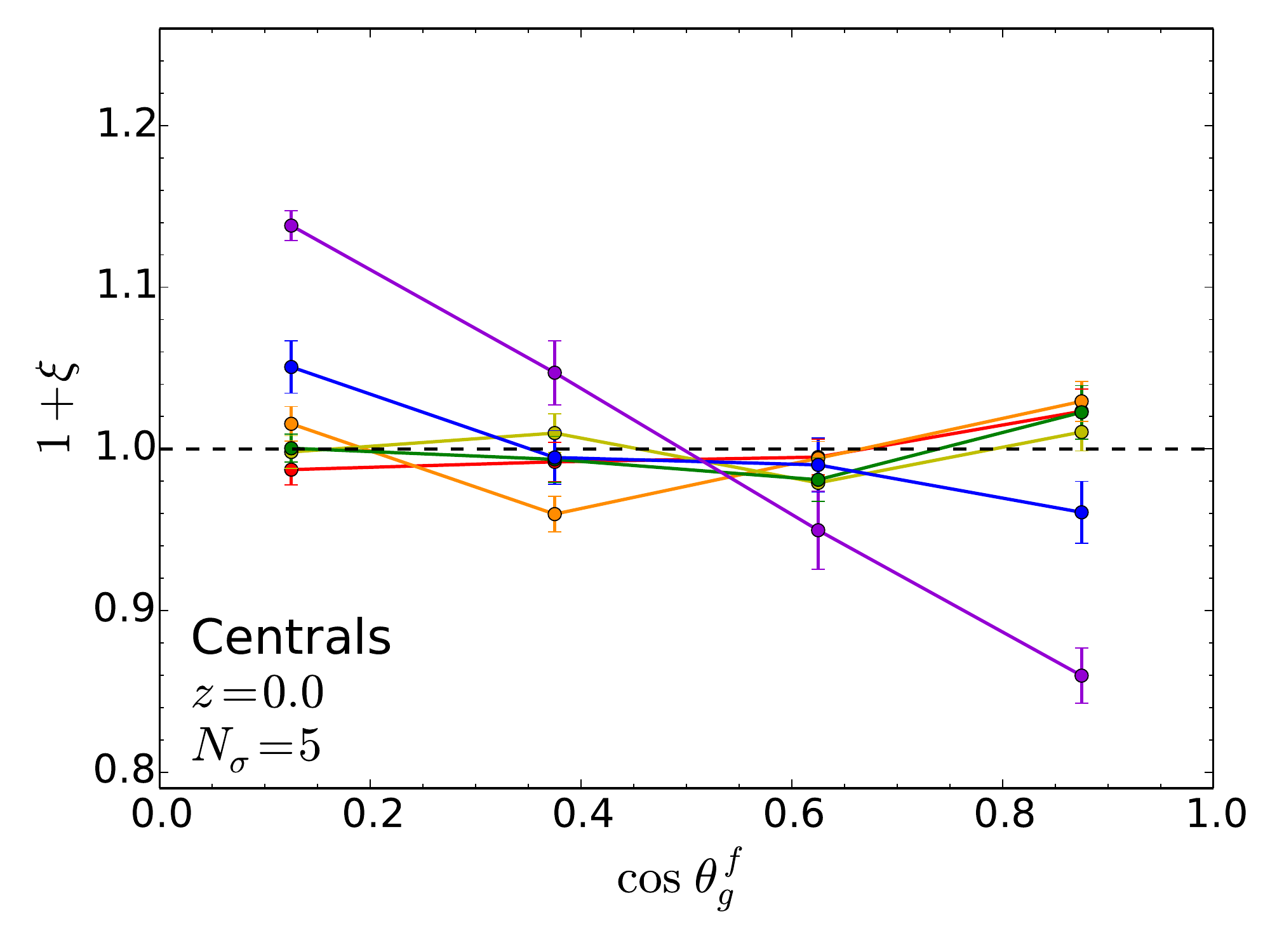}
\includegraphics[width=0.95\columnwidth]{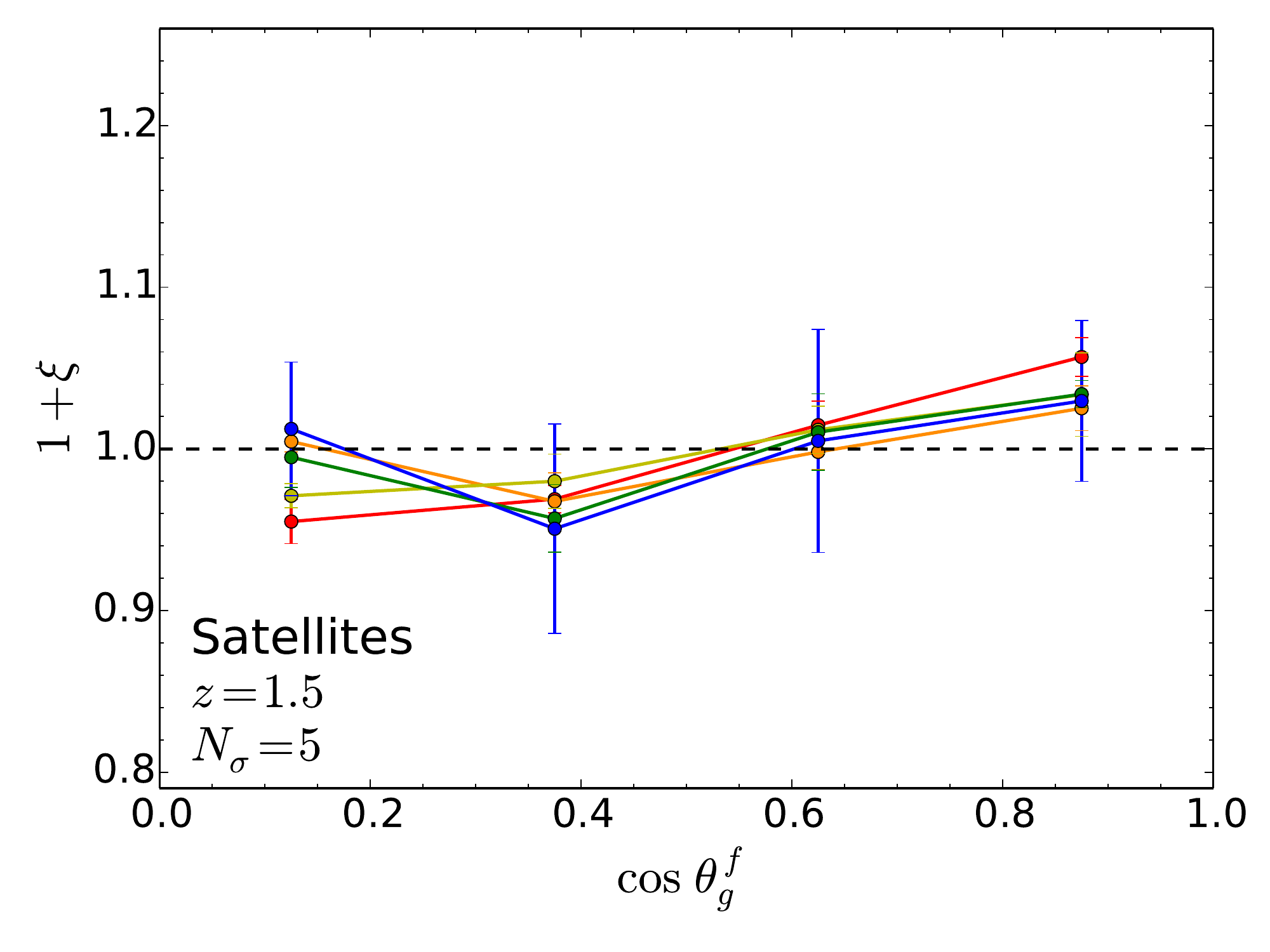}
\includegraphics[width=0.95\columnwidth]{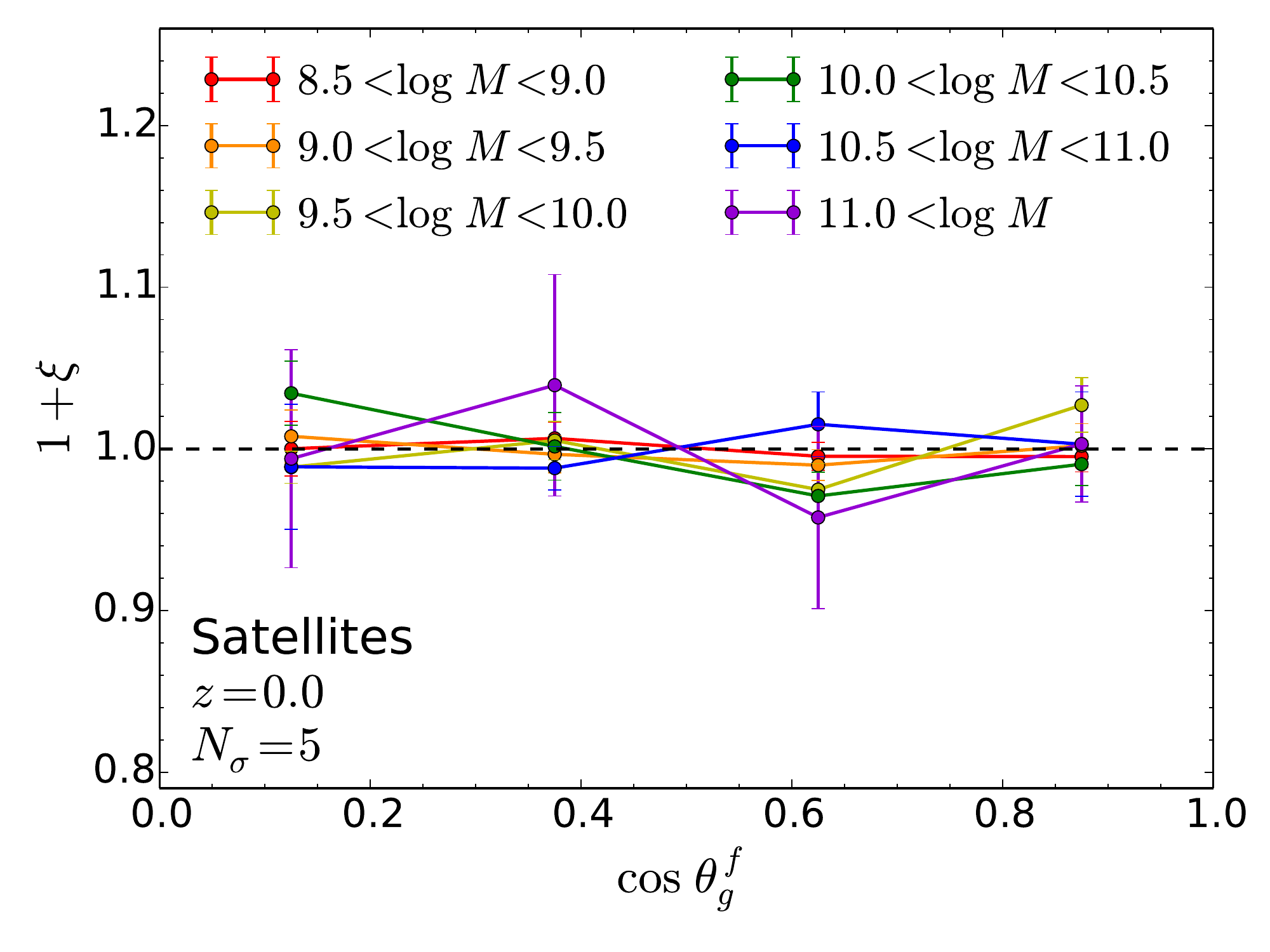}
\caption{PDF of the cosine of the angle between the spin of galaxies and nearby filament directions for either central galaxies (top panels) or satellites (bottom panels) for different redshifts: $z=1.5$ (left-hand panels) or $z=0$ (right-hand panels). Most of the alignment signal at low redshift comes from central galaxies. At higher redshift,   a clear alignment of the spin of satellites is detected.}
\label{fig:CentralVsSatellite}
\end{figure*}

The resulting PDF of the cosine of the angle between galactic spins and their host filaments is shown in Figure~\ref{fig:GalFilamentAlign} and the redshift and mass evolution of the mean angle is displayed on the top right panel of Figure~\ref{fig:CosThetaVsMass}.
As for haloes,  a robust clear spin flip of galaxies is detected,
consistently with the weaker detection of \cite{dubois14} at high redshift (which claimed a $\sim 0.7$ sigma detection at $z=1.8$).  An  almost 5 sigmas  detection\footnote{At redshift zero,   15\,199 cosines above 0.5 and 16\,042 below 0.5 are measured for galaxies with logarithmic mass above 10.3, which is a 4.8 sigmas deviation from the uniform case. Indeed, if N numbers are uniformly drawn between 0 and  1, one expects to find on average N/2 draws above 0.5 with a standard deviation given by sigma=$\sqrt N/2$. } shows here without ambiguity that the spin of low-mass galaxies tends to be aligned with the filament direction while high-mass galaxies tend to have a spin perpendicular to the axis of the closest filament.
The transition between those two regimes occurs at roughly $\log M_{\rm tr}^{g}/M_{\odot}\approx{10.1\pm0.3}$.  

No significant redshift evolution of the transition mass is detected, due to a lack of statistics. Note that the amplitude of the signal for the most massive galaxies ($\sim 10\%$) is similar to haloes while the low mass signal is weaker. For galaxies, the strength of the low-mass alignment diminishes towards low redshift from $\sim 5\%$ at $z=1.5$ to less than $2\%$ at redshift zero. A transition from a high redshift alignment of spins with filaments 
 (in agreement with the high redshift radial alignment of disk-like galaxies found in \citealp{Chisari16})
 to the building up of a population of massive galaxies with a spin perpendicular to filaments at lower redshift
  is therefore detected at the 9 sigmas level \footnote{At z=1.5,  125\,421 cosines above 0.5 and 121\,031 below 0.5 are measured for galaxies with logarithmic mass between 8.5 and 10.3, which is a 8.9 sigmas detection. Note that the total number of cosines  is twice the number of objects since for each one one looks for the two closest segments of filament.}.
  
{    Note that the resolution of the simulation is shown not to affect the spin-filament alignments detected here (see Appendix~\ref{sec:res} for details).}

\subsection{Link between morphology and spin flip}
Let us now investigate whether spin flips along filaments also depend on galaxy properties such as their morphology. 
The alignment between galactic spins and the closest filament is measured for different values of $V/\sigma$.
The top panel of Figure~\ref{fig:VoSResid} shows the mean alignment angle as a function of $V/\sigma$ to understand the morphology evolution. 
Since morphology (by means of $V/\sigma$ here) is correlated to stellar mass, the focus is now on quantifying  whether there is any residual alignment trend, 
beyond mass.
To do so,   $\cos\theta_{g}^{f}$ is computed for each galaxy, and  the corresponding interpolated average alignment is subtracted as a function of mass. This residual is denoted as $\Delta_{M}\cos{\theta_{g}^{f}}$. In practice,  a second order interpolation is used when calculating the residuals.
Note that the analysis is restricted to galaxies with more than 100 particles here, which roughly corresponds to a stellar mass of $\sim 3.4 \times10^{8}M_{\odot}$.
The result is shown in the bottom panel of Figure~\ref{fig:VoSResid} which shows significant residuals. At fixed mass, disk-like galaxies tend to align their spins with the axis of the filament while the spin of ellipticals is more likely to be perpendicular to the direction of the closest filament which means that massive ellipticals are driving the spin flip. This result is redshift-dependent. At high redshift, the residuals are very small meaning that all the information is contained in the mass but the excess correlation  increases toward low redshifts when spin flips occur. Appendix~\ref{sec:vsig} also displays the full PDF of the alignment between spins and filaments for the largest $V/\sigma$ on the one-hand and smallest $V/\sigma$ galaxies on the other hand, showing again that the alignment of spins perpendicular to filaments is driven by massive elliptical galaxies while the parallel alignment of less massive galaxies is due to the population of spirals.

\begin{figure*}
\includegraphics[width=0.95\columnwidth]{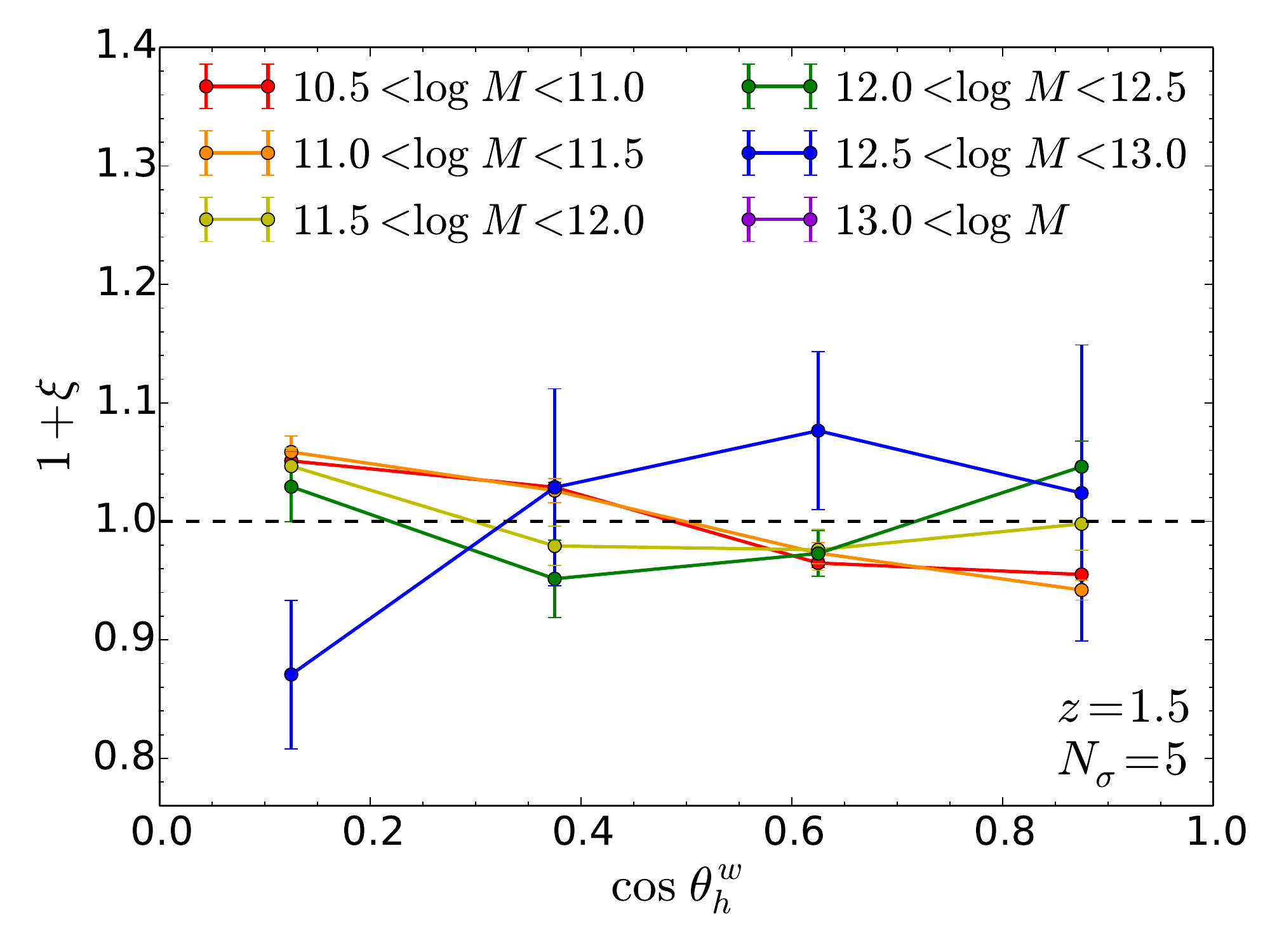}
\includegraphics[width=0.95\columnwidth]{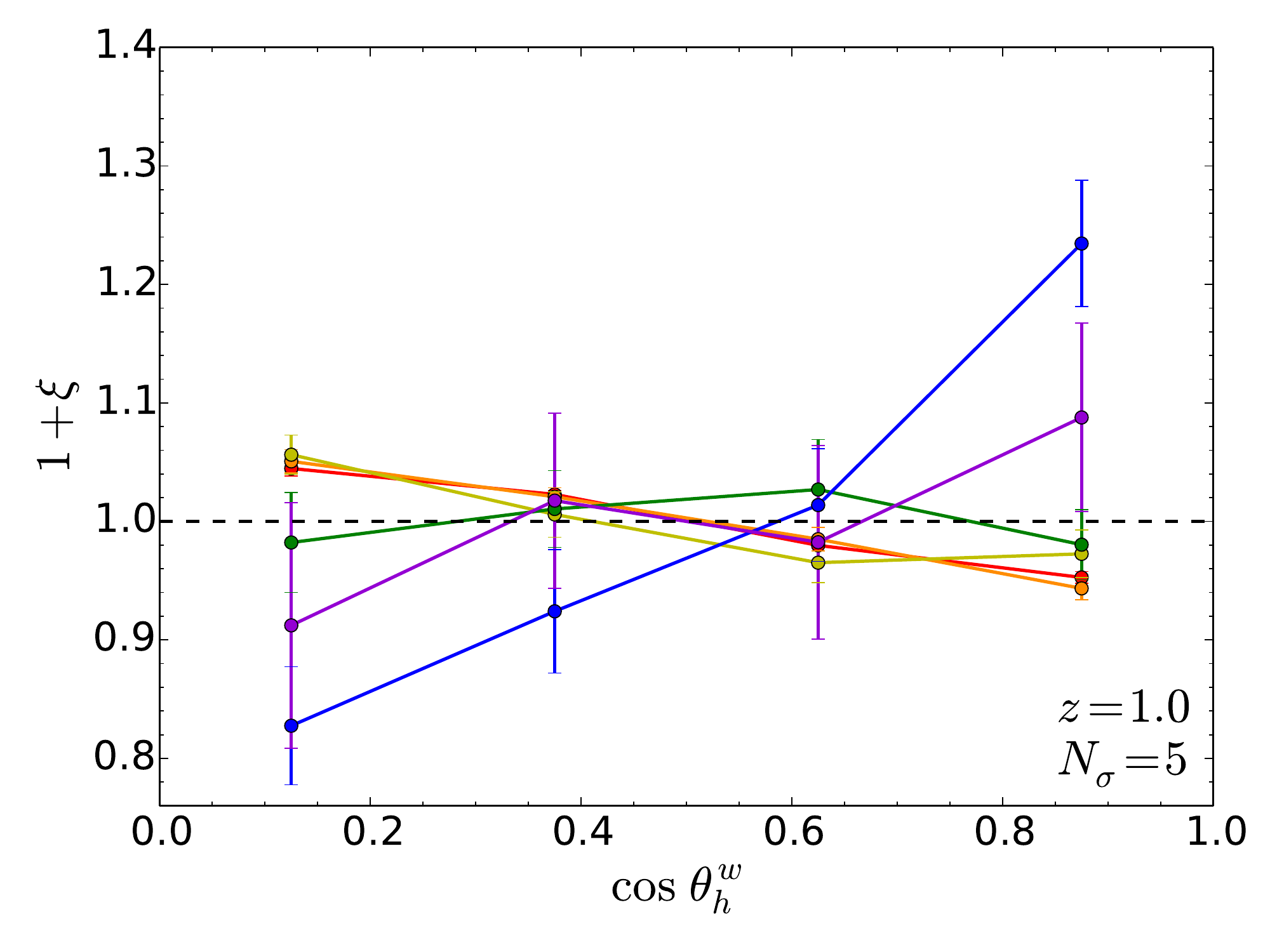}
\includegraphics[width=0.95\columnwidth]{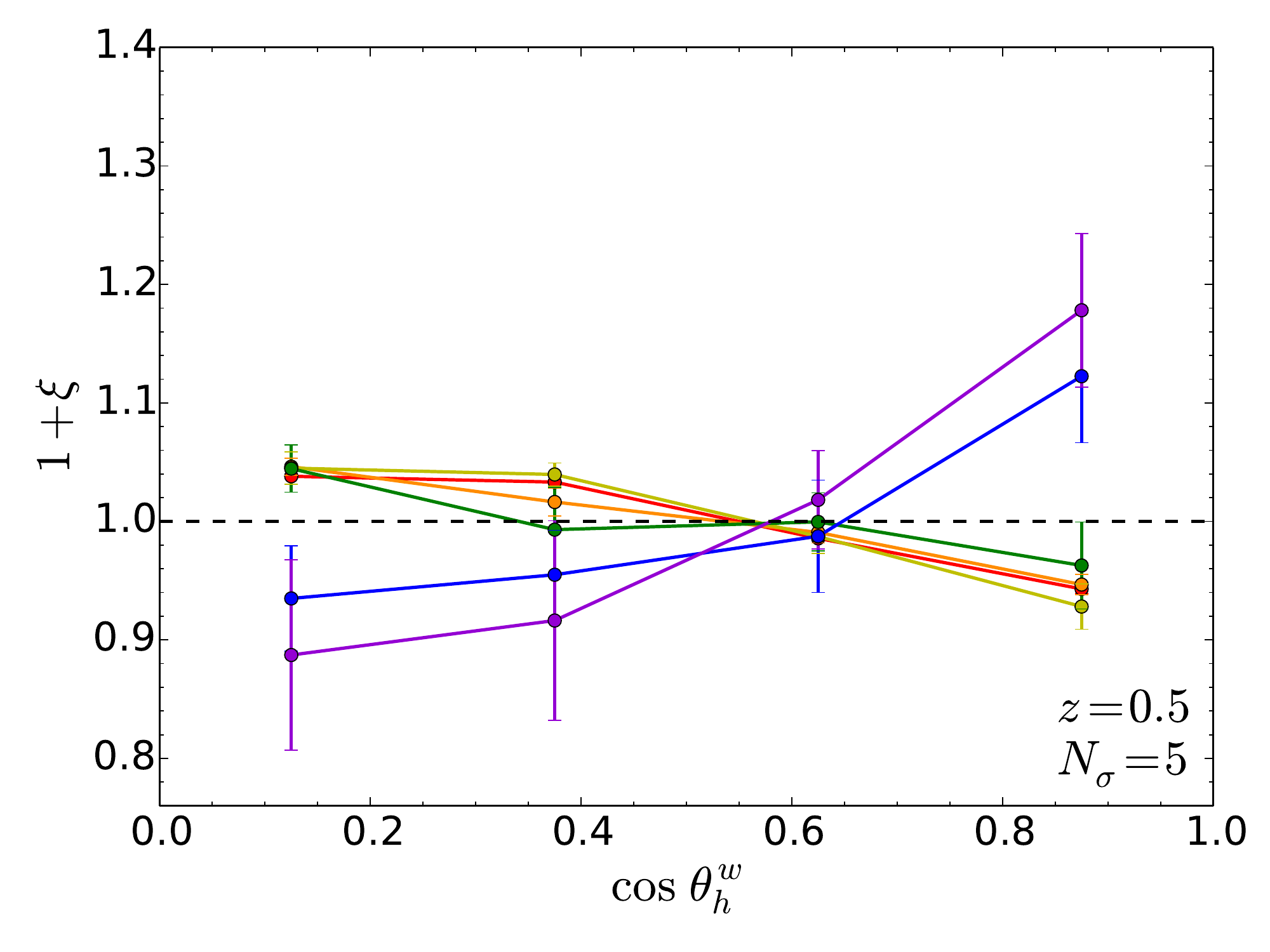}
\includegraphics[width=0.95\columnwidth]{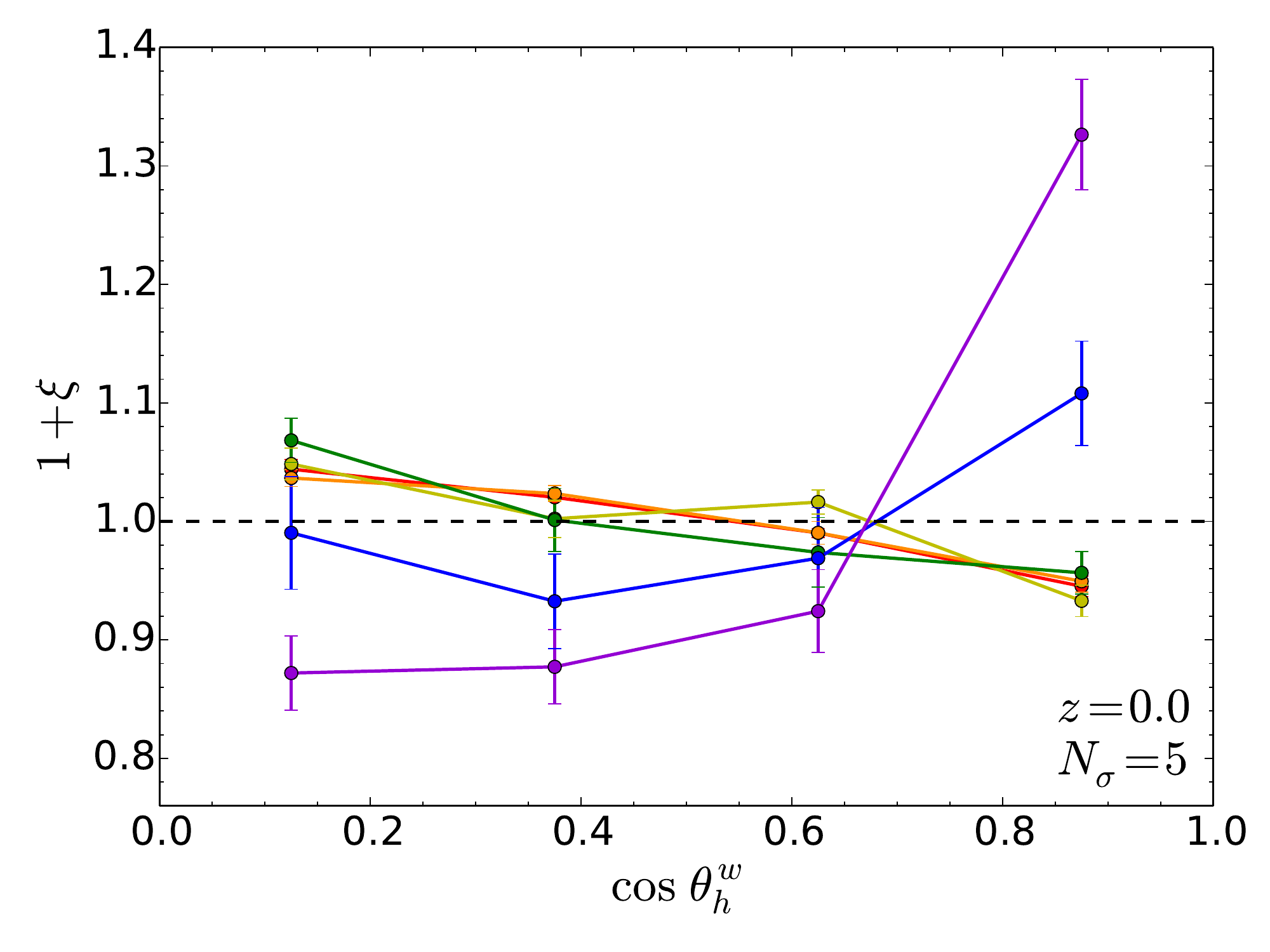}
\caption{PDF of the alignment angle between halo spins and the normal direction of walls for a persistence threshold of $N_{\sigma}=5$, for different redshifts from $z=1.5$ (top left-hand panel) to $z=0$ (bottom right-hand panel). Different mass bins are displayed from red to purple as labelled. Massive haloes tend to have a spin perpendicular to the sheets (aligned with the normal) while low-mass haloes are more likely to have a spin lying inside the walls of the cosmic web.}
\label{fig:HaloWallAlign}
\end{figure*}

\subsection{Centrals versus satellite alignments}
There is usually a massive galaxy near the centre of the halo, denoted here the `central' galaxy, and there can be several smaller `satellite' galaxies populating the halo. These satellite galaxies seem to be associated closely with the subhalo structures\footnote{Note that
\cite{2017MNRAS.472.1163C} defined satellites and centrals precisely depending on the level of their matched halo (satellites associated to subhaloes and centrals to haloes). }.
Let us therefore finally investigate the difference between centrals and satellites (see also Appendix~\ref{sec:halo-gal}). 

Referring to the top panels of Figure~\ref{fig:CentralVsSatellite} for centrals, very similar trends are found, with high mass centrals tending to have spins perpendicular to the filaments while low mass centrals are parallel. Note that the amplitude of the perpendicular signal at high mass is enhanced. A consistent transition mass of ${M}_{\rm tr}^{g}\approx10^{10.3}M_{\odot}$ is observed, with a signal of roughly $15\%$ for the more massive galaxies, which is enhanced compared to that of the full set  of  galaxies. On the other hand, for the satellite galaxies in the bottom panels, no transition to perpendicularity  is seen and no mass dependence. Interestingly, though satellites show a tendency to align their spins with filaments at high redshifts, this signal is not present at low redshifts which may indicate that satellites lose the memory of the filaments from which they emerge as virialisation occurs.
This result is consistent with the analysis performed in~\cite{welkeretal15,2017arXiv171207818W} on the same data: satellites, as they plunge into the center of dark matter haloes, tend to reorient their spin so that it is more aligned with that of the central galaxy.
 These measurements  confirm that intrinsic alignments should be modelled differently at large and small scales, where respectively the two- and one-halo contribution dominate.

\section{Spin flips inside walls}
\label{sec:spin-wall}

Having investigated spin flips along the axes of one-dimensional filamentary structures, let us now turn our investigation toward the two-dimensional walls. 
Walls are easier to observe than filaments in real surveys, which is why the study of the interplay between cosmic walls and galaxies has a longer history. Because the cosmic web is by nature multi-scale, walls also represent the loci of unresolved 
filaments. 
\subsection{Extracting walls}
As for the filaments, the positions of the galaxies are fed into the {\sc DisPerSE} algorithm which returns a tessellation of the walls by means of a set of triangles (the counterpart for 2D walls of segments for 1D filaments).  As illustrated in the right-hand panel of Figure~\ref{fig:Filaments}, the walls track closely the overdense envelopes of voids that  the eye can catch. With higher persistence levels, fewer wall structures are kept by  {\sc DisPerSE}. For this analysis, a persistence level of $N_{\sigma}=5$ is set in the main text (as for filaments). For each galaxy or halo,  the two closest triangles that tesselate the walls are sought  (as for filaments  where  the two closest segments were considered).  We then compute the normal vector to these triangles as the cross product of two of their sides.  The cosine of the angle, $\cos\theta_{h/g}^{w}$, between the spin of the structure (haloes and galaxies) and the normal vector to each wall triangle is then measured.

\subsection{Dark matter haloes}

For the dark matter haloes,  the cosine of the angle, $\cos\theta_{h}^{w}$, between their spins and the normal vector of the closest walls is measured, and the corresponding PDF is computed.
The result is displayed in Figure~\ref{fig:HaloWallAlign} for four different redshifts and different halo masses above $10^{10.5}M_{\odot}$ . 
The mean of the PDF against masses and redshifts is  also plotted in the bottom left panel of Figure~\ref{fig:CosThetaVsMass}.
High mass haloes tend to have a spin aligned with the normal vector of the walls, implying that their spin is perpendicular to the plane of the walls themselves. This effect is especially clear at low redshifts. Focusing on the blue and green curves, it is clear that the transition happens at a halo mass of roughly $\log(M/M_{\odot})=12.5\pm 0.3$. The transition mass is increasing with decreasing redshift,  as was observed for filaments in Section~\ref{sec:filhalo}. 
Below this transition mass, the spins lie preferentially inside the walls, a result in agreement with pure dark matter simulations \citep{2012MNRAS.421L.137L}.
Note that at high redshifts, the number of haloes in the most massive samples decreases  significantly,  hence error bars blow up and are not displayed for the purple line at $z=1.5$. 
As filaments are embedded in walls and as we do not make any restriction on the distance of the objects with respect to the filaments and walls, the signal and the value of transition mass are very similar for filaments and walls.
Noteworthily, the strength of the signal for walls goes up to $35\%$ at low redshift, which is stronger than the signal detected for filaments. 
Note also that the signal for low mass objects is almost constant with redshift here, while for filaments it was decreasing with cosmic time. A possible explanation might be the following. First note that we expect \citep{codisetal12} the low mass objects to have a spin either preferentially along the filament or uniformly distributed in the wall depending on their formation time. Indeed, first, one direction collapses to form a wall, and during this process haloes tend to acquire a spin perpendicular to the direction of collapse, i.e. within the wall. Then, the second direction collapses and haloes acquire again a spin perpendicular  to this direction of collapse, hence eventually aligned with the remaining axis which is that of  the filament-to-be. Overall, it is therefore expected that the spin of haloes remains coherently inside the walls across cosmic time but its likelihood to be aligned with the filament, which is one particular direction of the wall, may vary. 
In particular, from Figure~\ref{fig:FilSlices2}, it seems that the skeleton we use at low redshift probes slightly larger scales than its high redshift counterpart which means that a larger part of the small haloes form during the first collapse at low redshift and are therefore less correlated with the filamentary directions.

\begin{figure*}
\includegraphics[width=0.95\columnwidth]{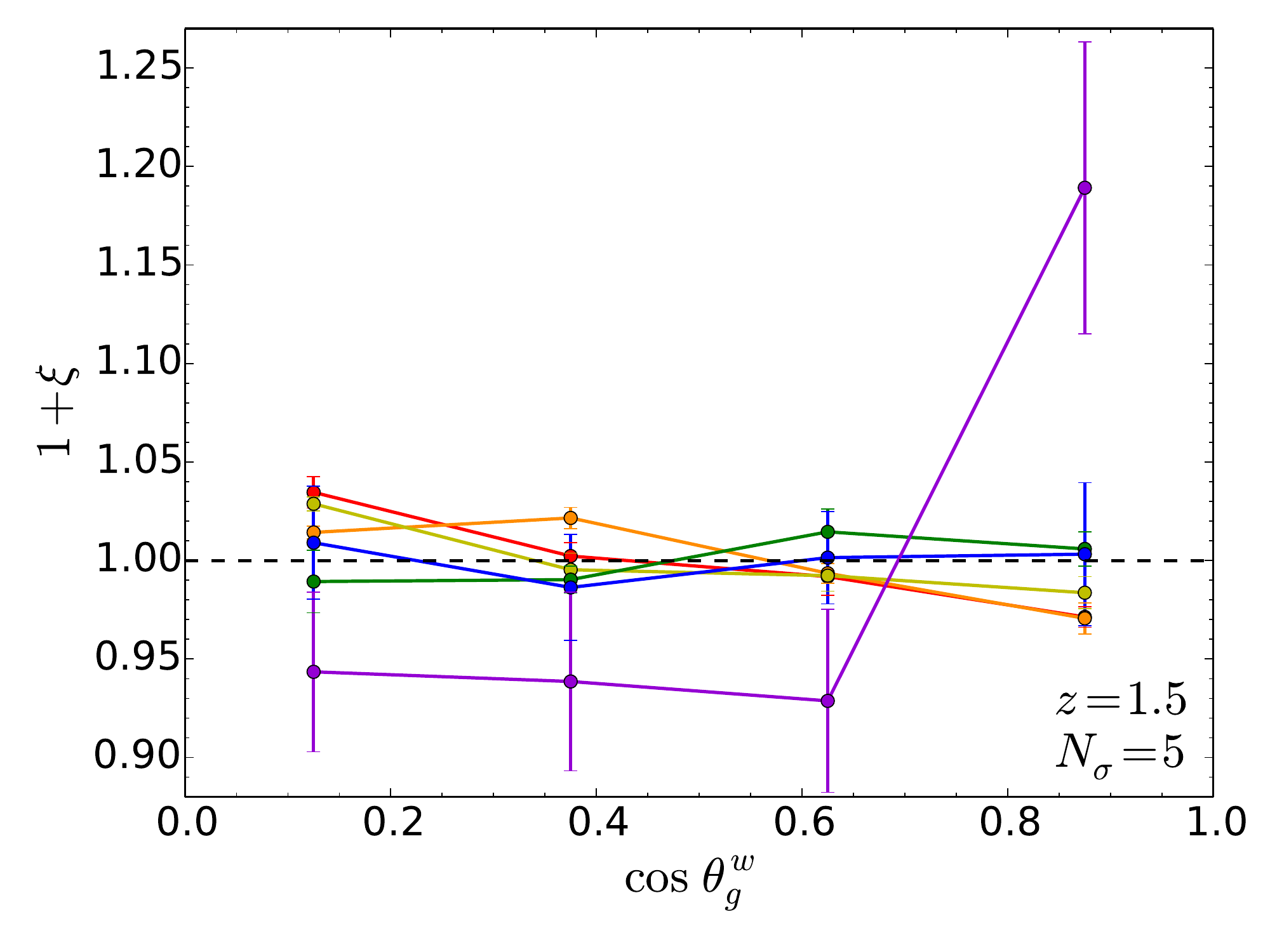}
\includegraphics[width=0.95\columnwidth]{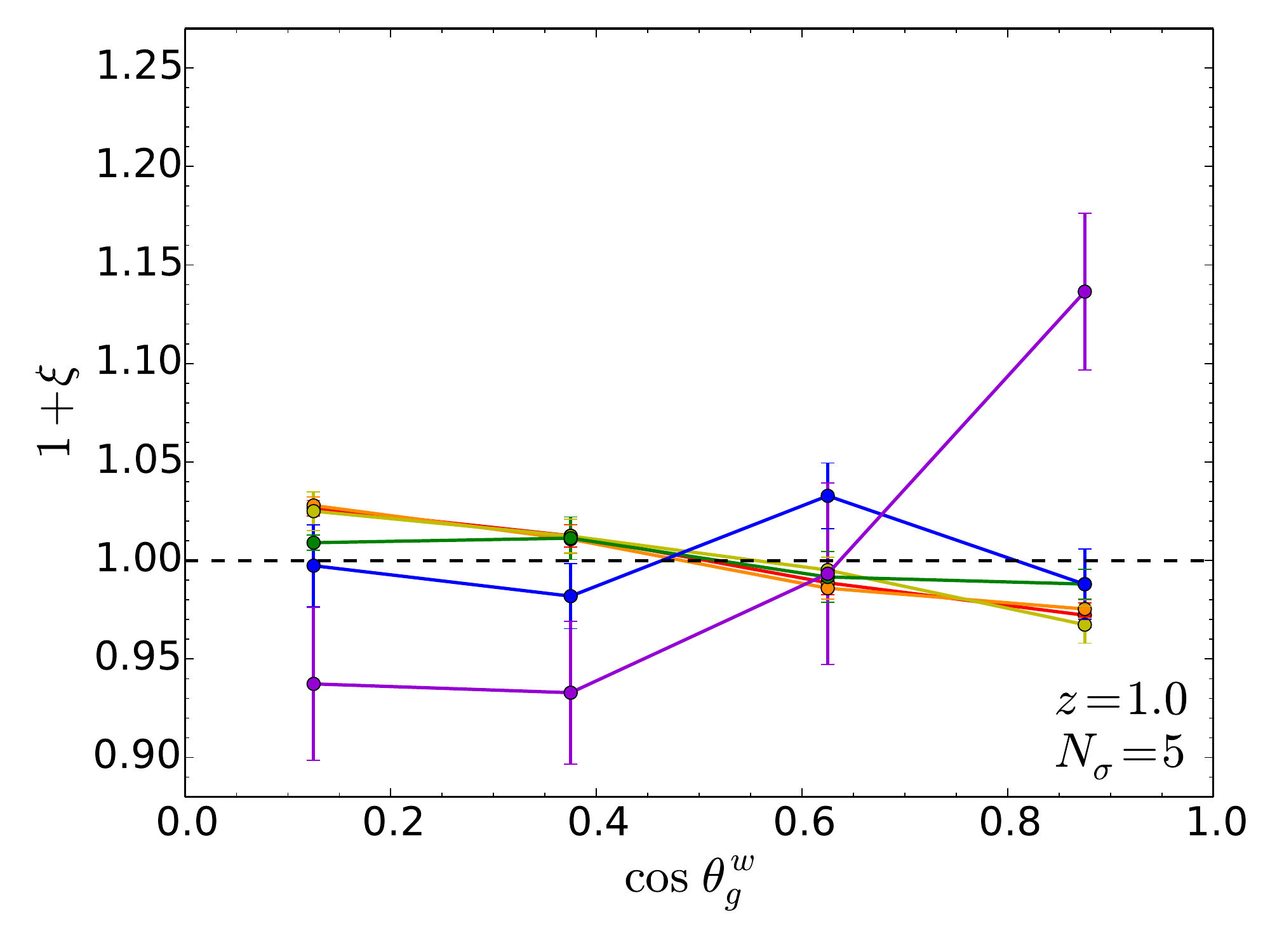}
\includegraphics[width=0.95\columnwidth]{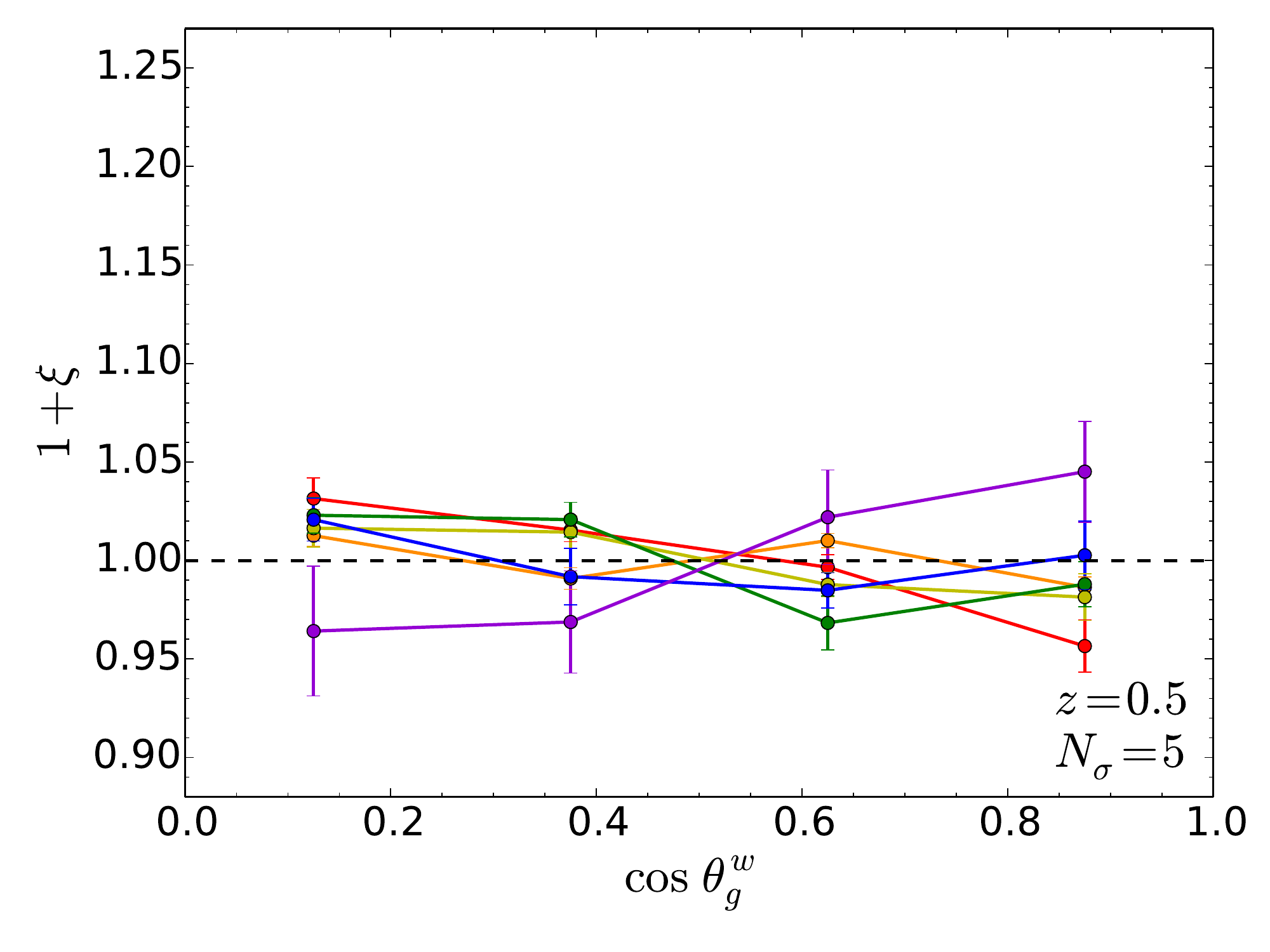}
\includegraphics[width=0.95\columnwidth]{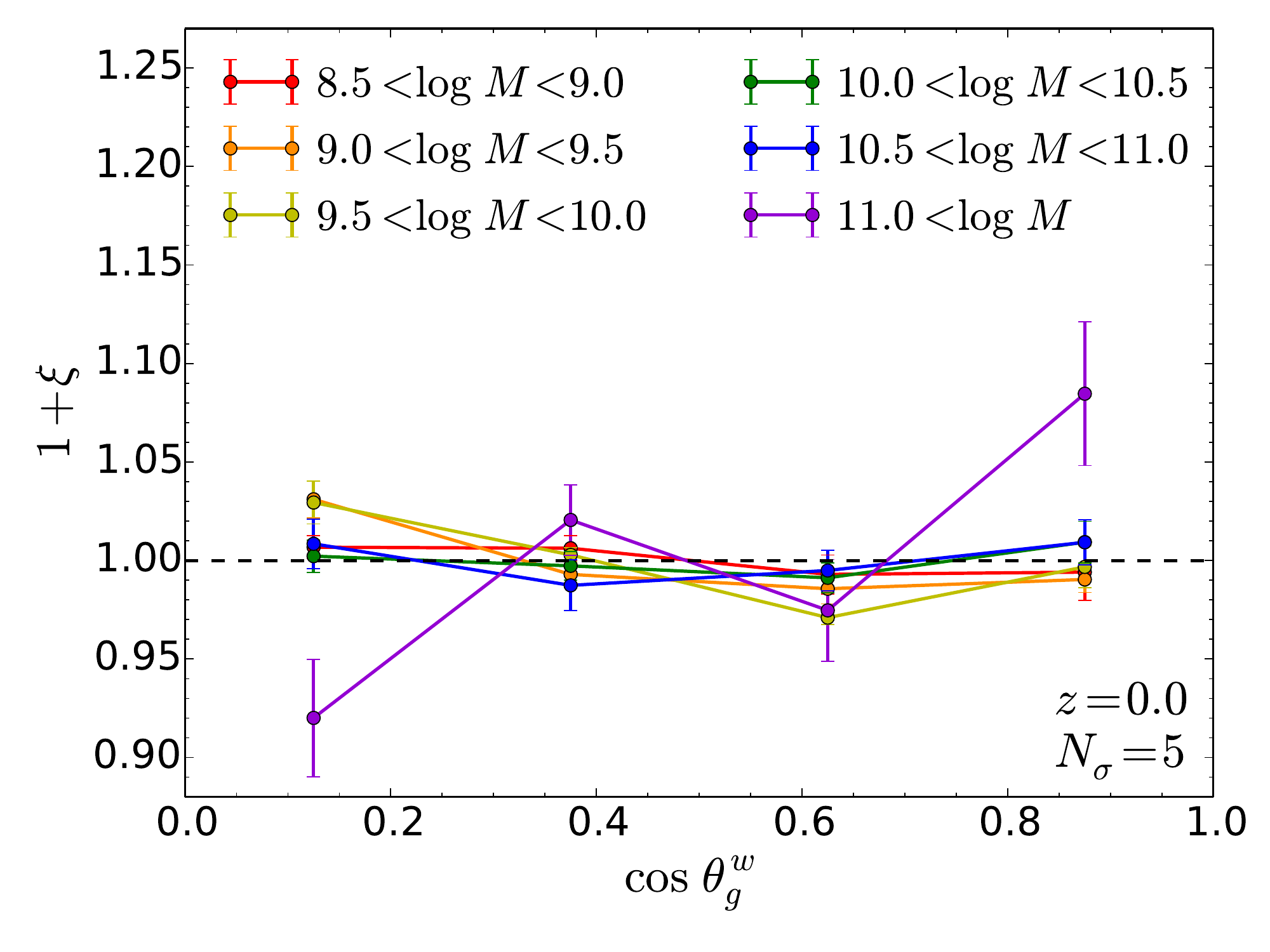}
\caption{PDF of the alignment angle between galaxy spins and the normal direction to the walls. Different redshifts are displayed from $z=1.5$ (top left-hand panel) to $z=0$ (bottom right-hand panel) and different stellar masses from red to purple. Massive galaxies are likely to have a spin perpendicular to the walls while less massive galaxies tend to have a spin lying inside the plane of the walls, in particular at high redshifts.}
\label{fig:GalWallAlign}
\end{figure*}

These results are consistent with pure dark matter simulations which have shown that the spin of small and intermediate mass haloes tends to lie within the plane of the walls \citep{hahnetal07,calvoetal07,2008MNRAS.385..867C,2009ApJ...706..747Z,2014MNRAS.440L..46A}.
The spin flip of massive haloes perpendicular to the plane of the walls is probably due to mergers of smaller haloes occurring within these walls (and their embedded filaments). Note that the transition mass in this case seems systematically slightly higher than for the filaments, which could be explained as follows.
Most mergers occur along filaments and therefore generate spin perpendicular to the filament axis. It is very likely that these mergers tend also to be more in the plane of the walls so that  eventually very massive objects will tend to get a spin aligned with the normal to the wall and not uniformly distributed in the plane perpendicular to the filament. However, because the probability of mergers is stronger in the direction of the filament (compared to the other dimension of the wall where the density is lower), the flip will typically occur faster with respect to the filament axis than the other direction of the wall for which more mergers are needed to re-orient the spin. 
Note that this picture is consistent with most massive objects being connected to 3 main filaments therefore defining a wall, as predicted by \citep{2018MNRAS.479..973C}. Accretion of matter within this sheet-like structure then typically generates an orthogonal spin.
A central ingredient here in making this prediction is the anisotropy and the flattening of walls and filaments, we refer to Kraljic et al. (in prep.) for further details.

\subsection{Galaxies}

Focusing now on the observable galaxies, let us plot the PDF of the cosine of the angle, $\cos\theta_{g}^{w}$, between the galaxy spins and the normal vectors of the closest wall structures. Figure~\ref{fig:GalWallAlign} displays the resulting PDFs for different stellar masses and redshifts. It shows that high mass galaxy spins tend to be parallel to these normals and therefore perpendicular to the plane of the walls, at all redshifts. There is a significant trend for the spin of low mass galaxies to have the opposite behaviour and lie inside the walls. As for filaments, this transition mass occurs at around $\log(M/M_{\odot})=11$.
Focussing on  the bottom right-hand panel of Figure~\ref{fig:CosThetaVsMass}, there is no evolution of the transition mass with redshift, though a clear transition does occur at each redshift.
The strength of the PDF signal varies from roughly $5\%$ to $20\%$, consistently  with the amplitude of alignments with filaments. 

\begin{figure}
\includegraphics[width=0.95\columnwidth]{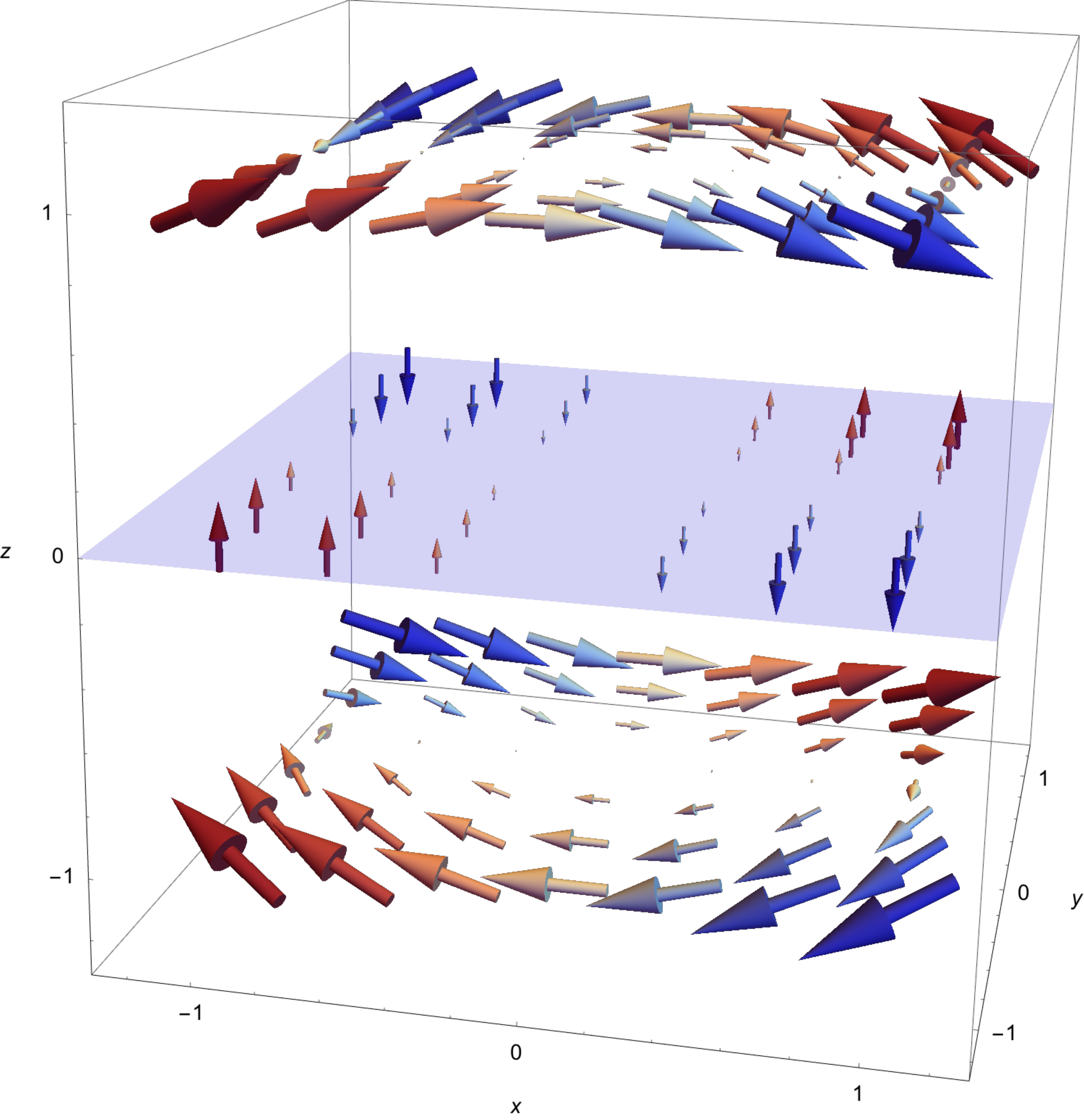}
\caption{Mean spin vectors in the vicinity of a wall (in the x-y plane) as predicted by anisotropic tidal torquing and colour coded from red to blue depending on their orientation along the normal to the wall, see Appendix~\ref{app:ATTT} for more details.}
\label{fig:spinwallATTT}
\end{figure}

Note that this signal is very similar to the alignment of the spin of dark matter haloes with walls and very consistent with 
prediction from tidal torque theory. Indeed, \cite{ATTT} showed that conditional tidal torque theory predicts a spin perpendicular to the plane of the walls close to that surface but quickly aligned with the surface of these sheets further away as can be seen on Figure~\ref{fig:spinwallATTT}. As we expect the most massive objects to be closer to the walls and less massive objects further in the field, this prediction is fully consistent with our measurements in the Horizon-AGN simulation. This transition is also seen in observations. Indeed, for instance, \cite{2018ApJ...860..127L} showed that in order to get a signal of alignment for galactic spins with the W-M sheet, it is necessary to exclude galaxies that are too close to the sheet (below 2 Mpc$/h$), an observation fully consistent with the picture predicted by 
\cite{ATTT} and measured in this section.

\section{Shape alignments}
\label{sec:shape-fil}

Previous sections have demonstrated that haloes and galaxies have a spin correlated with the large-scale cosmic web in a mass- and morphology-dependent way. 
Let us now investigate how the {\it shape} of haloes and galaxies align with filaments and walls.

For this purpose, let us use the minor axes of galaxies of the simple inertia tensor, following the convention of  \cite{2017MNRAS.472.1163C}. 
Although the minor axes are expected to be well aligned with the spin axes of spiral galaxies -- which is one of the building blocks of standard intrinsic alignment models \citep{Joachimi15}, the orientation of elliptical galaxies is much less dictated by their spin but rather by tidal stretching \citep{Catelan01}. 
It is therefore of interest to investigate a proxy other than the spin and study the actual orientation of all galaxies with respect to the cosmic web. Here, the same statistics as Sections~\ref{sec:spin-fil} and ~\ref{sec:spin-wall}
are used, replacing the spin vectors by the minor axis of the inertia tensors.

\subsection{Orientation of galaxies along filaments}

\begin{figure*}
\includegraphics[width=0.95\columnwidth]{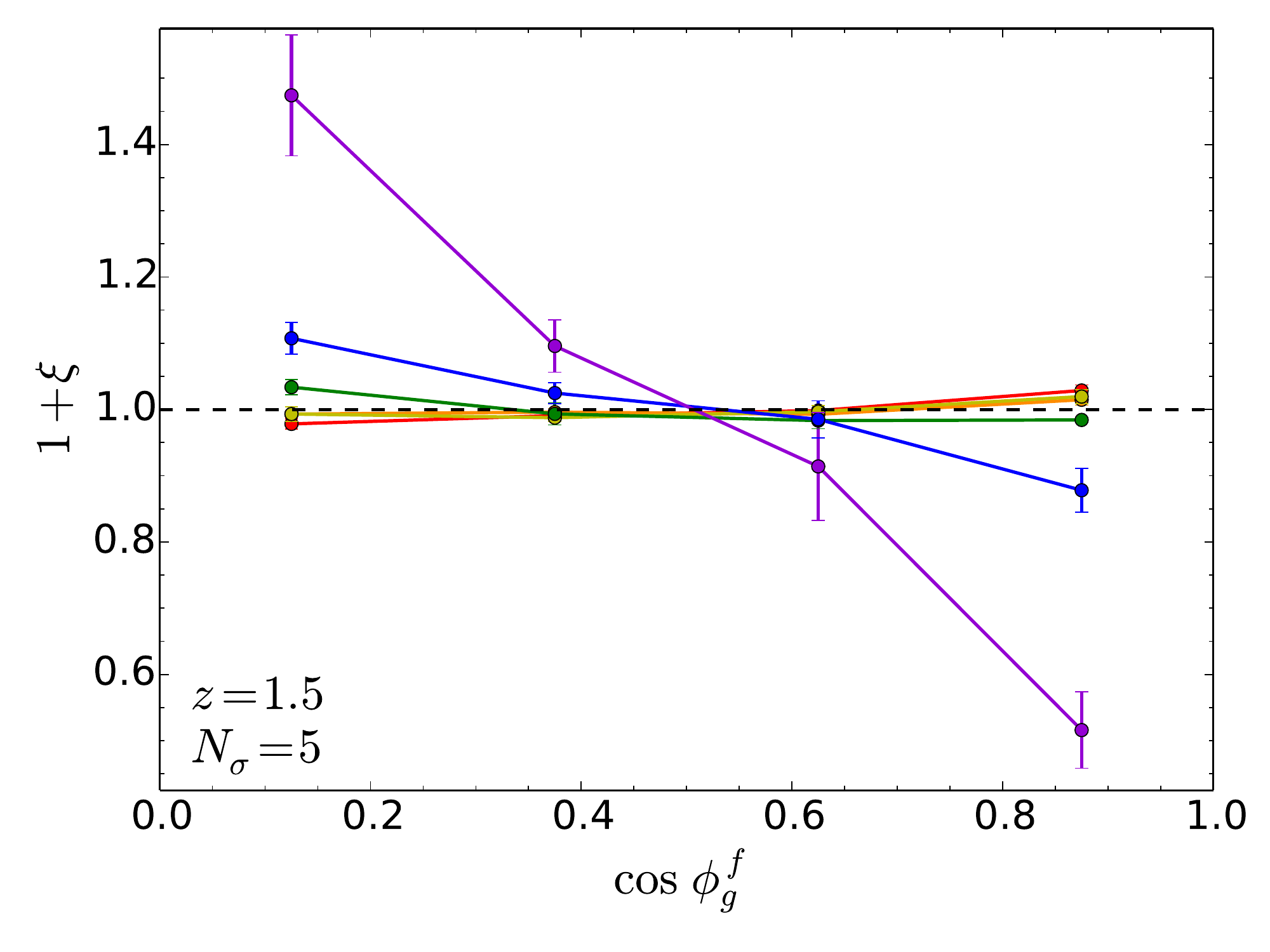}
\includegraphics[width=0.95\columnwidth]{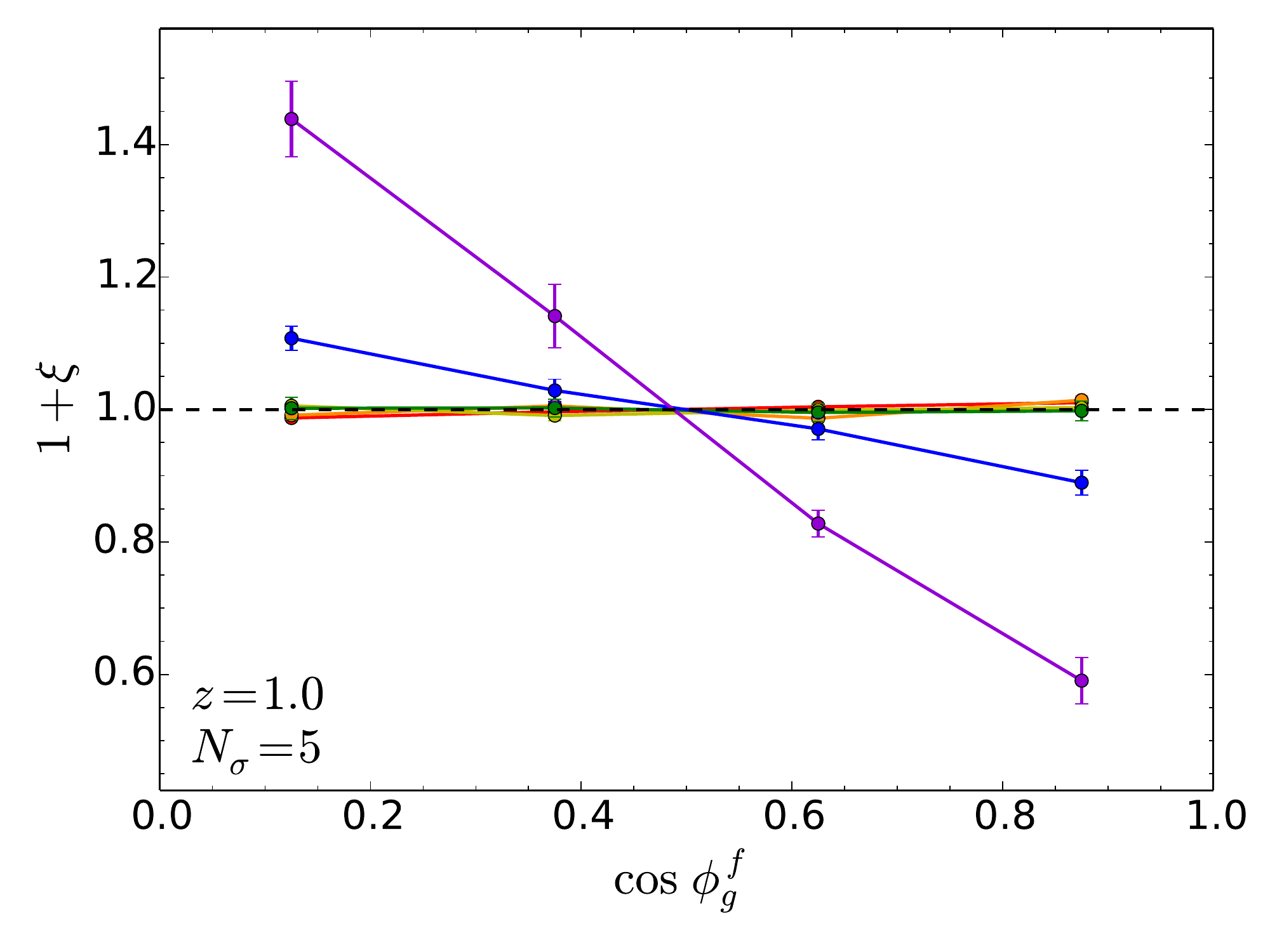}
\includegraphics[width=0.95\columnwidth]{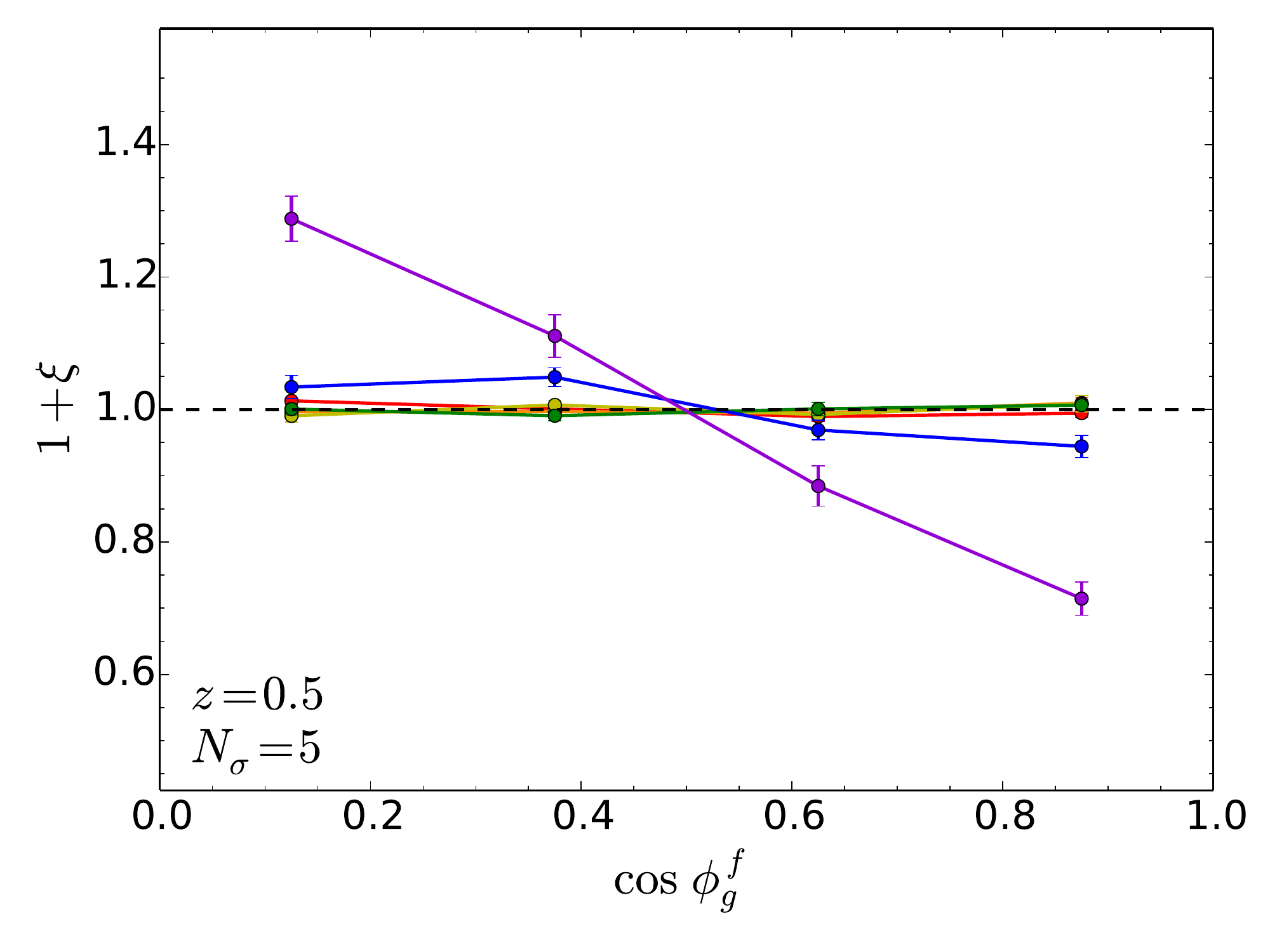}
\includegraphics[width=0.95\columnwidth]{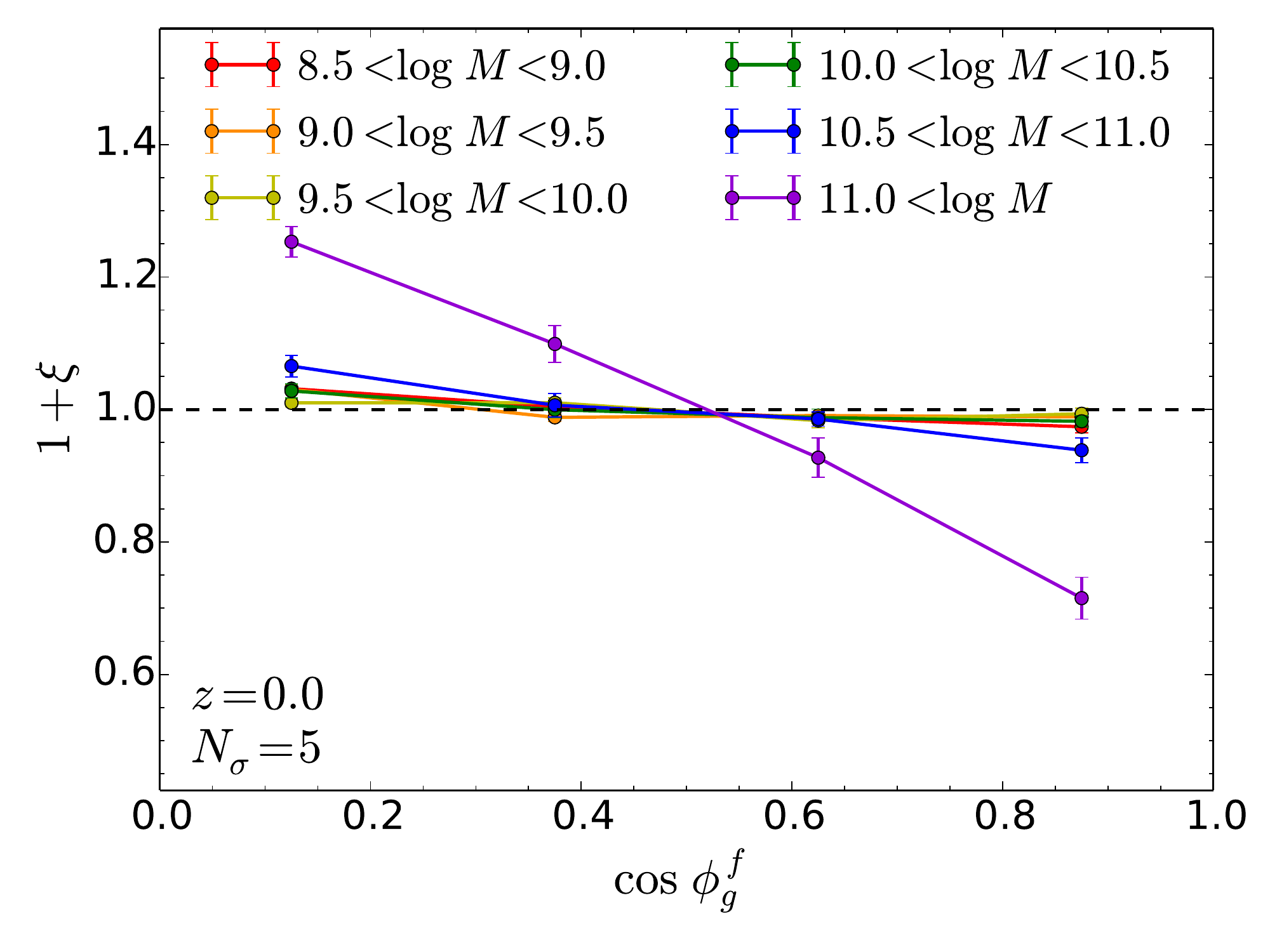}
\caption{PDF of the cosine of the angle between the minor axis of galaxies and the nearest filament direction for six different mass bins as labelled and redshifts from 1.5 (top left-hand panel) to 0 (bottom right-hand panel). Massive galaxies are strongly elongated in the direction of the filaments (minor axis perpendicular) while low-mass galaxies are marginally elongated perpendicular to the filaments at high redshifts.}
\label{fig:FilShapeAlign}
\end{figure*}

Let us first focus on the alignment of minor axes with respect to the filaments. The corresponding PDFs of $\cos\phi_{g}^{f}$ are plotted in Figure~\ref{fig:FilShapeAlign}. As expected,  a very clear trend for high mass galaxies to have minor axes perpendicular to the direction of filaments at all redshifts is found, which means that massive galaxies tend to align their shape (i.e their major axis) with nearby filaments. This result is in qualitative agreement with the findings of \cite{Chen15} based on the Massive-Black simulation. The amplitude of this signal is in fact roughly twice the amplitude of alignments using spins, ranging from $25-50\%$, suggesting that, for massive galaxies, shapes may be more sensitive to the tides generated by cosmic large-scale structures as already pointed out by \cite{2017MNRAS.472.1163C}. 
Interestingly, a strong orthogonal signal is also found for the second highest mass bin (blue curve) at all redshifts, unlike what was found for spins. 
Note that this alignment of galaxies with filaments has also been detected in current surveys such as the SDSS \citep{2018arXiv180500159C}.
This means that massive (central) galaxies inherit the alignment of their parent (massive) halo, for which this large scale orientation of the major axis is expected because of anisotropic collapse. Indeed massive haloes (whose scale is similar to the smoothing scale probed here) are expected to be elongated towards their neighbouring filament, which corresponds to the slowest collapse axis. 
 
At high redshifts, a clear trend is also detected for low mass galaxies to align their minor axis parallel to their host filament. However, the amplitude of the alignment is small and decreases with decreasing redshift, as was the case for spins. This alignment with filaments at high redshift could be the source of the high-redshift tangential alignment of disc-like galaxies found in \cite{Chisari16}. Note that no such alignments in SPH simulations have been reported so far \citep{Hilbert16,Tenneti16}.

To pin down how shapes connect to spins, Appendix~\ref{sec:vsig} shows how the morphology of galaxies affects their alignment with filaments.
For low $V/\sigma$ ratio (elliptical galaxies), shapes align very well with the axis of the filaments and the amplitude of the correlation increases with the prolateness. For these ellipticals, the spin is a poor tracer of the shape-filament alignment as expected. This is because elliptical galaxies are supported by the random motion of stars rather than the rotation and their spin is therefore not so well defined. In this case, the alignment of spins and shapes also depends on  triaxiality.
For high $V/\sigma$ ratio (spiral galaxies), spin alignments are very similar to shape alignments despite being of slightly lower amplitude and with a larger variance. Comparing now ellipticals and disks, shape alignments are stronger for elliptical galaxies as predicted by tidal stretching. These alignments can be as large as $40\%$ for the most massive elliptical galaxies explaining the detection of an intrinsic alignment signal of luminous red galaxies in observations \citep{2009ApJ...694..214O,2015MNRAS.450.2195S}.

The cleaner and stronger correlations found for galaxy shapes 
therefore suggest that the intrinsic alignments of elliptical galaxies should be modelled separately,  using an anisotropic tidal stretching model (Codis et al, in prep.). 
On the other hand, discs have similar spin and shape alignments which support  spin acquisition model to describe their intrinsic alignments. A detailed model should allow the alignment to change sign (parallel or perpendicular to filaments) depending on mass (and therefore luminosity), morphology and redshift. This could be built on an extension of \cite{ATTT}.

\subsection{Minor axis flips inside walls}

\begin{figure*}
\includegraphics[width=0.95\columnwidth]{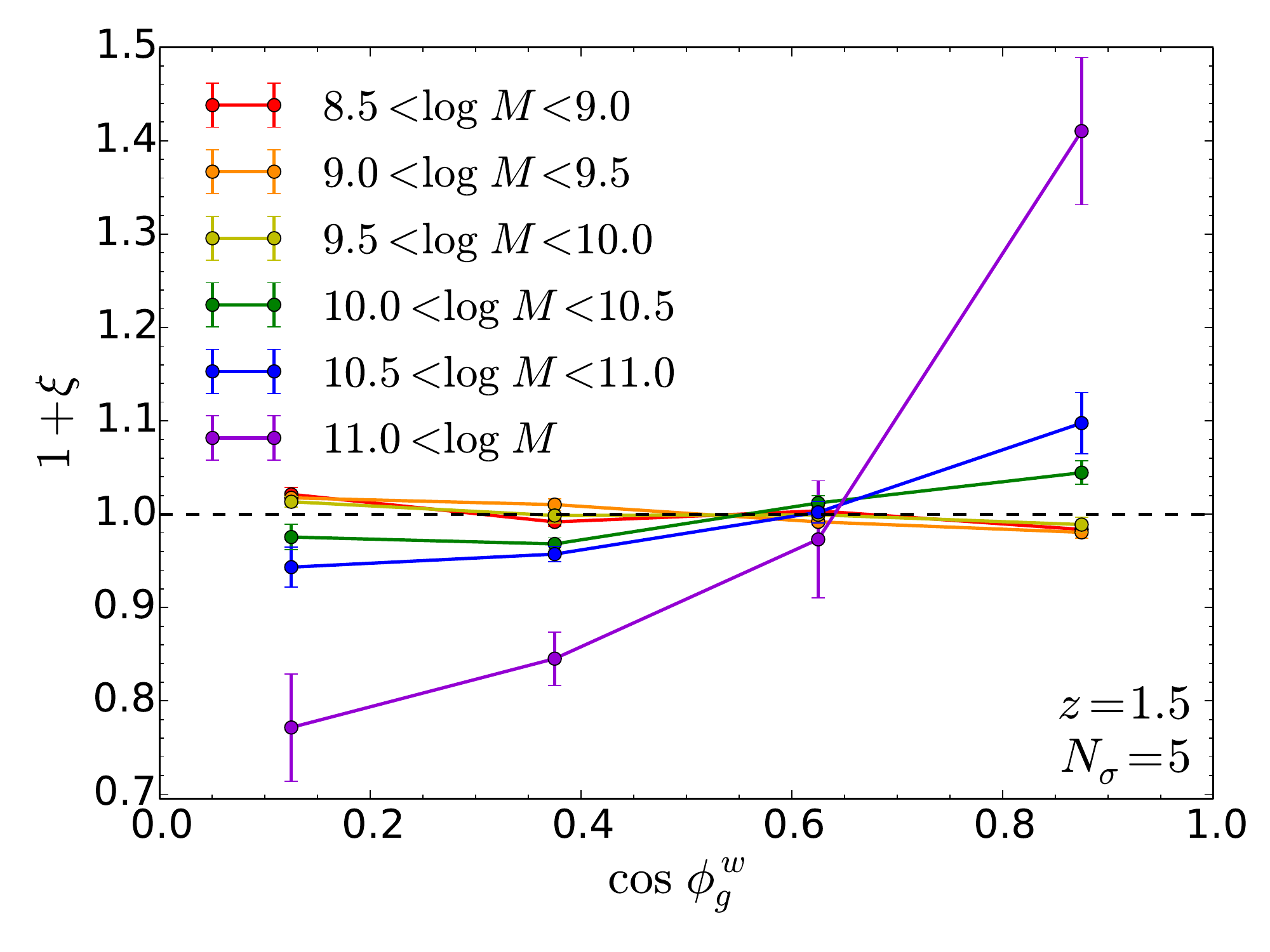}
\includegraphics[width=0.95\columnwidth]{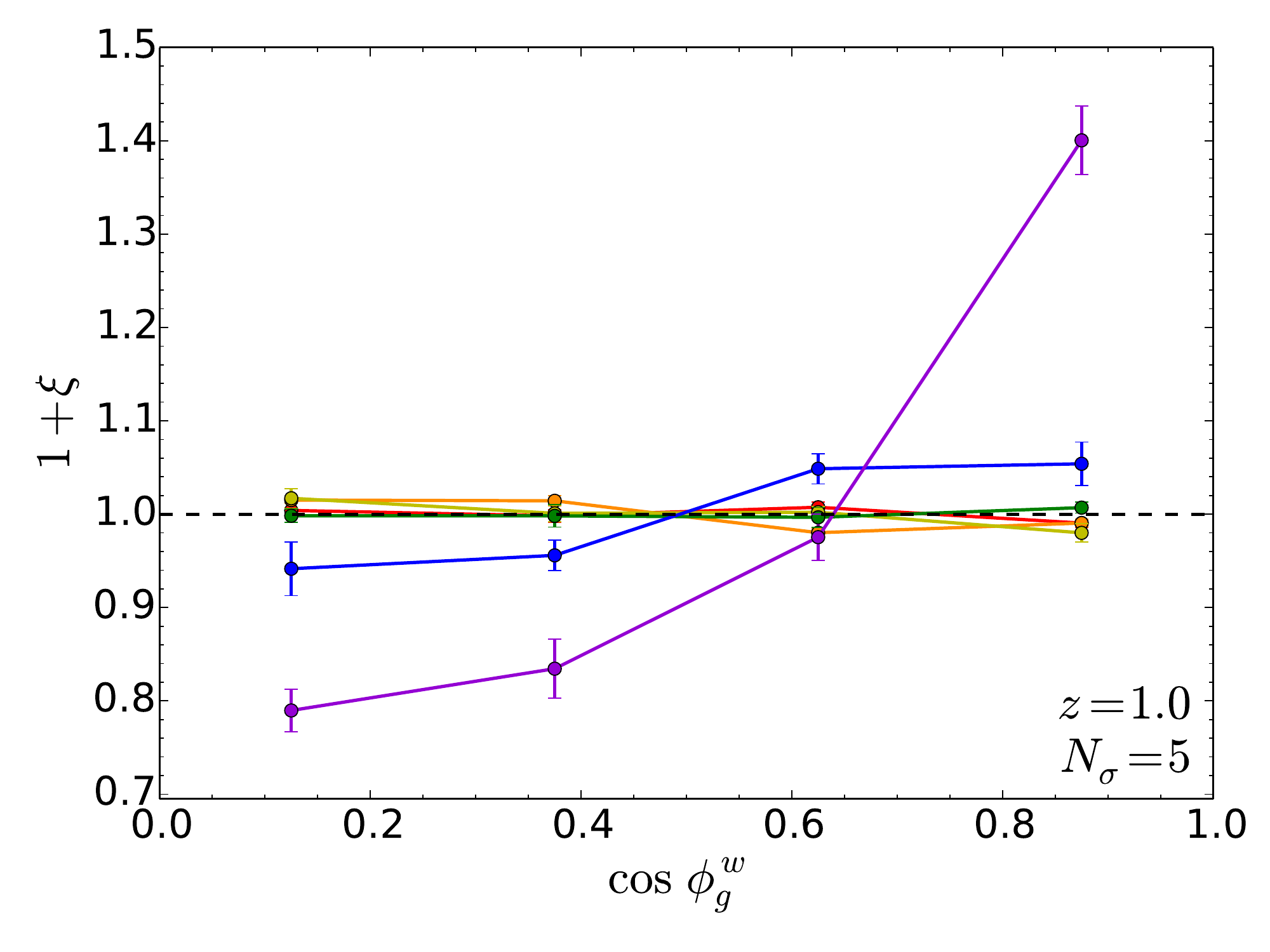}
\includegraphics[width=0.95\columnwidth]{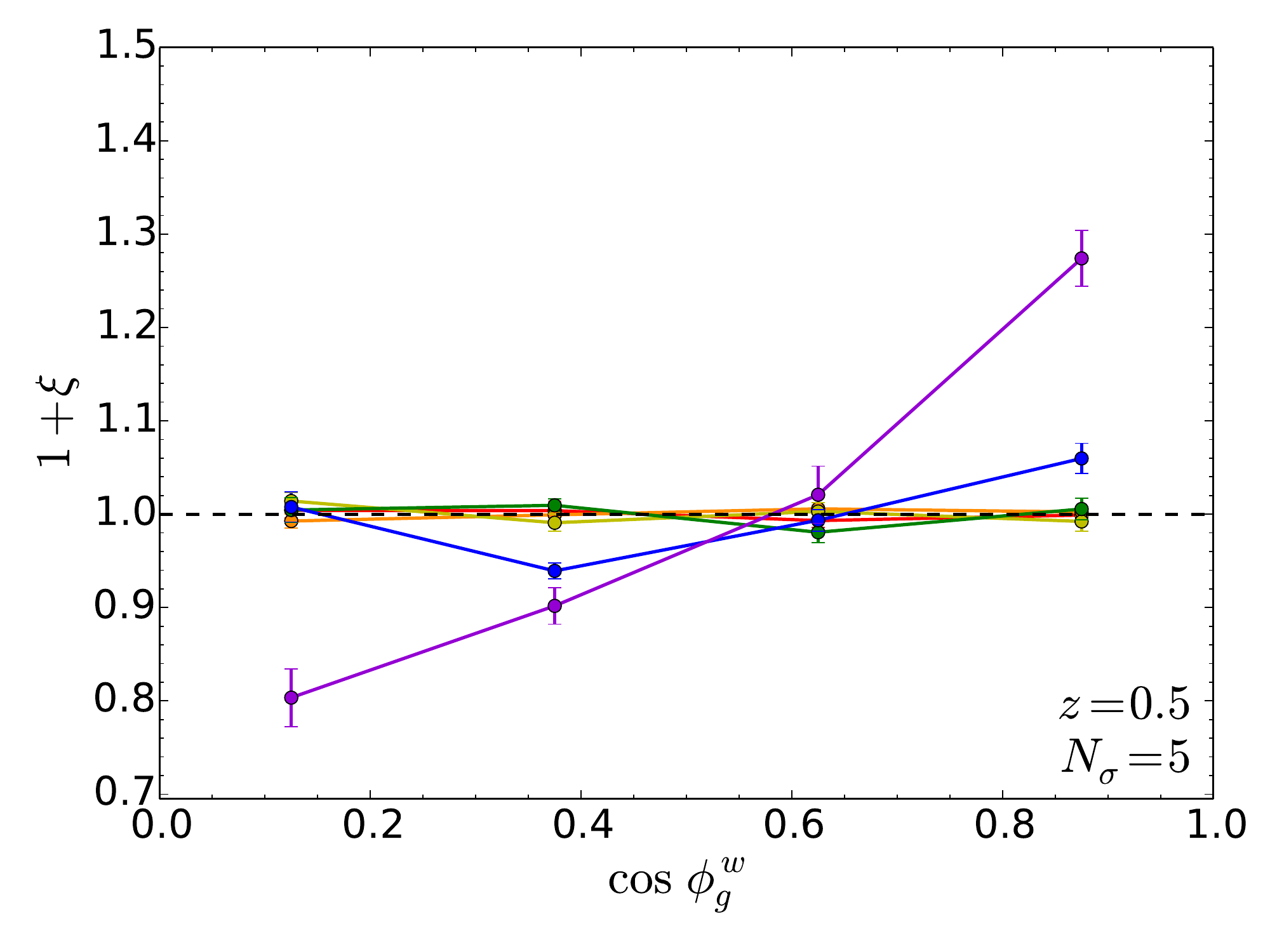}
\includegraphics[width=0.95\columnwidth]{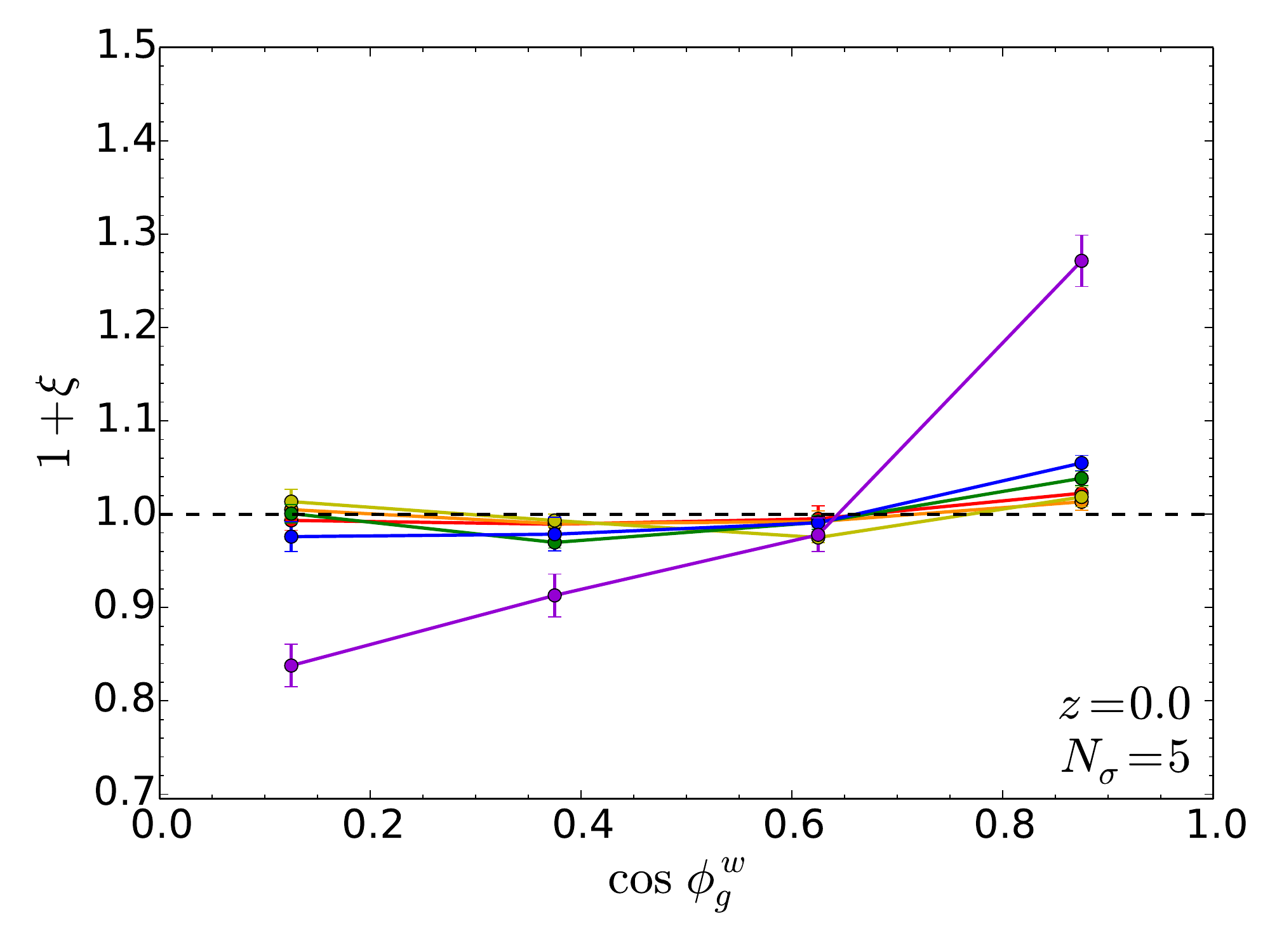}
\caption{The redshift evolution of the PDF of the cosine of the angle between the minor axis of galaxies and the normal vector of the nearest wall structure from $z=1.5$ (top left-hand panel) to $z=0$ (bottom right-hand panel). The data is separated into six mass bins as labelled. The shortest axis of the shape of galaxies tends to be inside the walls at high mass with a large associated probability and marginally perpendicular at small mass and high redshift.}
\label{fig:WallShapeAlign}
\end{figure*}

Let us finally focus on the PDFs of the cosine of the angle between minor axes and normal vectors of walls, $\cos\phi_{g}^{w}$. As shown in Figure~\ref{fig:WallShapeAlign}, very similar trends are found  compared to the filaments case, noting that the directions are flipped, since we focus on the normal vector to walls. Massive galaxies indeed show a strong tendency to align their minor axis perpendicular to the plane of the walls  and therefore to point their minor axis toward the centre of the voids at all redshifts, while the minor axis of low-mass galaxies at high redshift is more likely to lie within the cosmic sheets (although this trend disappears for decreasing redshift). Once again,  stronger signals are found for the orientation of high mass galaxies (at the level of $\sim 30-50\%$) compared to spin alignments ($\sim 5-20\%$).

\section{Conclusions and discussion}
\label{sec:conclusion}
 
\subsection{Results}
The state-of-the-art Horizon-AGN cosmological hydrodynamical simulation was analysed to investigate the imprint of the highly anisotropic cosmic web on some vectorial and tensorial galactic properties across cosmic time, 
an effect which can be straightforwardly probed by measuring correlations between the directionality of non-scalar quantities and the preferred directions set by the cosmic skeleton.
In particular, the orientation of haloes and galaxies with  respect to filaments and walls, as defined by the topological ridge extractor {\sc DisPerSE}, was investigated by means of their spin vectors
and shapes (minor axes of their inertia tensor).
Both spins and minor axes were shown to correlate with the direction of the closest filaments and walls, in a mass and redshift dependent way. 

Our results can be summarised as follows.
\begin{itemize}
\item Haloes undergo a spin flip from alignment with filaments and walls at low-mass to orthogonality at larger mass. The transition occurs around ${M}_{\rm tr}^{h}\approx10^{12.2\pm 0.1}M_{\odot}$, decreases with redshift, in agreement with pure dark matter simulations and theoretical works and is, as expected, slightly higher for walls compared to filaments.
\item Galaxies also clearly align their spin to the filaments but mainly at high redshift. Their spin becomes perpendicular to the axis of the filaments and perpendicular to walls at lower redshift and higher mass. This perpendicular signal is dominated by the contribution from massive elliptical and central galaxies and occurs for stellar masses above $\log M_{\rm tr}^{g}/M_{\odot}\approx{10.1\pm0.3}$. This spin flip is detected at the 9 sigmas level and therefore confirms the tentative detection presented in \cite{dubois14} for $z=1.8$ and extends their results to a redshift regime more easily accessible to current galaxy surveys.
\item The shapes of galaxies are found to correlate more  strongly than spins with the cosmic web structures. Massive galaxies tend to be elongated along the filament's directions and to lie within the sheets, while less massive ones tend to be oriented perpendicular to the axis of the host filament and wall and therefore pointing toward the centre of the void. 
\item The spin of central galaxies tends to align with the spin of their host halo with a mean misalignment angle of 45 to $55\degree$ depending on mass and redshift. Similar alignments happen for satellite galaxies as soon as they are compared to their host subhalo. The alignment signal increases towards low redshift for the most massive galaxies in agreement with the analysis of galaxy shapes in \cite{2017MNRAS.472.1163C}. This phenomenon should be accounted for in semi-analytical models of galaxy formation. 
\end{itemize}

\begin{figure}
\includegraphics[width=0.95\columnwidth]{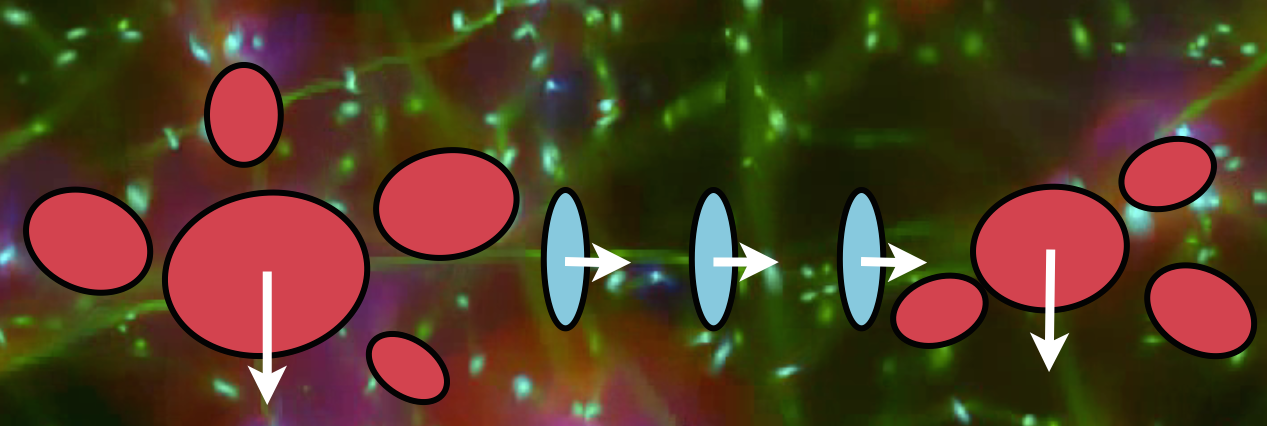}
\caption{Sketch of the galaxy alignments found in this work. Massive elliptical galaxies(red) tend to be elongated along filaments and inside the sheets while less massive disc-like galaxies (blue) are more likely to have a spin lying in the plane of the walls and in filaments.
}
\label{fig:cartoon}
\end{figure}

Figure~\ref{fig:cartoon} sketches the main alignment signal found between galaxies and filaments.
It also applies to haloes to some extent as less massive haloes in the middle of filaments tend to have a spin aligned with the axis of filaments while more massive haloes are found closer to the nodes of the cosmic web with a spin perpendicular to filaments and walls.

The present work extends the first attempts of \cite{dubois14} on the spin-filament alignments at $z=1.8$ to all redshifts, to cosmic sheets and to minor axes of the inertia tensor and with an improved detection of the cosmic web. These results confirm the trends seen in pure dark matter simulations \citep[e.g][and references in introduction]{codisetal12}, therefore showing that  alignments with the cosmic web are not completely erased by baryonic physics happening on small and intermediate scales. In particular,
the measured spin alignments  are likely due to cold flows feeding galaxies with coherent angular momentum accreted from the cosmic web via secondary infall \citep{2011MNRAS.418.2493P}. As they move along filaments, they merge and create a population of massive galaxies which  convert their orbital angular momentum into spin perpendicular to the filament axis \citep{codisetal12,2014MNRAS.445L..46W}. 
Low mass objects tend to have a spin aligned with the filaments because they were formed earlier in the vorticity rich filaments \citep{laigle2014}.

Our results are also consistent with prediction from tidal torque theory. Indeed, \cite{ATTT} theoretically studied the effect of the large scale structure (in particular filaments and walls) on spin acquisition by tidal torquing. They showed that the spin-filament alignment can be explained from first principles  when taking into account the anisotropy of the large-scale structure in the standard tidal torque theory picture. They also investigated the effect of the cosmic walls on spins and found that the spin of small mass objects is typically parallel to the plane of the wall while more massive objects form closer to the walls with a spin perpendicular to it and therefore pointing toward the centre of the void. Our findings are therefore totally consistent with this anisotropic tidal torque theory, demonstrating in particular that local baryonic physics does not completely erase memory of the large scale flows from which their host haloes emerged {  \citep[as also discussed in][]{2018MNRAS.473.1562W}}\footnote{
The alignment of the shapes (i.e minor axes) with the filaments and walls could also be predicted using a tidal stretching model together with the anisotropy of the cosmic web but this is left for future works.
}.

The main difference we observe between haloes in pure DM runs and virtual galaxies in Horizon-AGN is the significant decrease in amplitude of the alignment of low-mass galaxies with filaments towards low redshift. This effect seems to be due to the population of (small) satellites which decorrelate from their large-scale environment as they get accreted on a cluster and progressively get closer to the center of their host DM halo.
{  Further works using other cosmological hydrodynamical simulations would be useful in order to assess the role of gas dynamics and subgrid physics on galactic spin-filament alignments but are beyond the scope of this paper.}

Our novel predictions for the alignment between virtual galaxies and the cosmic web based on Horizon-AGN
are finally also in good agreement with observational results that tend to find disks with a spin lying in the plane of walls and parallel to filaments, as well as ellipticals elongated along filaments and walls implying their spins are instead perpendicular to them \citep[see e.g][]{leeetal07,2013ApJ...770L..12L,2013ApJ...779..160Z,2016MNRAS.463..222H,2017A&A...599A..31H,2018ApJ...860..127L}, despite other conflicting observational results \citep[e.g][]{varela11}. 
In addition, based on tidal torque theory and results from \citep{ATTT}, we were able to explain why removing galaxies too close to walls typically increases the alignment signal between spins and the plane of the walls as observed by \citep{2018ApJ...860..127L} since those objects tend to transition from a parallel to an orthogonal alignment which therefore reduces the alignment signal. Our predictions can potentially already be more thoroughly tested with current surveys like SDSS, GAMA or SAMI or with future experiments like HECTOR \citep{2016SPIE.9908E..1FB}.

\subsection{Implication for weak lensing studies}  
The alignment of galaxy shapes with the large-scale cosmic web is of prime importance in cosmology as it can severely contaminate weak lensing observables \citep{Kirk10,Krause15}. Current intrinsic alignments models are based on linear theory \citep{Catelan01}, perturbation theory \citep{Blazek15,2017arXiv170809247B} or the `halo model', informed by current observational or numerical constraints \citep{Schneider10,joachimi13a,Joachimi13b}. In light of the desired accuracy of dark energy equation of state constraints from the next generation of weak lensing surveys, it is important to consider potential extensions to these models that account for environmental dependence. Firstly, our work has shown qualitatively different trends in the spin alignments of discs and the shape alignments of galaxies, implying that these two populations should be modelled separately. Spin alignments of disk-like galaxies with filaments are responsible for the high redshift tangential alignment of blue galaxies described in \cite{Chisari16} while tidal stretching by the cosmic web engenders a population of massive elliptical galaxies elongated along filaments at low redshift and responsible for the late-time radial alignment of red galaxies. It would thus be of  interest to investigate whether the knowledge of the cosmic web could be used in order to mitigate weak lensing contamination by intrinsic alignments.

Other groups have  studied intrinsic alignments in cosmological hydrodynamical simulations  using different numerical techniques and baryonic processes prescriptions. Among those works, \citet{Chen15} studied galaxy shape-filament alignment in the MassiveBlack II simulation, a smoothed-particle-hydrodynamics simulation with similar volume and resolution compared to Horizon-AGN. In this work, no transition was detected in the relative alignment with respect to subhalo mass. Galaxy shape alignments in MassiveBlack II preserve a parallel trend between the minor axis and the filamentary axis regardless of subhalo mass at $z=0$. While this is at odds with our results (see e.g Figure~\ref{fig:CosThetaVsMass}), there are known discrepancies between galaxy alignments among smoothed-particle-hydrodynamics and adaptive-mesh-refinement simulations \citep{Tenneti16}. In particular, although the alignments of massive ellipticals are in qualitative agreement between the different simulations and with observational results \citep{2009ApJ...694L..83O,2011A&A...527A..26J,2015MNRAS.450.2195S}, the alignments of spirals have opposite signs (tangential versus radial alignments towards overdensities) in the two types of simulations. This discrepancy should be the object of future study {  as alignments of disc galaxies have been shown to be a potential source notably of GI contamination for lensing \citep{2018ApJ...853...25W}}. 
Notwithstanding, let us emphasize that the results presented in this work are in agreement with the predictions of tidal torque theory \citep{ATTT} and with current observed trends. 

\section*{Acknowledgments}
This work is partially supported by the Spin(e) grant {ANR-13-BS05-0005} (\href{http://cosmicorigin.org}{http://cosmicorigin.org}) of the French {\sc Agence Nationale de la Recherche},  
and by the SURP program at CITA. 
SC is partially supported by a research grant from Fondation MERAC.
We thank S. Rouberol for smoothly running the {\sc Horizon} cluster for us, and T. Sousbie for his help with {\sc DisPerSE} on walls.
SC thanks Katarina Kraljic for fruitful comments and discussions while this work was carried out, J.~R.~Bond for his co-supervision of AJ and all the members of the Spin(e) project. CP thanks SUPA and the ERC grant 670193 for funding.

\bibliographystyle{mn2e}
\bibliography{spin}

\begin{thebibliography}{131}
\expandafter\ifx\csname natexlab\endcsname\relax\def\natexlab#1{#1}\fi

\bibitem[{{Alonso}, {Eardley} \& {Peacock}(2015){Alonso}, {Eardley}, \&
  {Peacock}}]{2015MNRAS.447.2683A}
{Alonso} D., {Eardley} E., {Peacock} J.~A., 2015, \mnras, 447, 2683

\bibitem[{{Alpaslan} {et~al}\mbox{.}(2015){Alpaslan}, {Driver}, {Robotham},
  {Obreschkow}, {Andrae}, {Cluver}, {Kelvin}, {Lange}, {Owers}, {Taylor},
  {Andrews}, {Bamford}, {Bland-Hawthorn}, {Brough}, {Brown}, {Colless},
  {Davies}, {Eardley}, {Grootes}, {Hopkins}, {Kennedy}, {Liske},
  {Lara-L{\'o}pez}, {L{\'o}pez-S{\'a}nchez}, {Loveday}, {Madore}, {Mahajan},
  {Meyer}, {Moffett}, {Norberg}, {Penny}, {Pimbblet}, {Popescu}, {Seibert}, \&
  {Tuffs}}]{2015MNRAS.451.3249A}
{Alpaslan} M. {et~al.}, 2015, \mnras, 451, 3249

\bibitem[{{Alpaslan} {et~al}\mbox{.}(2016){Alpaslan}, {Grootes}, {Marcum},
  {Popescu}, {Tuffs}, {Bland-Hawthorn}, {Brough}, {Brown}, {Davies}, {Driver},
  {Holwerda}, {Kelvin}, {Lara-L{\'o}pez}, {L{\'o}pez-S{\'a}nchez}, {Loveday},
  {Moffett}, {Taylor}, {Owers}, \& {Robotham}}]{2016MNRAS.457.2287A}
{Alpaslan} M. {et~al.}, 2016, \mnras, 457, 2287

\bibitem[{{Arag{\'o}n-Calvo} {et~al}\mbox{.}(2007){Arag{\'o}n-Calvo}, {van de
  Weygaert}, {Jones}, \& {van der Hulst}}]{calvoetal07}
{Arag{\'o}n-Calvo} M.~A., {van de Weygaert} R., {Jones} B.~J.~T., {van der
  Hulst} J.~M., 2007, \apjl, 655, L5

\bibitem[{{Aragon-Calvo} \& {Yang}(2014)}]{2014MNRAS.440L..46A}
{Aragon-Calvo} M.~A., {Yang} L.~F., 2014, \mnras, 440, L46

\bibitem[{{Arnold}, {Shandarin} \& {Zeldovich}(1982){Arnold}, {Shandarin}, \&
  {Zeldovich}}]{1982GApFD..20..111A}
{Arnold} V.~I., {Shandarin} S.~F., {Zeldovich} I.~B., 1982, Geophysical and
  Astrophysical Fluid Dynamics, 20, 111

\bibitem[{Aubert, Pichon \& Colombi(2004)Aubert, Pichon, \&
  Colombi}]{aubertetal04}
Aubert D., Pichon C., Colombi S., 2004, \mnras, 352, 376

\bibitem[{{Bailin} \& {Steinmetz}(2005)}]{2005ApJ...627..647B}
{Bailin} J., {Steinmetz} M., 2005, \apj, 627, 647

\bibitem[{{Blazek} {et~al}\mbox{.}(2017){Blazek}, {MacCrann}, {Troxel}, \&
  {Fang}}]{2017arXiv170809247B}
{Blazek} J., {MacCrann} N., {Troxel} M.~A., {Fang} X., 2017, ArXiv e-prints

\bibitem[{{Blazek}, {Vlah} \& {Seljak}(2015){Blazek}, {Vlah}, \&
  {Seljak}}]{Blazek15}
{Blazek} J., {Vlah} Z., {Seljak} U., 2015, \jcap, 8, 015

\bibitem[{{Bond}, {Kofman} \& {Pogosyan}(1996){Bond}, {Kofman}, \&
  {Pogosyan}}]{bkp96}
{Bond} J.~R., {Kofman} L., {Pogosyan} D., 1996, \nat, 380, 603

\bibitem[{{Brunino} {et~al}\mbox{.}(2007){Brunino}, {Trujillo}, {Pearce}, \&
  {Thomas}}]{2007MNRAS.375..184B}
{Brunino} R., {Trujillo} I., {Pearce} F.~R., {Thomas} P.~A., 2007, \mnras, 375,
  184

\bibitem[{{Bryant} {et~al}\mbox{.}(2016){Bryant}, {Bland-Hawthorn}, {Lawrence},
  {Croom}, {Brown}, {Venkatesan}, {Gillingham}, {Zhelem}, {Content},
  {Saunders}, {Staszak}, {van de Sande}, {Couch}, {Leon-Saval}, {Tims},
  {McDermid}, \& {Schaefer}}]{2016SPIE.9908E..1FB}
{Bryant} J.~J. {et~al.}, 2016, in \procspie, Vol. 9908, Ground-based and
  Airborne Instrumentation for Astronomy VI, p. 99081F

\bibitem[{{Catelan}, {Kamionkowski} \& {Blandford}(2001){Catelan},
  {Kamionkowski}, \& {Blandford}}]{Catelan01}
{Catelan} P., {Kamionkowski} M., {Blandford} R.~D., 2001, \mnras, 320, L7

\bibitem[{{Chen} {et~al}\mbox{.}(2016){Chen}, {Wang}, {Mo}, \&
  {Shi}}]{2016ApJ...825...49C}
{Chen} S., {Wang} H., {Mo} H.~J., {Shi} J., 2016, \apj, 825, 49

\bibitem[{{Chen} {et~al}\mbox{.}(2018){Chen}, {Ho}, {Blazek}, {He},
  {Mandelbaum}, {Melchior}, \& {Singh}}]{2018arXiv180500159C}
{Chen} Y.-C., {Ho} S., {Blazek} J., {He} S., {Mandelbaum} R., {Melchior} P.,
  {Singh} S., 2018, ArXiv e-prints

\bibitem[{{Chen} {et~al}\mbox{.}(2015){Chen}, {Ho}, {Tenneti}, {Mandelbaum},
  {Croft}, {DiMatteo}, {Freeman}, {Genovese}, \& {Wasserman}}]{Chen15}
{Chen} Y.-C. {et~al.}, 2015, \mnras, 454, 3341

\bibitem[{{Chisari} {et~al}\mbox{.}(2015){Chisari}, {Codis}, {Laigle},
  {Dubois}, {Pichon}, \& {Devriendt}}]{Chisari15}
{Chisari} N., {Codis} S., {Laigle} C., {Dubois} Y., {Pichon} C., {Devriendt}
  J., 2015, \mnras, 454, 2736

\bibitem[{{Chisari} {et~al}\mbox{.}(2016){Chisari}, {Laigle}, {Codis},
  {Dubois}, {Devriendt}, {Miller}, {Benabed}, {Slyz}, {Gavazzi}, \&
  {Pichon}}]{Chisari16}
{Chisari} N. {et~al.}, 2016, \mnras, 461, 2702

\bibitem[{{Chisari} {et~al}\mbox{.}(2017){Chisari}, {Koukoufilippas}, {Jindal},
  {Peirani}, {Beckmann}, {Codis}, {Devriendt}, {Miller}, {Dubois}, {Laigle},
  {Slyz}, \& {Pichon}}]{2017MNRAS.472.1163C}
{Chisari} N.~E. {et~al.}, 2017, \mnras, 472, 1163

\bibitem[{{Codis} {et~al}\mbox{.}(2015){Codis}, {Gavazzi}, {Dubois}, {Pichon},
  {Benabed}, {Desjacques}, {Pogosyan}, {Devriendt}, \& {Slyz}}]{codis14}
{Codis} S. {et~al.}, 2015, \mnras, 448, 3391

\bibitem[{{Codis} {et~al}\mbox{.}(2012){Codis}, {Pichon}, {Devriendt}, {Slyz},
  {Pogosyan}, {Dubois}, \& {Sousbie}}]{codisetal12}
{Codis} S., {Pichon} C., {Devriendt} J., {Slyz} A., {Pogosyan} D., {Dubois} Y.,
  {Sousbie} T., 2012, \mnras, 427, 3320

\bibitem[{{Codis}, {Pichon} \& {Pogosyan}(2015){Codis}, {Pichon}, \&
  {Pogosyan}}]{ATTT}
{Codis} S., {Pichon} C., {Pogosyan} D., 2015, \mnras, 452, 3369

\bibitem[{{Codis}, {Pogosyan} \& {Pichon}(2018){Codis}, {Pogosyan}, \&
  {Pichon}}]{2018MNRAS.479..973C}
{Codis} S., {Pogosyan} D., {Pichon} C., 2018, \mnras, 479, 973

\bibitem[{{Crittenden} {et~al}\mbox{.}(2001){Crittenden}, {Natarajan}, {Pen},
  \& {Theuns}}]{Cri++01}
{Crittenden} R.~G., {Natarajan} P., {Pen} U.-L., {Theuns} T., 2001, \apj, 559,
  552

\bibitem[{{Cuesta} {et~al}\mbox{.}(2008){Cuesta}, {Betancort-Rijo},
  {Gottl{\"o}ber}, {Patiri}, {Yepes}, \& {Prada}}]{2008MNRAS.385..867C}
{Cuesta} A.~J., {Betancort-Rijo} J.~E., {Gottl{\"o}ber} S., {Patiri} S.~G.,
  {Yepes} G., {Prada} F., 2008, \mnras, 385, 867

\bibitem[{{Dalal} {et~al}\mbox{.}(2008){Dalal}, {White}, {Bond}, \&
  {Shirokov}}]{2008ApJ...687...12D}
{Dalal} N., {White} M., {Bond} J.~R., {Shirokov} A., 2008, \apj, 687, 12

\bibitem[{{Doroshkevich}(1970)}]{doroshkevich70}
{Doroshkevich} A.~G., 1970, Astrophysics, 6, 320

\bibitem[{{Dressler}(1980)}]{1980ApJ...236..351D}
{Dressler} A., 1980, \apj, 236, 351

\bibitem[{{Dubois} {et~al}\mbox{.}(2012){Dubois}, {Devriendt}, {Slyz}, \&
  {Teyssier}}]{duboisetal12agnmodel}
{Dubois} Y., {Devriendt} J., {Slyz} A., {Teyssier} R., 2012, \mnras, 420, 2662

\bibitem[{{Dubois} {et~al}\mbox{.}(2016){Dubois}, {Peirani}, {Pichon},
  {Devriendt}, {Gavazzi}, {Welker}, \& {Volonteri}}]{2016MNRAS.463.3948D}
{Dubois} Y., {Peirani} S., {Pichon} C., {Devriendt} J., {Gavazzi} R., {Welker}
  C., {Volonteri} M., 2016, \mnras, 463, 3948

\bibitem[{{Dubois} {et~al}\mbox{.}(2014){Dubois}, {Pichon}, {Welker}, {Le
  Borgne}, {Devriendt}, {Laigle}, {Codis}, {Pogosyan}, {Arnouts}, {Benabed},
  {Bertin}, {Blaizot}, {Bouchet}, {Cardoso}, {Colombi}, {de Lapparent},
  {Desjacques}, {Gavazzi}, {Kassin}, {Kimm}, {McCracken}, {Milliard},
  {Peirani}, {Prunet}, {Rouberol}, {Silk}, {Slyz}, {Sousbie}, {Teyssier},
  {Tresse}, {Treyer}, {Vibert}, \& {Volonteri}}]{dubois14}
{Dubois} Y. {et~al.}, 2014, \mnras, 444, 1453

\bibitem[{{Dubois} \& {Teyssier}(2008)}]{dubois&teyssier08winds}
{Dubois} Y., {Teyssier} R., 2008, \aap, 477, 79

\bibitem[{{Flin} \& {Godlowski}(1986)}]{flin86}
{Flin} P., {Godlowski} W., 1986, \mnras, 222, 525

\bibitem[{{Flin} \& {Godlowski}(1990)}]{flin90}
{Flin} P., {Godlowski} W., 1990, Soviet Astronomy Letters, 16, 209

\bibitem[{{Greggio} \& {Renzini}(1983)}]{greggio&renzini83}
{Greggio} L., {Renzini} A., 1983, \aap, 118, 217

\bibitem[{{Haardt} \& {Madau}(1996)}]{haardt&madau96}
{Haardt} F., {Madau} P., 1996, \apj, 461, 20

\bibitem[{{Hahn} {et~al}\mbox{.}(2007){Hahn}, {Porciani}, {Carollo}, \&
  {Dekel}}]{hahnetal07}
{Hahn} O., {Porciani} C., {Carollo} C.~M., {Dekel} A., 2007, \mnras, 375, 489

\bibitem[{{Hahn}, {Teyssier} \& {Carollo}(2010){Hahn}, {Teyssier}, \&
  {Carollo}}]{hahn10}
{Hahn} O., {Teyssier} R., {Carollo} C.~M., 2010, \mnras, 405, 274

\bibitem[{{Heymans} {et~al}\mbox{.}(2004){Heymans}, {Brown}, {Heavens},
  {Meisenheimer}, {Taylor}, \& {Wolf}}]{Hey++04}
{Heymans} C., {Brown} M., {Heavens} A., {Meisenheimer} K., {Taylor} A., {Wolf}
  C., 2004, \mnras, 347, 895

\bibitem[{{Hilbert} {et~al}\mbox{.}(2017){Hilbert}, {Xu}, {Schneider},
  {Springel}, {Vogelsberger}, \& {Hernquist}}]{Hilbert16}
{Hilbert} S., {Xu} D., {Schneider} P., {Springel} V., {Vogelsberger} M.,
  {Hernquist} L., 2017, \mnras, 468, 790

\bibitem[{{Hirv} {et~al}\mbox{.}(2017){Hirv}, {Pelt}, {Saar}, {Tago}, {Tamm},
  {Tempel}, \& {Einasto}}]{2017A&A...599A..31H}
{Hirv} A., {Pelt} J., {Saar} E., {Tago} E., {Tamm} A., {Tempel} E., {Einasto}
  M., 2017, \aap, 599, A31

\bibitem[{Hoyle(1949)}]{hoyle49}
Hoyle F., 1949, Problems of Cosmical Aerodynamics, Central Air Documents,
  Office, Dayton, OH. Central Air Documents Office, Dayton, OH, p. 195

\bibitem[{{Huang} {et~al}\mbox{.}(2016){Huang}, {Mandelbaum}, {Freeman},
  {Chen}, {Rozo}, {Rykoff}, \& {Baxter}}]{2016MNRAS.463..222H}
{Huang} H.-J., {Mandelbaum} R., {Freeman} P.~E., {Chen} Y.-C., {Rozo} E.,
  {Rykoff} E., {Baxter} E.~J., 2016, \mnras, 463, 222

\bibitem[{{Joachimi} {et~al}\mbox{.}(2015){Joachimi}, {Cacciato}, {Kitching},
  {Leonard}, {Mandelbaum}, {Sch{\"a}fer}, {Sif{\'o}n}, {Hoekstra}, {Kiessling},
  {Kirk}, \& {Rassat}}]{Joachimi15}
{Joachimi} B. {et~al.}, 2015, \ssr, 193, 1

\bibitem[{{Joachimi} {et~al}\mbox{.}(2011){Joachimi}, {Mandelbaum}, {Abdalla},
  \& {Bridle}}]{2011A&A...527A..26J}
{Joachimi} B., {Mandelbaum} R., {Abdalla} F.~B., {Bridle} S.~L., 2011, \aap,
  527, A26

\bibitem[{{Joachimi} {et~al}\mbox{.}(2013{\natexlab{a}}){Joachimi},
  {Semboloni}, {Bett}, {Hartlap}, {Hilbert}, {Hoekstra}, {Schneider}, \&
  {Schrabback}}]{joachimi13a}
{Joachimi} B., {Semboloni} E., {Bett} P.~E., {Hartlap} J., {Hilbert} S.,
  {Hoekstra} H., {Schneider} P., {Schrabback} T., 2013{\natexlab{a}}, \mnras,
  431, 477

\bibitem[{{Joachimi} {et~al}\mbox{.}(2013{\natexlab{b}}){Joachimi},
  {Semboloni}, {Hilbert}, {Bett}, {Hartlap}, {Hoekstra}, \&
  {Schneider}}]{Joachimi13b}
{Joachimi} B., {Semboloni} E., {Hilbert} S., {Bett} P.~E., {Hartlap} J.,
  {Hoekstra} H., {Schneider} P., 2013{\natexlab{b}}, \mnras, 436, 819

\bibitem[{{Kac}(1943)}]{Kac1943}
{Kac} M., 1943, Bull.~Am.~Math.~Soc., 49, 938

\bibitem[{{Kaiser}(1984)}]{1984ApJ...284L...9K}
{Kaiser} N., 1984, \apjl, 284, L9

\bibitem[{{Kang} \& {Wang}(2015)}]{2015ApJ...813....6K}
{Kang} X., {Wang} P., 2015, \apj, 813, 6

\bibitem[{{Kennicutt}(1998)}]{kennicutt98}
{Kennicutt}, Jr. R.~C., 1998, \apj, 498, 541

\bibitem[{{Kiessling} {et~al}\mbox{.}(2015){Kiessling}, {Cacciato}, {Joachimi},
  {Kirk}, {Kitching}, {Leonard}, {Mandelbaum}, {Sch{\"a}fer}, {Sif{\'o}n},
  {Brown}, \& {Rassat}}]{Kiessling15}
{Kiessling} A. {et~al.}, 2015, \ssr, 193, 67

\bibitem[{{Kirk}, {Bridle} \& {Schneider}(2010){Kirk}, {Bridle}, \&
  {Schneider}}]{Kirk10}
{Kirk} D., {Bridle} S., {Schneider} M., 2010, \mnras, 408, 1502

\bibitem[{{Kirk} {et~al}\mbox{.}(2015){Kirk}, {Brown}, {Hoekstra}, {Joachimi},
  {Kitching}, {Mandelbaum}, {Sif{\'o}n}, {Cacciato}, {Choi}, {Kiessling},
  {Leonard}, {Rassat}, \& {Sch{\"a}fer}}]{2015SSRv..193..139K}
{Kirk} D. {et~al.}, 2015, \ssr, 193, 139

\bibitem[{{Klypin} \& {Shandarin}(1983)}]{klypin&shandarin83}
{Klypin} A.~A., {Shandarin} S.~F., 1983, \mnras, 204, 891

\bibitem[{{Komatsu} {et~al}\mbox{.}(2011){Komatsu}, {Smith}, {Dunkley}, \&
  et~al.}]{komatsuetal11}
{Komatsu} E., {Smith} K.~M., {Dunkley} J., et~al., 2011, \apjs, 192, 18

\bibitem[{{Kraljic} {et~al}\mbox{.}(2018){Kraljic}, {Arnouts}, {Pichon},
  {Laigle}, {de la Torre}, {Vibert}, {Cadiou}, {Dubois}, {Treyer}, {Schimd},
  {Codis}, {de Lapparent}, {Devriendt}, {Hwang}, {Le Borgne}, {Malavasi},
  {Milliard}, {Musso}, {Pogosyan}, {Alpaslan}, {Bland-Hawthorn}, \&
  {Wright}}]{Kraljic2018}
{Kraljic} K. {et~al.}, 2018, \mnras, 474, 547

\bibitem[{{Krause}, {Eifler} \& {Blazek}(2016){Krause}, {Eifler}, \&
  {Blazek}}]{Krause15}
{Krause} E., {Eifler} T., {Blazek} J., 2016, \mnras, 456, 207

\bibitem[{{Krumholz} \& {Tan}(2007)}]{krumholz&tan07}
{Krumholz} M.~R., {Tan} J.~C., 2007, \apj, 654, 304

\bibitem[{{Laigle} {et~al}\mbox{.}(2015){Laigle}, {Pichon}, {Codis}, {Dubois},
  {Le Borgne}, {Pogosyan}, {Devriendt}, {Peirani}, {Prunet}, {Rouberol},
  {Slyz}, \& {Sousbie}}]{laigle2014}
{Laigle} C. {et~al.}, 2015, \mnras, 446, 2744

\bibitem[{{Lazeyras}, {Musso} \& {Schmidt}(2017){Lazeyras}, {Musso}, \&
  {Schmidt}}]{2017JCAP...03..059L}
{Lazeyras} T., {Musso} M., {Schmidt} F., 2017, \jcap, 3, 059

\bibitem[{{Lee}(2013)}]{2013arXiv1301.0348L}
{Lee} J., 2013, ArXiv e-prints

\bibitem[{Lee \& Erdogdu(2007)}]{leeetal07}
Lee J., Erdogdu P., 2007, \apj, 671, 1248

\bibitem[{{Lee}, {Kim} \& {Rey}(2018){Lee}, {Kim}, \&
  {Rey}}]{2018ApJ...860..127L}
{Lee} J., {Kim} S., {Rey} S.-C., 2018, \apj, 860, 127

\bibitem[{Lee \& Pen(2000)}]{lee&pen00}
Lee J., Pen U., 2000, \apj, 532, L5

\bibitem[{{Leitherer} {et~al}\mbox{.}(2010){Leitherer}, {Ortiz Ot{\'a}lvaro},
  {Bresolin}, {Kudritzki}, {Lo Faro}, {Pauldrach}, {Pettini}, \&
  {Rix}}]{leithereretal10}
{Leitherer} C., {Ortiz Ot{\'a}lvaro} P.~A., {Bresolin} F., {Kudritzki} R.-P.,
  {Lo Faro} B., {Pauldrach} A.~W.~A., {Pettini} M., {Rix} S.~A., 2010, \apjs,
  189, 309

\bibitem[{{Leitherer} {et~al}\mbox{.}(1999){Leitherer}, {Schaerer}, {Goldader},
  \& et~al.}]{leithereretal99}
{Leitherer} C., {Schaerer} D., {Goldader} J.~D., et~al., 1999, \apjs, 123, 3

\bibitem[{{Li} {et~al}\mbox{.}(2013){Li}, {Jing}, {Faltenbacher}, \&
  {Wang}}]{2013ApJ...770L..12L}
{Li} C., {Jing} Y.~P., {Faltenbacher} A., {Wang} J., 2013, \apjl, 770, L12

\bibitem[{{Libeskind} {et~al}\mbox{.}(2012){Libeskind}, {Hoffman}, {Knebe},
  {Steinmetz}, {Gottl{\"o}ber}, {Metuki}, \& {Yepes}}]{2012MNRAS.421L.137L}
{Libeskind} N.~I., {Hoffman} Y., {Knebe} A., {Steinmetz} M., {Gottl{\"o}ber}
  S., {Metuki} O., {Yepes} G., 2012, \mnras, 421, L137

\bibitem[{{Libeskind} {et~al}\mbox{.}(2013){Libeskind}, {Hoffman}, {Steinmetz},
  {Gottl{\"o}ber}, {Knebe}, \& {Hess}}]{libeskind13}
{Libeskind} N.~I., {Hoffman} Y., {Steinmetz} M., {Gottl{\"o}ber} S., {Knebe}
  A., {Hess} S., 2013, \apjl, 766, L15

\bibitem[{{Malavasi} {et~al}\mbox{.}(2017){Malavasi}, {Arnouts}, {Vibert}, {de
  la Torre}, {Moutard}, {Pichon}, {Davidzon}, {Kraljic}, {Bolzonella}, {Guzzo},
  {Garilli}, {Scodeggio}, {Granett}, {Abbas}, {Adami}, {Bottini}, {Cappi},
  {Cucciati}, {Franzetti}, {Fritz}, {Iovino}, {Krywult}, {Le Brun}, {Le
  F{\`e}vre}, {Maccagni}, {Ma{\l}ek}, {Marulli}, {Polletta}, {Pollo}, {Tasca},
  {Tojeiro}, {Vergani}, {Zanichelli}, {Bel}, {Branchini}, {Coupon}, {De Lucia},
  {Dubois}, {Hawken}, {Ilbert}, {Laigle}, {Moscardini}, {Sousbie}, {Treyer}, \&
  {Zamorani}}]{2017MNRAS.465.3817M}
{Malavasi} N. {et~al.}, 2017, \mnras, 465, 3817

\bibitem[{{Montero-Dorta} {et~al}\mbox{.}(2017){Montero-Dorta}, {P{\'e}rez},
  {Prada}, {Rodr{\'{\i}}guez-Torres}, {Favole}, {Klypin}, {Cid Fernandes},
  {Gonz{\'a}lez Delgado}, {Dom{\'{\i}}nguez}, {Bolton}, {Garc{\'{\i}}a-Benito},
  {Jullo}, \& {Niemiec}}]{2017ApJ...848L...2M}
{Montero-Dorta} A.~D. {et~al.}, 2017, \apjl, 848, L2

\bibitem[{{Musso} {et~al}\mbox{.}(2018){Musso}, {Cadiou}, {Pichon}, {Codis},
  {Kraljic}, \& {Dubois}}]{Musso2018}
{Musso} M., {Cadiou} C., {Pichon} C., {Codis} S., {Kraljic} K., {Dubois} Y.,
  2018, \mnras, 476, 4877

\bibitem[{{Navarro}, {Abadi} \& {Steinmetz}(2004){Navarro}, {Abadi}, \&
  {Steinmetz}}]{navarro04}
{Navarro} J.~F., {Abadi} M.~G., {Steinmetz} M., 2004, \apjl, 613, L41

\bibitem[{{Novikov}, {Colombi} \& {Dor{\'e}}(2006){Novikov}, {Colombi}, \&
  {Dor{\'e}}}]{novikov}
{Novikov} D., {Colombi} S., {Dor{\'e}} O., 2006, \mnras, 366, 1201

\bibitem[{{Obreschkow} \& {Glazebrook}(2014)}]{2014ApJ...784...26O}
{Obreschkow} D., {Glazebrook} K., 2014, \apj, 784, 26

\bibitem[{{Okumura} \& {Jing}(2009)}]{2009ApJ...694L..83O}
{Okumura} T., {Jing} Y.~P., 2009, \apjl, 694, L83

\bibitem[{{Okumura}, {Jing} \& {Li}(2009){Okumura}, {Jing}, \&
  {Li}}]{2009ApJ...694..214O}
{Okumura} T., {Jing} Y.~P., {Li} C., 2009, \apj, 694, 214

\bibitem[{{Paranjape} \& {Padmanabhan}(2017)}]{2017MNRAS.468.2984P}
{Paranjape} A., {Padmanabhan} N., 2017, \mnras, 468, 2984

\bibitem[{{Patiri} {et~al}\mbox{.}(2006){Patiri}, {Cuesta}, {Prada},
  {Betancort-Rijo}, \& {Klypin}}]{2006ApJ...652L..75P}
{Patiri} S.~G., {Cuesta} A.~J., {Prada} F., {Betancort-Rijo} J., {Klypin} A.,
  2006, \apjl, 652, L75

\bibitem[{{Peebles}(1969)}]{peebles69}
{Peebles} P.~J.~E., 1969, \apj, 155, 393

\bibitem[{{Pichon} \& {Bernardeau}(1999)}]{pichon99}
{Pichon} C., {Bernardeau} F., 1999, \aap, 343, 663

\bibitem[{{Pichon} {et~al}\mbox{.}(2011){Pichon}, {Pogosyan}, {Kimm}, {Slyz},
  {Devriendt}, \& {Dubois}}]{2011MNRAS.418.2493P}
{Pichon} C., {Pogosyan} D., {Kimm} T., {Slyz} A., {Devriendt} J., {Dubois} Y.,
  2011, \mnras, 418, 2493

\bibitem[{{Pogosyan}, {Bond} \& {Kofman}(1998){Pogosyan}, {Bond}, \&
  {Kofman}}]{pogosyanetal1998}
{Pogosyan} D., {Bond} J.~R., {Kofman} L., 1998, \jrasc, 92, 313

\bibitem[{{Pogosyan} {et~al}\mbox{.}(2009){Pogosyan}, {Pichon}, {Gay},
  {Prunet}, {Cardoso}, {Sousbie}, \& {Colombi}}]{pogo09}
{Pogosyan} D., {Pichon} C., {Gay} C., {Prunet} S., {Cardoso} J.~F., {Sousbie}
  T., {Colombi} S., 2009, \mnras, 396, 635

\bibitem[{Porciani, Dekel \& Hoffman(2002)Porciani, Dekel, \&
  Hoffman}]{porcianietal02}
Porciani C., Dekel A., Hoffman Y., 2002, \mnras, 332, 325

\bibitem[{{Power} {et~al}\mbox{.}(2003){Power}, {Navarro}, {Jenkins}, {Frenk},
  {White}, {Springel}, {Stadel}, \& {Quinn}}]{Power03}
{Power} C., {Navarro} J.~F., {Jenkins} A., {Frenk} C.~S., {White} S.~D.~M.,
  {Springel} V., {Stadel} J., {Quinn} T., 2003, \mnras, 338, 14

\bibitem[{{Rasera} \& {Teyssier}(2006)}]{rasera&teyssier06}
{Rasera} Y., {Teyssier} R., 2006, \aap, 445, 1

\bibitem[{{Rice}(1945)}]{Rice1945}
{Rice} S.~O., 1945, Bell~System~Tech.~J., 25, 46

\bibitem[{{Roberts} \& {Haynes}(1994)}]{1994ARA&A..32..115R}
{Roberts} M.~S., {Haynes} M.~P., 1994, \araa, 32, 115

\bibitem[{Salpeter(1955)}]{salpeter55}
Salpeter E.~E., 1955, \apj, 121, 161

\bibitem[{{Sandage}, {Freeman} \& {Stokes}(1970){Sandage}, {Freeman}, \&
  {Stokes}}]{1970ApJ...160..831S}
{Sandage} A., {Freeman} K.~C., {Stokes} N.~R., 1970, \apj, 160, 831

\bibitem[{{Schaefer}(2009)}]{schaefer09}
{Schaefer} B.~M., 2009, International Journal of Modern Physics D, 18, 173

\bibitem[{{Schneider} \& {Bridle}(2010)}]{Schneider10}
{Schneider} M.~D., {Bridle} S., 2010, \mnras, 402, 2127

\bibitem[{{Shandarin} \& {Klypin}(1984)}]{1984AZh....61..837S}
{Shandarin} S.~F., {Klypin} A.~A., 1984, \azh, 61, 837

\bibitem[{{Singh}, {Mandelbaum} \& {More}(2015){Singh}, {Mandelbaum}, \&
  {More}}]{2015MNRAS.450.2195S}
{Singh} S., {Mandelbaum} R., {More} S., 2015, \mnras, 450, 2195

\bibitem[{{Slosar} \& {White}(2009)}]{slosar09}
{Slosar} A., {White} M., 2009, \jcap, 6, 9

\bibitem[{{Smargon} {et~al}\mbox{.}(2012){Smargon}, {Mandelbaum}, {Bahcall}, \&
  {Niederste-Ostholt}}]{2012MNRAS.423..856S}
{Smargon} A., {Mandelbaum} R., {Bahcall} N., {Niederste-Ostholt} M., 2012,
  \mnras, 423, 856

\bibitem[{{Sousbie}(2011)}]{2011MNRAS.414..350S}
{Sousbie} T., 2011, \mnras, 414, 350

\bibitem[{{Sousbie}(2013)}]{2013arXiv1302.6221S}
{Sousbie} T., 2013, ArXiv e-prints

\bibitem[{{Sousbie}, {Colombi} \& {Pichon}(2009){Sousbie}, {Colombi}, \&
  {Pichon}}]{sousbie09}
{Sousbie} T., {Colombi} S., {Pichon} C., 2009, \mnras, 393, 457

\bibitem[{Sousbie {et~al}\mbox{.}(2008)Sousbie, Pichon, Colombi, \&
  Pogosyan}]{sousbie08}
Sousbie T., Pichon C., Colombi S., Pogosyan D., 2008, \mnras, 383, 1655

\bibitem[{{Sutherland} \& {Dopita}(1993)}]{sutherland&dopita93}
{Sutherland} R.~S., {Dopita} M.~A., 1993, \apjs, 88, 253

\bibitem[{{Tempel} \& {Libeskind}(2013)}]{tempel&libeskind13}
{Tempel} E., {Libeskind} N.~I., 2013, \apjl, 775, L42

\bibitem[{Tempel, Stoica \& Saar(2013)Tempel, Stoica, \& Saar}]{Tempel13}
Tempel E., Stoica R.~S., Saar E., 2013, \mnras, 428, 1827

\bibitem[{{Tenneti}, {Mandelbaum} \& {Di Matteo}(2016){Tenneti}, {Mandelbaum},
  \& {Di Matteo}}]{Tenneti16}
{Tenneti} A., {Mandelbaum} R., {Di Matteo} T., 2016, \mnras, 462, 2668

\bibitem[{{Tenneti} {et~al}\mbox{.}(2014){Tenneti}, {Mandelbaum}, {Di Matteo},
  {Feng}, \& {Khandai}}]{Ten++14}
{Tenneti} A., {Mandelbaum} R., {Di Matteo} T., {Feng} Y., {Khandai} N., 2014,
  \mnras, 441, 470

\bibitem[{{Tenneti} {et~al}\mbox{.}(2015){Tenneti}, {Singh}, {Mandelbaum},
  {Matteo}, {Feng}, \& {Khandai}}]{Tenneti15a}
{Tenneti} A., {Singh} S., {Mandelbaum} R., {Matteo} T.~D., {Feng} Y., {Khandai}
  N., 2015, \mnras, 448, 3522

\bibitem[{{Teyssier}(2002)}]{teyssier02}
{Teyssier} R., 2002, \aap, 385, 337

\bibitem[{Teyssier {et~al}\mbox{.}(2009)Teyssier, Pires, Prunet, Aubert,
  Pichon, Amara, Benabed, Colombi, Refregier, \& Starck}]{teyssieretal09}
Teyssier R. {et~al.}, 2009, \aa, 497, 335

\bibitem[{{Trowland}, {Lewis} \& {Bland-Hawthorn}(2013){Trowland}, {Lewis}, \&
  {Bland-Hawthorn}}]{2013ApJ...762...72T}
{Trowland} H.~E., {Lewis} G.~F., {Bland-Hawthorn} J., 2013, \apj, 762, 72

\bibitem[{{Troxel} \& {Ishak}(2015)}]{troxel15}
{Troxel} M.~A., {Ishak} M., 2015, \physrep, 558, 1

\bibitem[{{Trujillo}, {Carretero} \& {Patiri}(2006){Trujillo}, {Carretero}, \&
  {Patiri}}]{trujillo06}
{Trujillo} I., {Carretero} C., {Patiri} S.~G., 2006, \apjl, 640, L111

\bibitem[{{van Uitert} \& {Joachimi}(2017)}]{2017MNRAS.468.4502V}
{van Uitert} E., {Joachimi} B., 2017, \mnras, 468, 4502

\bibitem[{{Varela} {et~al}\mbox{.}(2012){Varela}, {Betancort-Rijo}, {Trujillo},
  \& {Ricciardelli}}]{varela11}
{Varela} J., {Betancort-Rijo} J., {Trujillo} I., {Ricciardelli} E., 2012, \apj,
  744, 82

\bibitem[{{Velliscig} {et~al}\mbox{.}(2015{\natexlab{a}}){Velliscig},
  {Cacciato}, {Schaye}, {Crain}, {Bower}, {van Daalen}, {Dalla Vecchia},
  {Frenk}, {Furlong}, {McCarthy}, {Schaller}, \&
  {Theuns}}]{2015MNRAS.453..721V}
{Velliscig} M. {et~al.}, 2015{\natexlab{a}}, \mnras, 453, 721

\bibitem[{{Velliscig} {et~al}\mbox{.}(2015{\natexlab{b}}){Velliscig},
  {Cacciato}, {Schaye}, {Hoekstra}, {Bower}, {Crain}, {van Daalen}, {Furlong},
  {McCarthy}, {Schaller}, \& {Theuns}}]{Velliscig15b}
{Velliscig} M. {et~al.}, 2015{\natexlab{b}}, \mnras, 454, 3328

\bibitem[{{Wang} \& {Kang}(2017)}]{2017MNRAS.468L.123W}
{Wang} P., {Kang} X., 2017, \mnras, 468, L123

\bibitem[{{Wang} \& {Kang}(2018)}]{2018MNRAS.473.1562W}
{Wang} P., {Kang} X., 2018, \mnras, 473, 1562

\bibitem[{{Wang} {et~al}\mbox{.}(2014){Wang}, {Szalay}, {Arag{\'o}n-Calvo},
  {Neyrinck}, \& {Eyink}}]{2014ApJ...793...58W}
{Wang} X., {Szalay} A., {Arag{\'o}n-Calvo} M.~A., {Neyrinck} M.~C., {Eyink}
  G.~L., 2014, \apj, 793, 58

\bibitem[{{Wei} {et~al}\mbox{.}(2018){Wei}, {Li}, {Kang}, {Luo}, {Xia}, {Wang},
  {Yang}, {Wang}, {Jing}, {Mo}, {Lin}, {Wang}, {Li}, {Lu}, {Zhang}, {Lim},
  {Tweed}, \& {Cui}}]{2018ApJ...853...25W}
{Wei} C. {et~al.}, 2018, \apj, 853, 25

\bibitem[{{Welker} {et~al}\mbox{.}(2014){Welker}, {Devriendt}, {Dubois},
  {Pichon}, \& {Peirani}}]{2014MNRAS.445L..46W}
{Welker} C., {Devriendt} J., {Dubois} Y., {Pichon} C., {Peirani} S., 2014,
  \mnras, 445, L46

\bibitem[{{Welker} {et~al}\mbox{.}(2015){Welker}, {Dubois}, {Pichon},
  {Devriendt}, \& {Chisari}}]{welkeretal15}
{Welker} C., {Dubois} Y., {Pichon} C., {Devriendt} J., {Chisari} E.~N., 2015,
  ArXiv e-prints

\bibitem[{{Welker} {et~al}\mbox{.}(2017){Welker}, {Power}, {Pichon}, {Dubois},
  {Devriendt}, \& {Codis}}]{2017arXiv171207818W}
{Welker} C., {Power} C., {Pichon} C., {Dubois} Y., {Devriendt} J., {Codis} S.,
  2017, ArXiv e-prints

\bibitem[{{White}(1984)}]{white84}
{White} S.~D.~M., 1984, \apj, 286, 38

\bibitem[{{White}, {Tully} \& {Davis}(1988){White}, {Tully}, \&
  {Davis}}]{1988ApJ...333L..45W}
{White} S.~D.~M., {Tully} R.~B., {Davis} M., 1988, \apjl, 333, L45

\bibitem[{{Zhang} {et~al}\mbox{.}(2009){Zhang}, {Yang}, {Faltenbacher},
  {Springel}, {Lin}, \& {Wang}}]{2009ApJ...706..747Z}
{Zhang} Y., {Yang} X., {Faltenbacher} A., {Springel} V., {Lin} W., {Wang} H.,
  2009, \apj, 706, 747

\bibitem[{{Zhang} {et~al}\mbox{.}(2015){Zhang}, {Yang}, {Wang}, {Wang}, {Luo},
  {Mo}, \& {van den Bosch}}]{2015ApJ...798...17Z}
{Zhang} Y., {Yang} X., {Wang} H., {Wang} L., {Luo} W., {Mo} H.~J., {van den
  Bosch} F.~C., 2015, \apj, 798, 17

\bibitem[{{Zhang} {et~al}\mbox{.}(2013){Zhang}, {Yang}, {Wang}, {Wang}, {Mo},
  \& {van den Bosch}}]{2013ApJ...779..160Z}
{Zhang} Y., {Yang} X., {Wang} H., {Wang} L., {Mo} H.~J., {van den Bosch} F.~C.,
  2013, \apj, 779, 160

\bibitem[{{Zhu} \& {Feng}(2017)}]{2017ApJ...838...21Z}
{Zhu} W., {Feng} L.-L., 2017, \apj, 838, 21

\end{thebibliography}

\appendix

\section{Galactic and halo alignment}
\label{sec:halo-gal}

\begin{figure*}
\includegraphics[width=0.95\columnwidth]{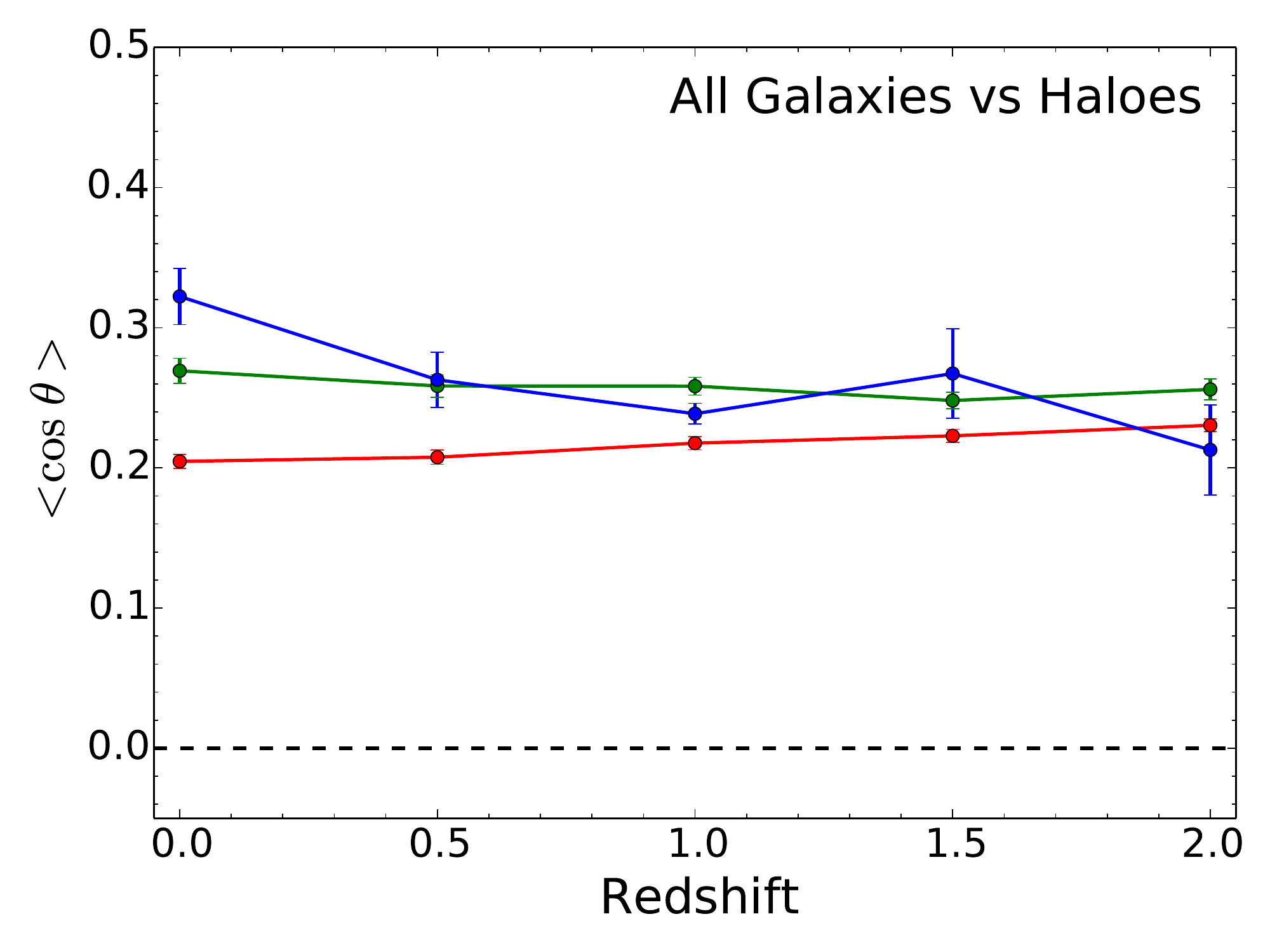}
\includegraphics[width=0.95\columnwidth]{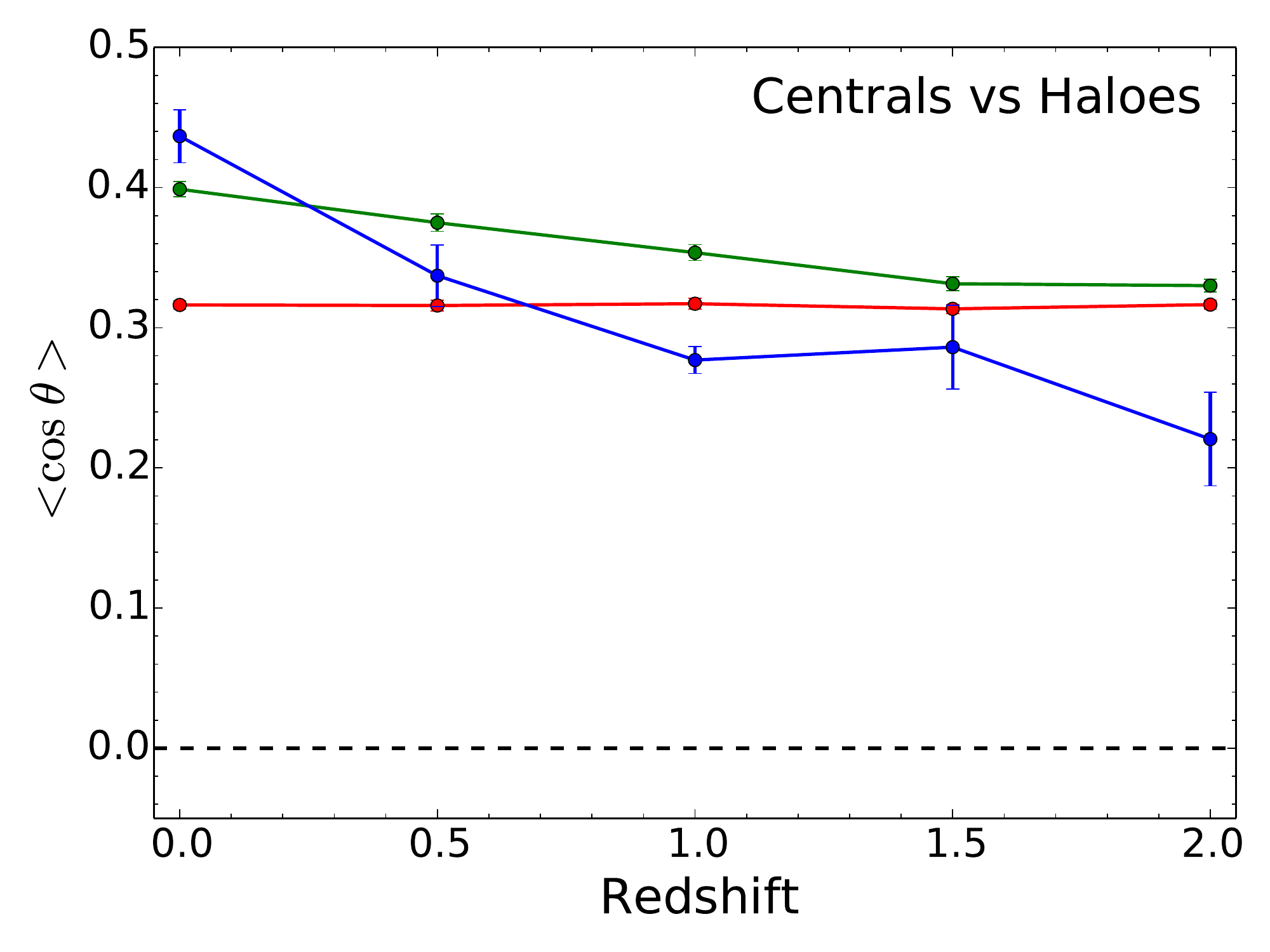}
\includegraphics[width=0.95\columnwidth]{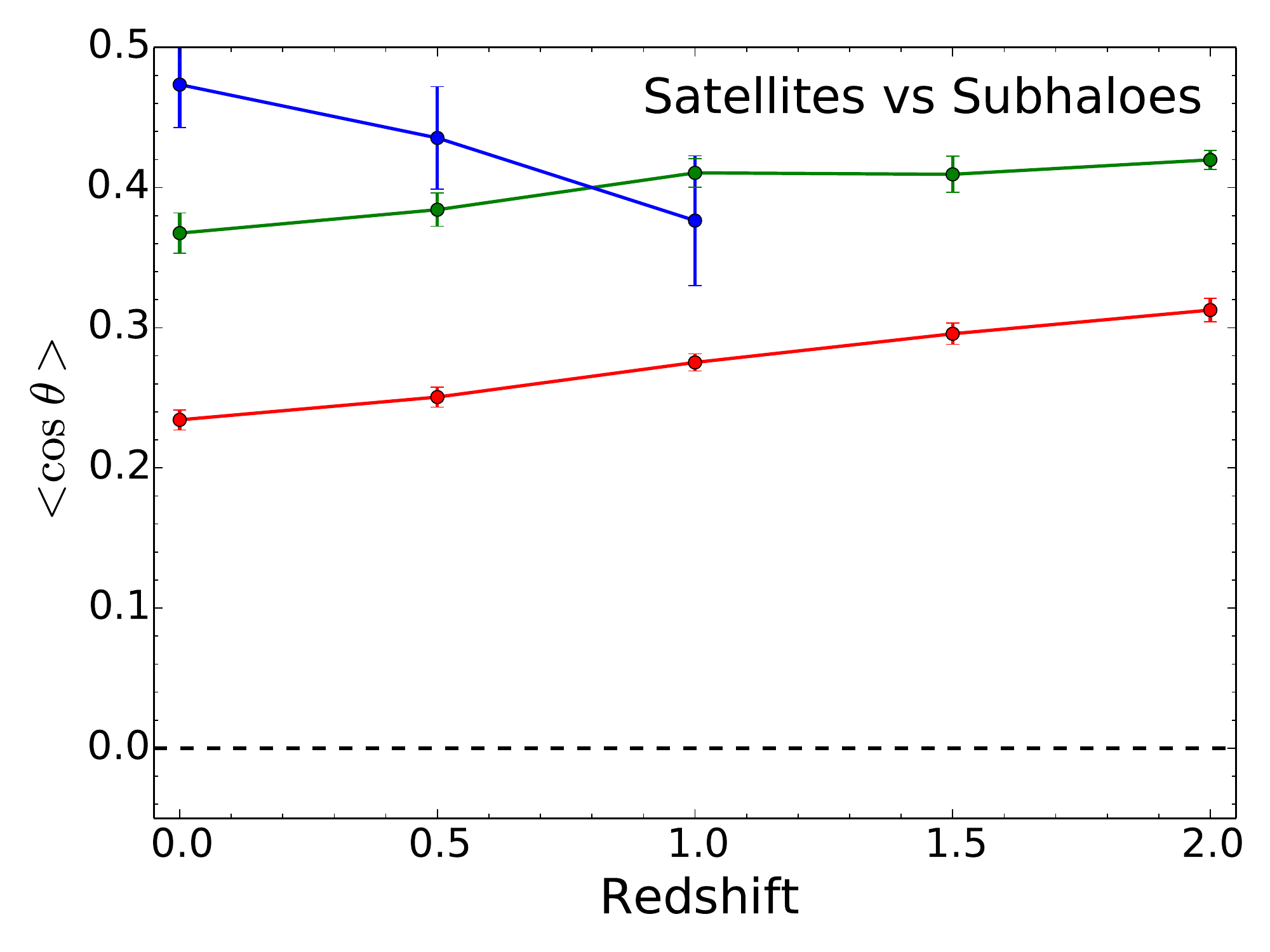}
\includegraphics[width=0.95\columnwidth]{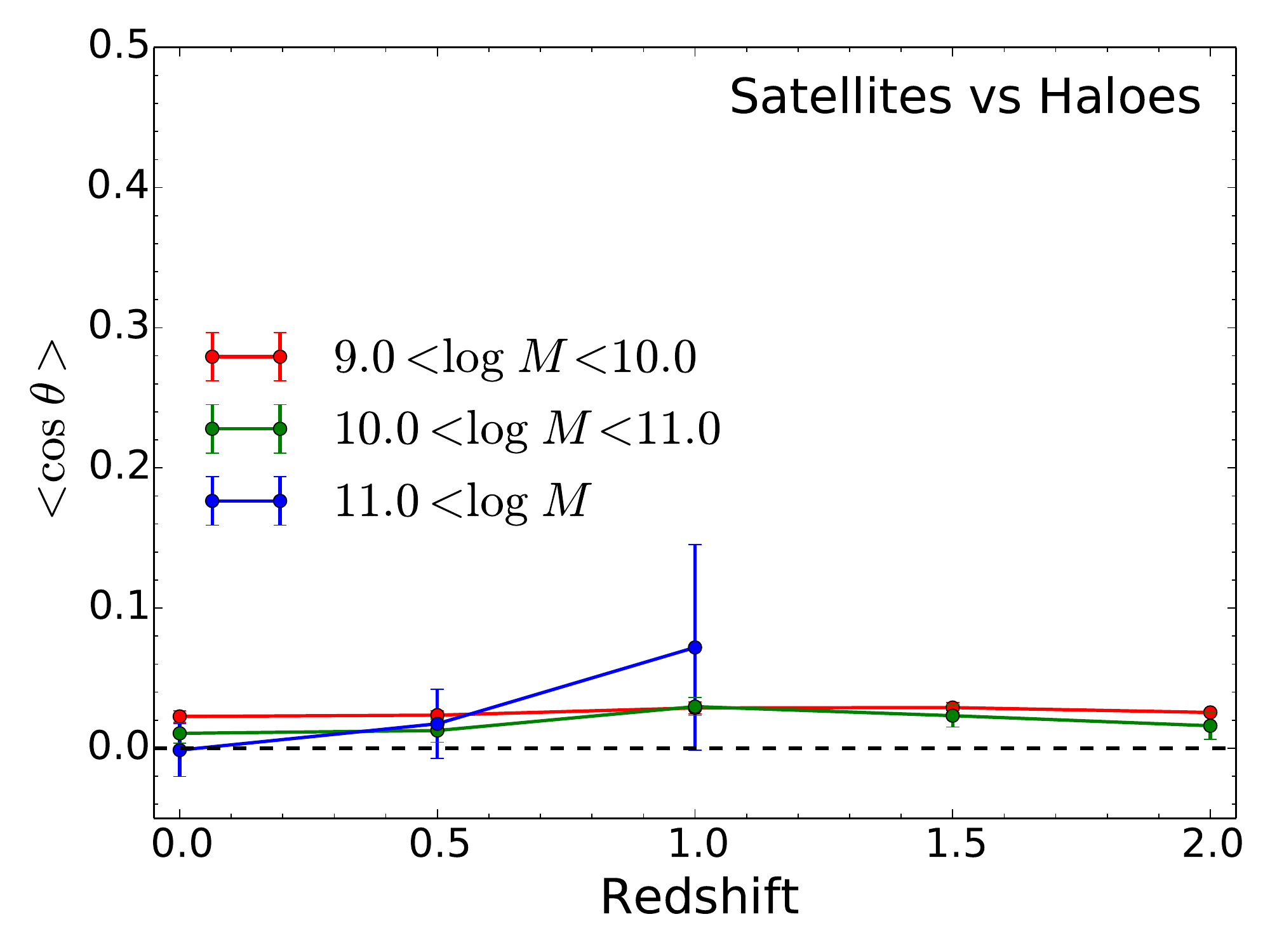}
\caption{The redshift evolution of the mean cosine of the angle between the spins of central and satellite galaxies and the nearest halo or subhalo as labelled. The data is separated into three stellar mass bins as labelled. The alignment of the spin of central galaxies and their host haloes is rather good and increases towards low redshift. Satellite galaxies align very well their spin with the spin of their host subhalo but very little correlation is found with the spin of the parent halo.
}
\label{fig:MeanCosThetas}
\end{figure*}

Let us characterize the relation between galaxy orientation and the host's halo. This relation is of interest for constructing and testing `halo models' of intrinsic alignments. 

In order to model how halo alignments translate into galaxy alignments  it is often assumed in the context of the halo model  that galaxy and halo spins align so that one can use a semi-analytical model in which galaxies are assigned to haloes with a  spin and shape orientation that is drawn from a PDF centred on halo's spin and orientation and with some width \citep{Hey++04,Joachimi13b}.
Let us therefore check how much galaxy and halo align their spin in order to  predict galactic alignments to be  compared to observations.
\cite{2017MNRAS.472.1163C} analysed the Horizon-AGN and quantified how the shape of galaxies and haloes are correlated. They found a mean misalignment angle between minor vectors of $\sim 45-55\degree$ depending on mass and redshift. Let us therefore first investigate if this shape alignment between galaxies and haloes also holds for spin alignments. 
To investigate this galaxy-halo connection,  the alignment of the galactic spins with the spin of their associated halo or subhalo is measured in Horizon-AGN.

The mean of the cosine of the angle between the spin of galaxies and their corresponding haloes or subhaloes is then computed. This cosine can take values between -1 and 1 and its mean is zero if there is no correlation. Error bars are estimated as the square root of the measured variance $\left\langle \cos ^{2}\theta\right\rangle_{c}$ divided by the the square root of the number of objects. Figure~\ref{fig:MeanCosThetas} shows that central galaxies tend to be aligned with their host haloes and satellite galaxies with their host subhaloes rather than the main halo for which there is no significant signal. Note that error bars for the high mass satellite galaxies are large given their small number. Interestingly, the amplitude of the alignment is similar for satellites and centrals. Overall, low mass satellite galaxies  become less aligned towards low redshifts while high mass (usually central) galaxies become more aligned with their respective haloes. The redshift evolution of the most massive objects is probably mainly due to the fact that masses evolve with redshift and therefore a fixed level of non-linearity corresponds to larger masses at lower redshift. To investigate this point further, one would need to use the merger history of galaxies but this is beyond the scope of this work.

The results presented here are consistent with \cite{2017MNRAS.472.1163C} who found similar alignments between the spins of galaxies and their matched haloes in a pure dark matter twin simulation but remain quite different from the alignment of the shapes of galaxies and haloes as the result of the redshift evolution of the fraction of disks for which  a better alignment of the spin and shape is expected.

In order to quantitatively compare our results with the literature, let us compute the mean misalignment angle (and not mean cosine of the angle).
Overall, it is found that the spin of galaxies and haloes are on average misaligned by $\sim 45-55\degree$ depending on mass and redshift, a result which is similar to the misalignment of the shapes found by \cite{2015MNRAS.453..721V,2017MNRAS.472.1163C}. This is to be compared with the expectation for uncorrelated directions $180/\pi\degree=57.3\degree$\footnote{Indeed, the mean misalignment angle in the uniform case is given by $\int_0^ {\pi/2}\theta\sin\theta {\rm d}\theta=1$ rad $\sim 57.3\degree$. }. This result holds if one does not account for the orientation of the spin vectors (i.e the angle goes from 0 to 90$\degree$). When the orientation is taken into account (i.e the angle goes from 0 to 180$\degree$), the mean misalignment angle becomes $\sim 55-75\degree$ (to be compared to 90$\degree$).

\section{Similarity of the skeletons}
\label{app:skel}
\begin{figure}
\includegraphics[width=0.95\columnwidth]{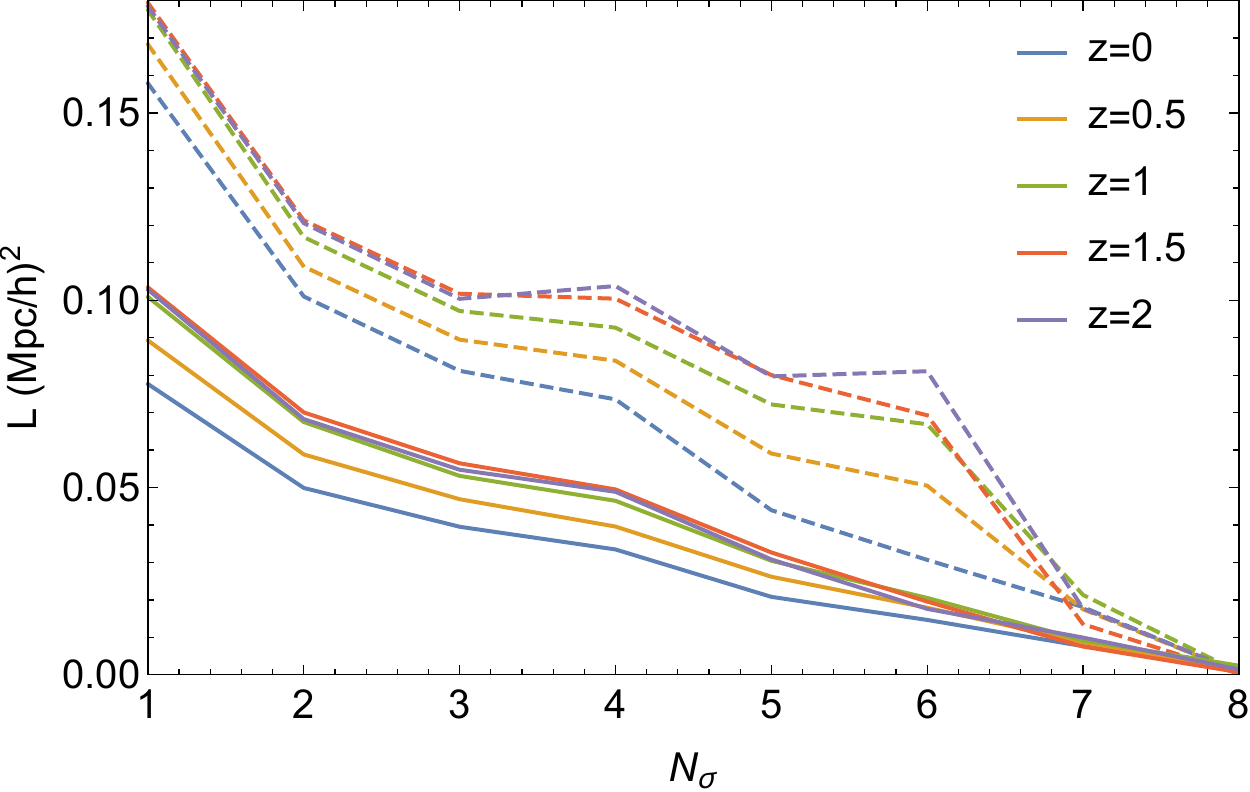}
\caption{Total length of filaments per unit volume extracted from the galaxy skeleton (solid lines) or halo skeleton (dashed lines) as a function of persistence level and for different redshifts as labelled.}
\label{fig:LengthVsSigma}
\end{figure}

\begin{table}
\centering
\begin{tabular}{|c|c|c|c|c|c|c|c|}
\hline
\!\!\!Galaxy Persistence :       \!  \! \! \!\!& 1   & 2   & 3   & 4   & 5   & 6   \\\hline
\!\!\!Halo Persistence, z=0\!\!\! \!\!& 3.5 & 4.8 & 5.3 & 5.7 & 6.8 & 7.2 \\\hline
\!\!\!Halo Persistence, z=0.5\!\!\! \!\!& 3.0 & 5.0 & 6.1 & 6.3 & 6.7 & 7.0 \\\hline
\!\!\!Halo Persistence, z=1\!\!\! \!\!& 2.8 & 5.9 & 6.3 & 6.4 & 6.8 & 7.0 \\\hline
\!\!\!Halo Persistence, z=1.5\!\!\! \!\!& 2.9 & 5.9 & 6.2 & 6.4 & 6.7 & 6.9 \\\hline
\!\!\!Halo Persistence, z=2\!\!\!\! \!& 2.8 & 6.2 & 6.4 & 6.5 & 6.8 & 7.0 \\\hline
\end{tabular}%
\caption{Persistence levels of galaxies and haloes matched for identical total length of filaments.}
\label{tab:SigmaMatch}
\end{table}

This appendix compares skeletons based on  halo distribution and those constructed from the galaxy population.  
To quantify these differences the total lengths of all filament  extracted by  {\sc DisPerSE}  is used, and compared   for both the galaxy and halo catalogues at various levels of persistence. Generally, one expects the total length of the filaments to decrease with decreasing numbers of objects in the catalogue, and decrease with increasing values of persistence as only the most prominent filaments are kept. In this section, $L$  is the density of filaments i.e the total length divided by the volume.

Figure~\ref{fig:LengthVsSigma} displays the density of filaments obtained from the galaxy distribution (solid lines) or halo distribution (dashed lines) as a function of the persistence level and for different redshifts between 0 and 2.
The overall trend is for the total length of filaments per unit volume to decrease with the persistence level as expected. The number of nodes generally also increases with redshift,  hence the  overall increase of the total length of filaments with redshift at a fixed value of persistence. Furthermore, the total length of halo filaments is always larger than for galaxies, given the relative abundance of haloes compared to galaxies (see also Section~\ref{sec:defgal}).

Due to the larger number of halo filaments, measurements done using a galaxy skeleton are consistent with a halo skeleton for persistence levels larger than their galaxy counterpart. To quantify this effect,  let us use the total length of filaments (as shown in  Figure~\ref{fig:LengthVsSigma}) as a ruler, and match the total lengths of filaments to find a mapping between the persistence levels of galaxies and haloes. The result for $z=0$ and $z=2$ is shown in Table~\ref{tab:SigmaMatch}. Varying the persistence level of galaxies from 1 to 7 is shown to correspond to a smaller range of halo persistence which is shown to vary from 3 and 8. As expected, the same total lengths of filaments occur for haloes at a systematically higher value of the persistence due to the larger number of haloes compared to galaxies. Note that the correspondence between halo and galaxy persistence is slightly redshift-dependent, mainly due to the  time-evolution of the relative number of haloes and galaxies.

\section{Evolution with persistence}
\label{sec:persistence}

\begin{figure*}
\includegraphics[width=0.66\columnwidth]{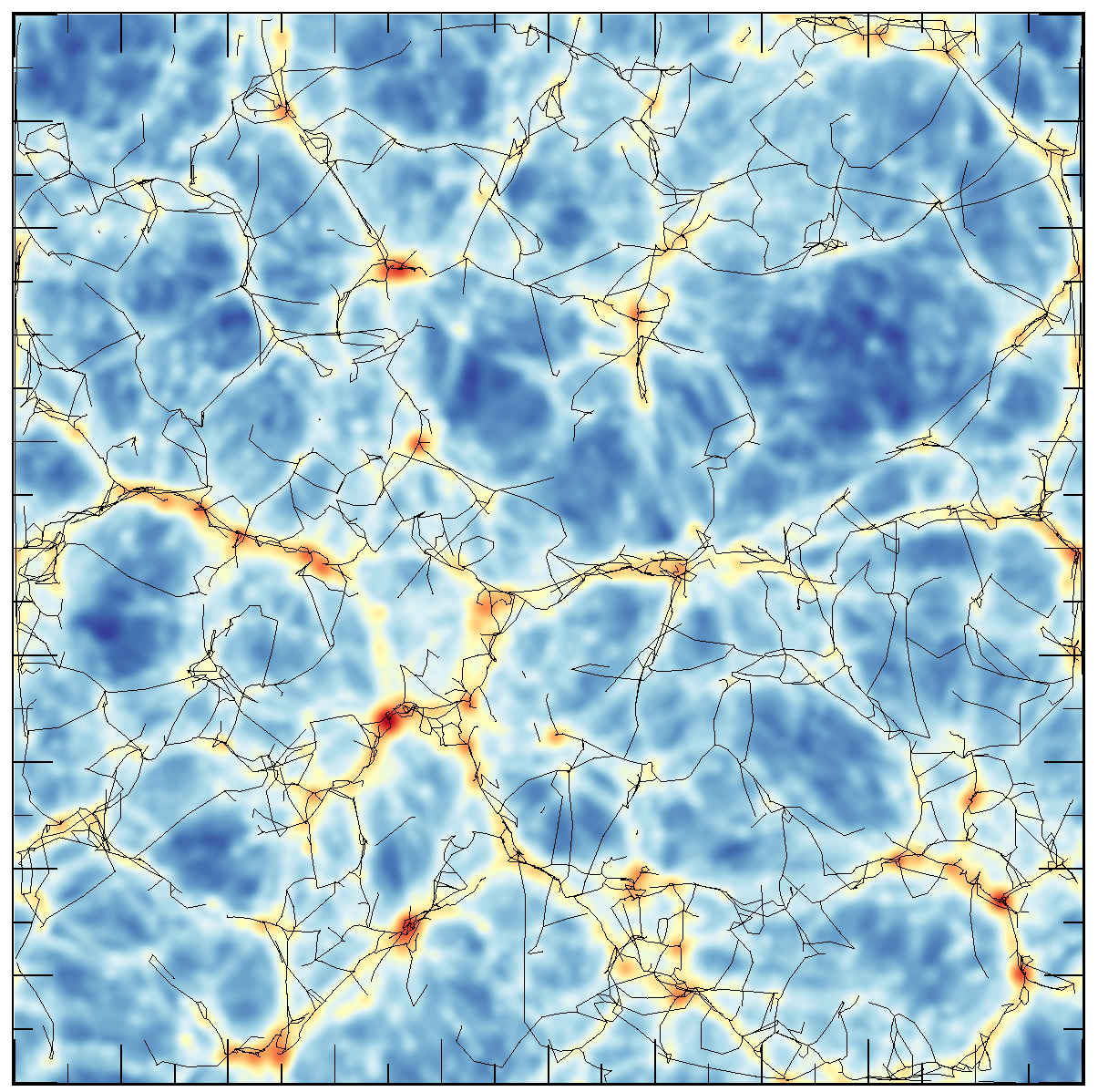}
\includegraphics[width=0.66\columnwidth]{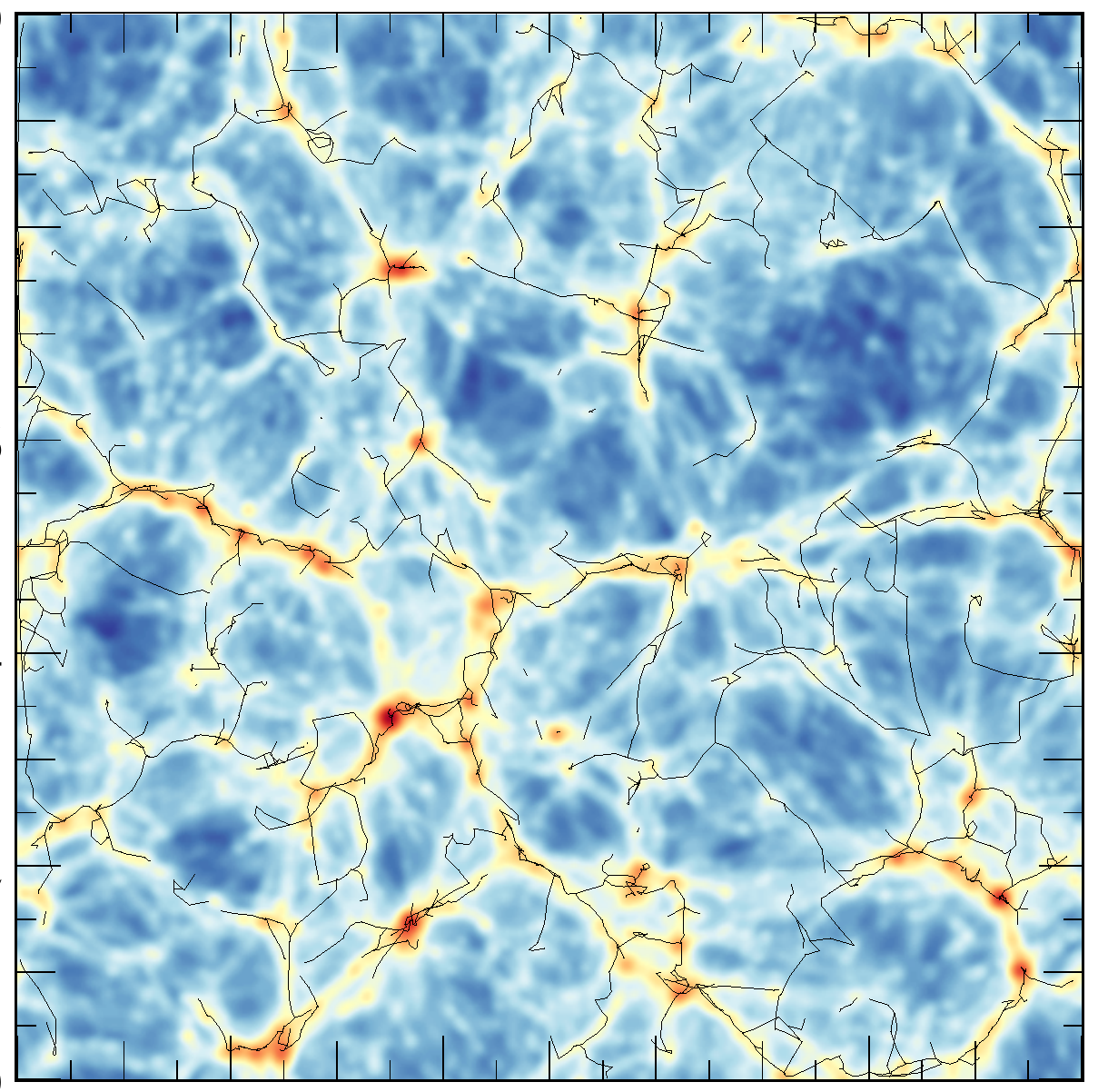}
\includegraphics[width=0.66\columnwidth]{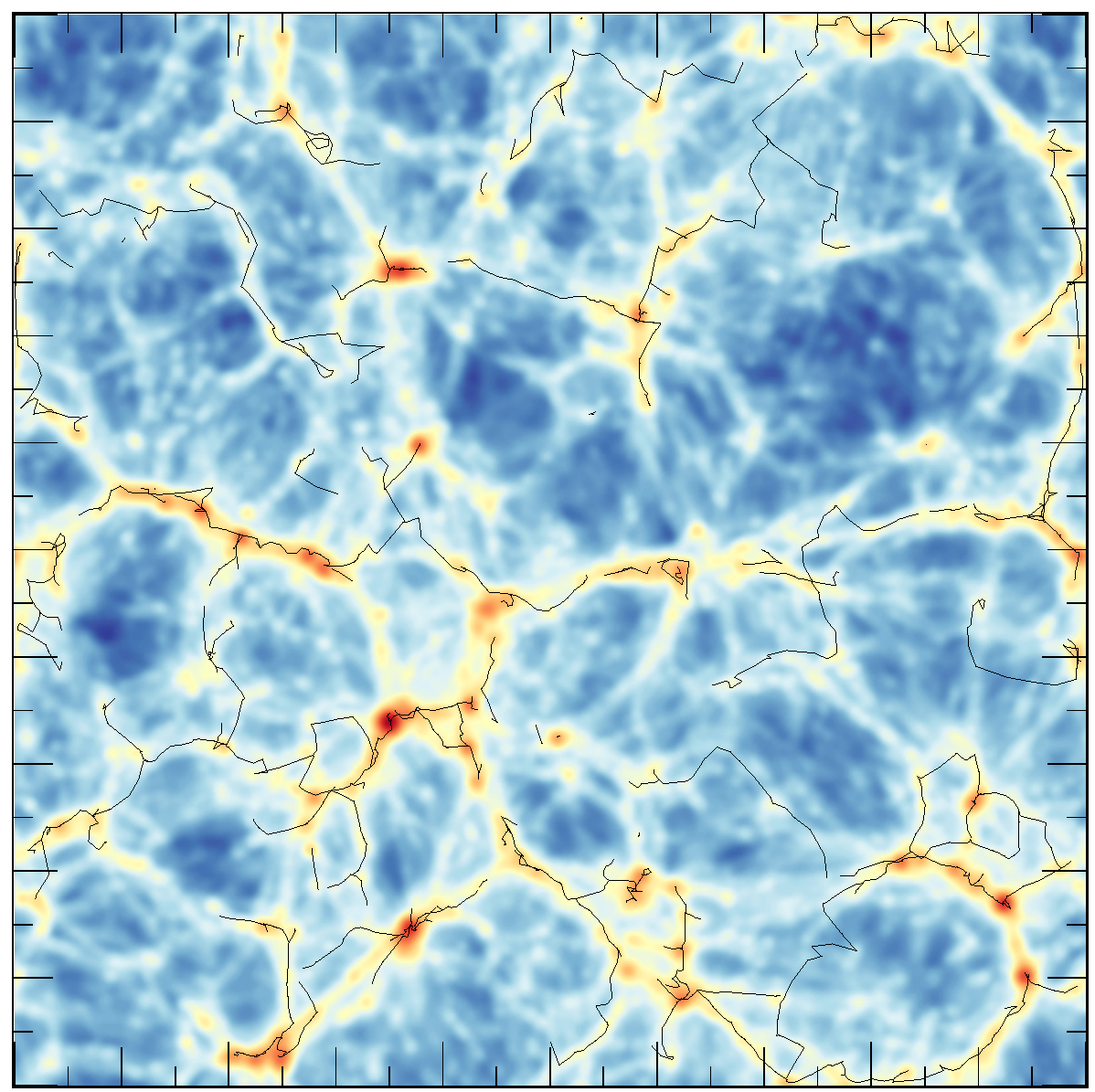}
\caption{Same as the left-hand panel of Figure~\ref{fig:Filaments} for different persistence threshold $N_{\sigma}=1,3,5$  from left to right-hand panels and redshift $z=0$. Note the absence of many filament features in the higher persistence level to be contrasted with low persistence skeleton which displays lots of structures. As expected, this persistence level allows us to filter out the noise and to focus on the most prominent filaments.
 }
\label{fig:FilSlices}
\end{figure*}
\begin{figure*}
\includegraphics[width=0.66\columnwidth]{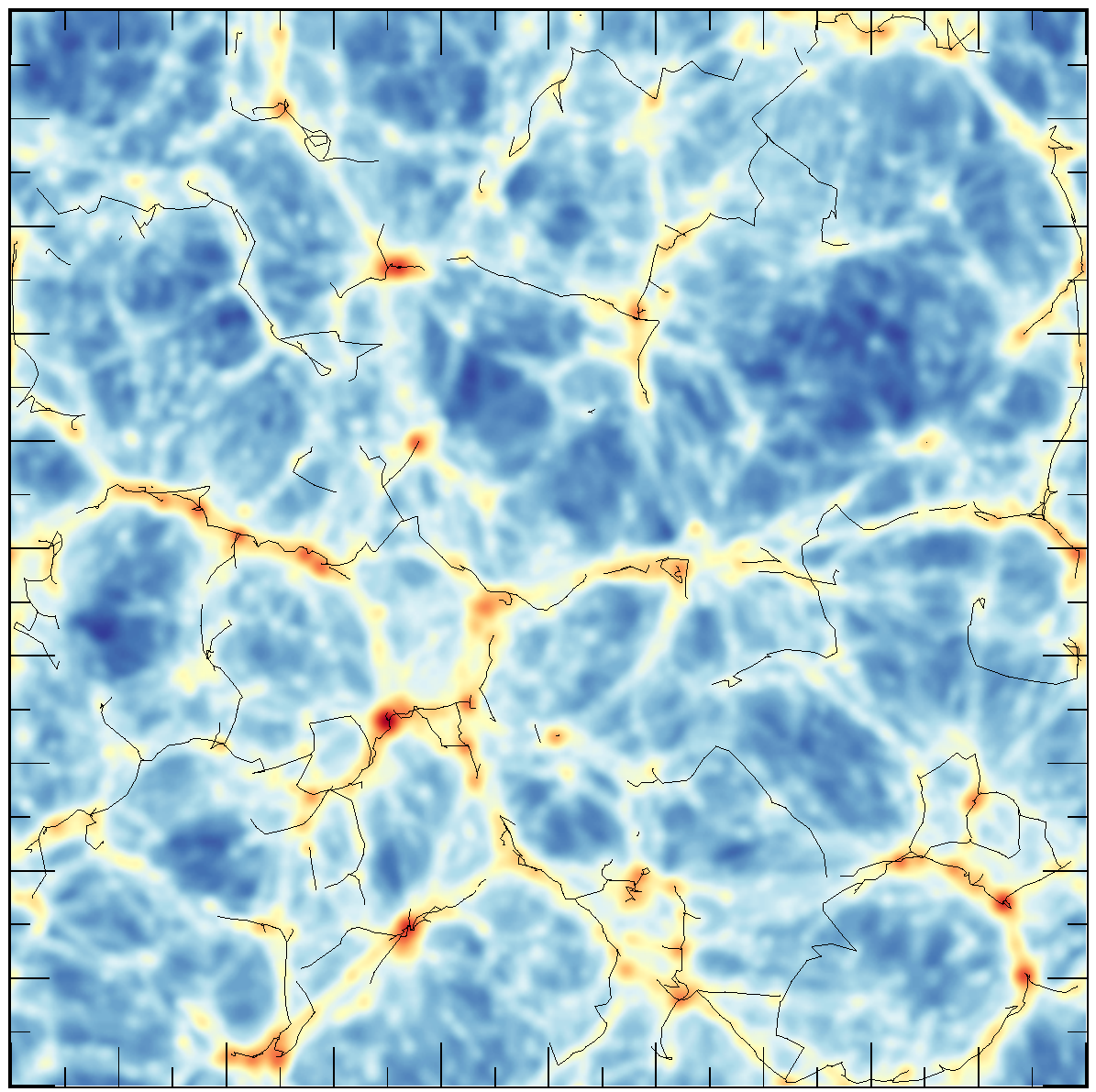}
\includegraphics[width=0.66\columnwidth]{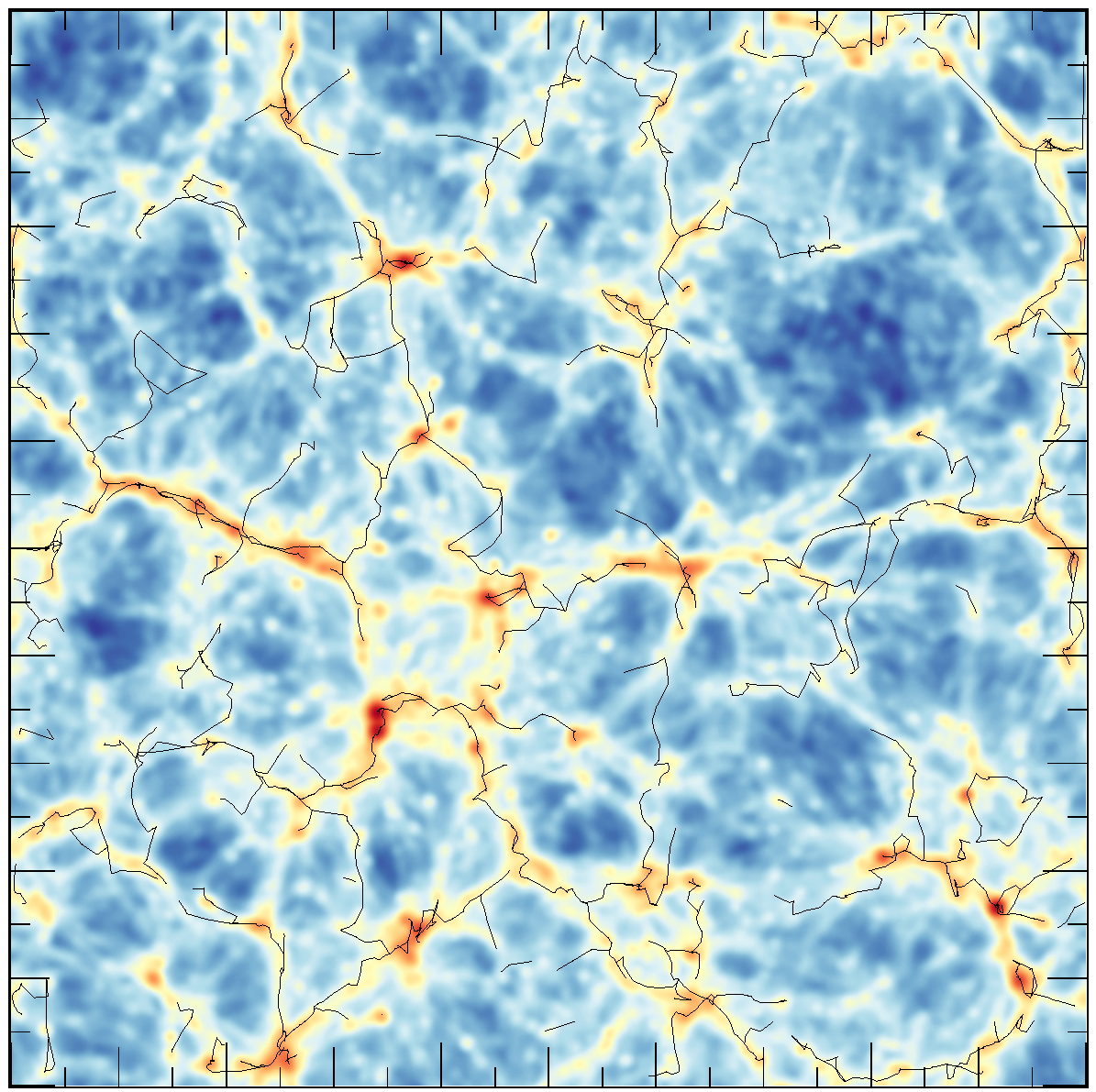}
\includegraphics[width=0.66\columnwidth]{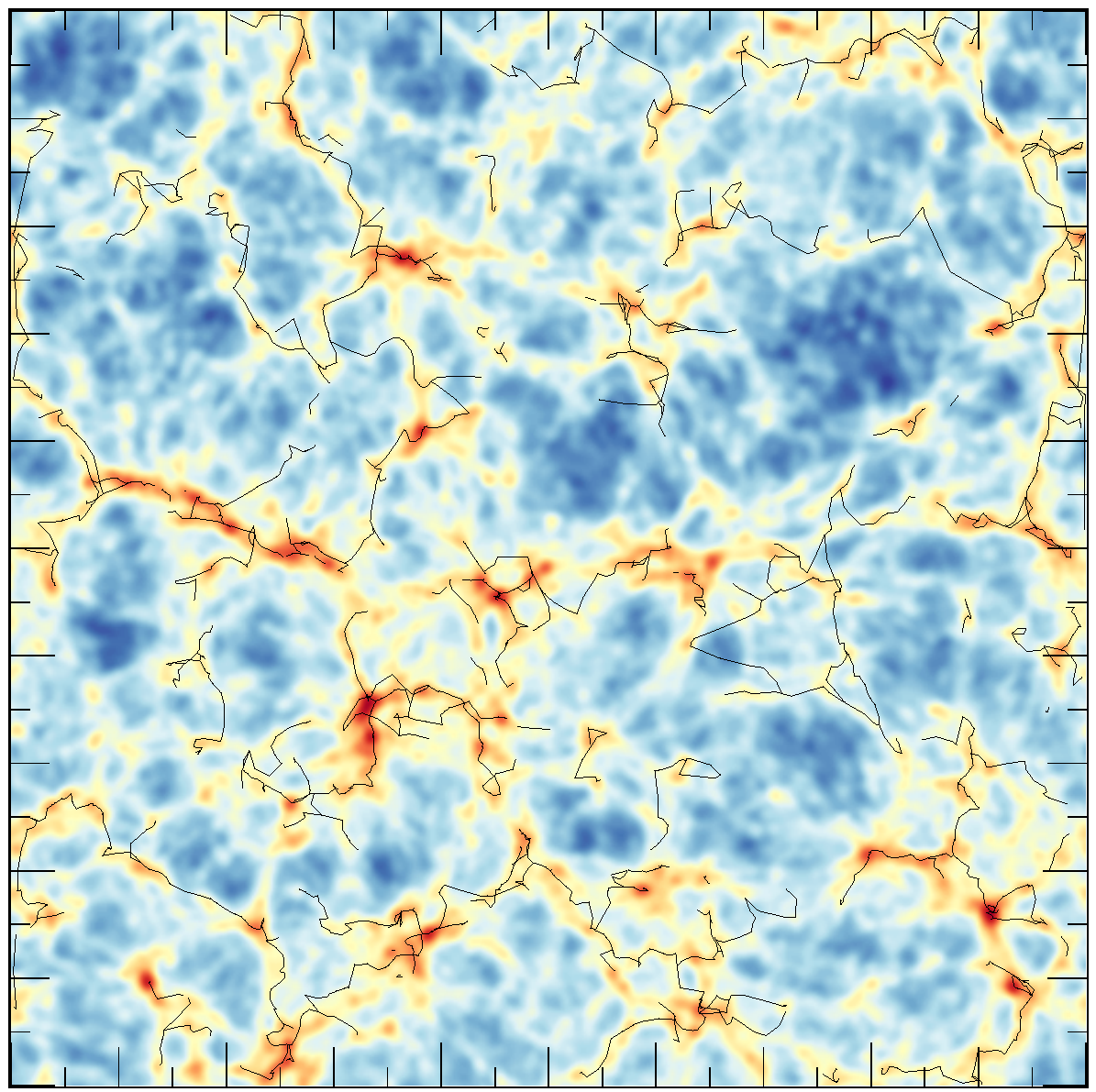}
\caption{Same as Figure~\ref{fig:FilSlices} for $N_{\sigma}=5$ and different redshifts $z=0,1,2$ from left to right. The most prominent filaments remain the same across cosmic time, in agreement with the early findings of \citep{bkp96}.
 }
\label{fig:FilSlices2}
\end{figure*}

\begin{figure}
\includegraphics[width=0.95\columnwidth]{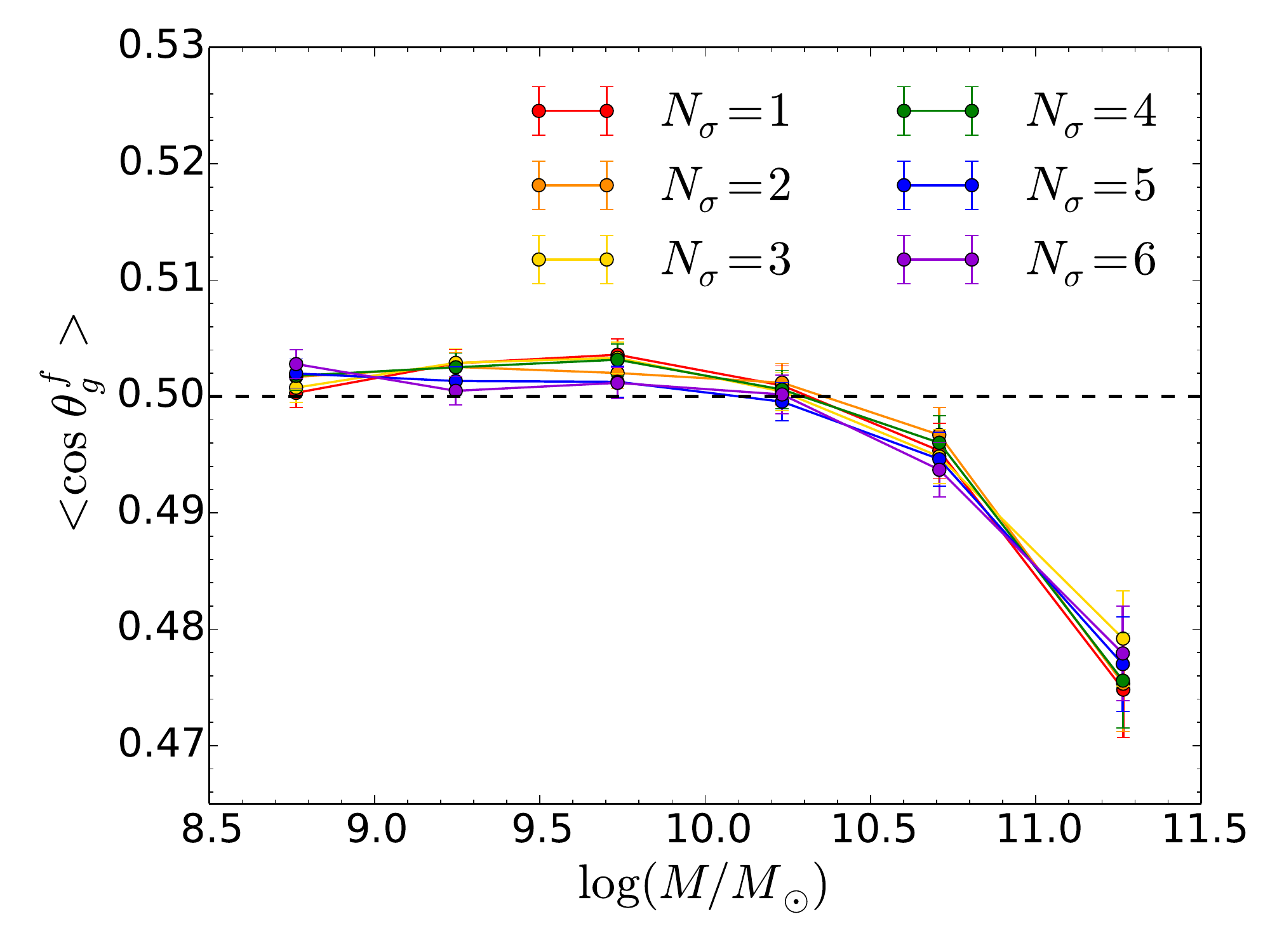}\\
\includegraphics[width=0.95\columnwidth]{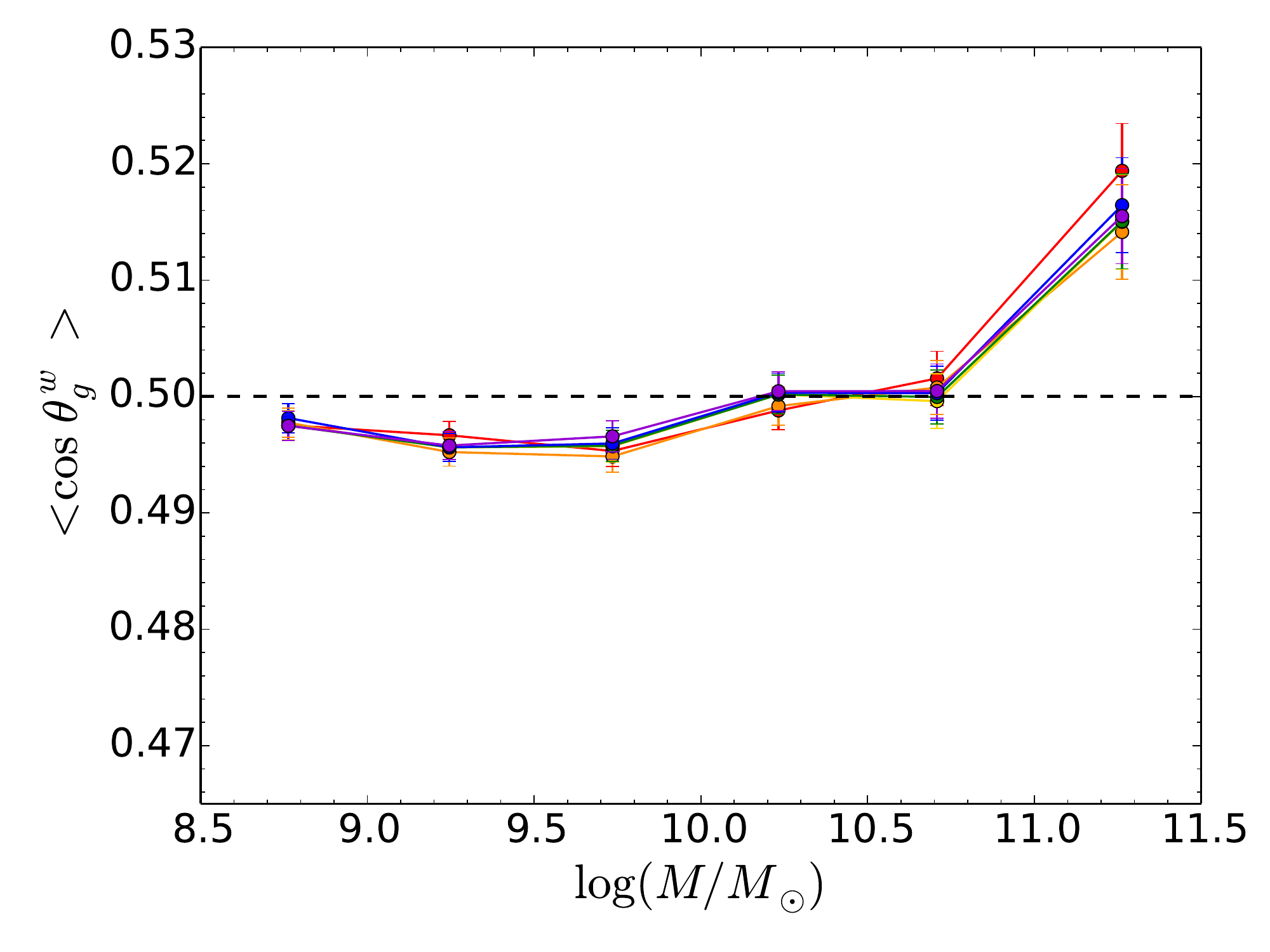}
\caption{Average misalignment of galaxies with filaments (top panel) and normal vector of walls (bottom panel) at redshift 0 for different persistence thresholds. No significant dependence with persistence is detected. }
\label{fig:SigmaDep}
\end{figure}

Let us first briefly investigate the dependence of the skeleton on persistence and redshift. To do so, we focus on the same slice of the simulation (which was already displayed in Figure~\ref{fig:Filaments}) but we vary the persistence threshold and the redshift.

Figure~\ref{fig:FilSlices} shows the filaments of the galaxy distribution in this slice of depth $10$ Mpc$/h$ for different persistence levels. As expected, for increasing values of the persistence level $N_\sigma$, the smallest filamentary structures disappear. Thus, we can use this variable to tune the robustness of the filaments to consider for analysis. 

At a constant persistence level, there is a small redshift evolution as illustrated in Figure~\ref{fig:FilSlices2}. 
This is expected since a constant persistence level is very similar to a constant level of non-linearity for which we expect almost no redshift evolution.
These small departures  are probably due to the evolving number of tracers together with the residual dynamical evolution of the cosmic web \footnote{The source of this residual evolution is diverse: scale-dependence of the power spectrum, small mismatch between persistence and non-linearity level, etc.}.

Let us now investigate the dependence of the alignments with varying persistence levels. Figure~\ref{fig:SigmaDep} shows the mean misalignment angle as a function of stellar mass for different persistence thresholds. The misalignment signal is prominent regardless of the persistence level chosen. This is true for both filaments  and  walls, as seen on the left and right panels respectively. Furthermore, the critical mass at which the transition occurs is independent of the persistence level chosen. The expected (but small) scale-dependence cannot be quantified significantly here because of the small range of persistence levels that we can probe  given the relatively small volume of the simulation. 

Hence the alignment signal described in the main text is robust and the scale-dependence small, as expected from pure dark matter studies.

\section{Scaling of transition mass}
\label{sec:Mtr} 

\begin{figure}
\includegraphics[width=0.95\columnwidth]{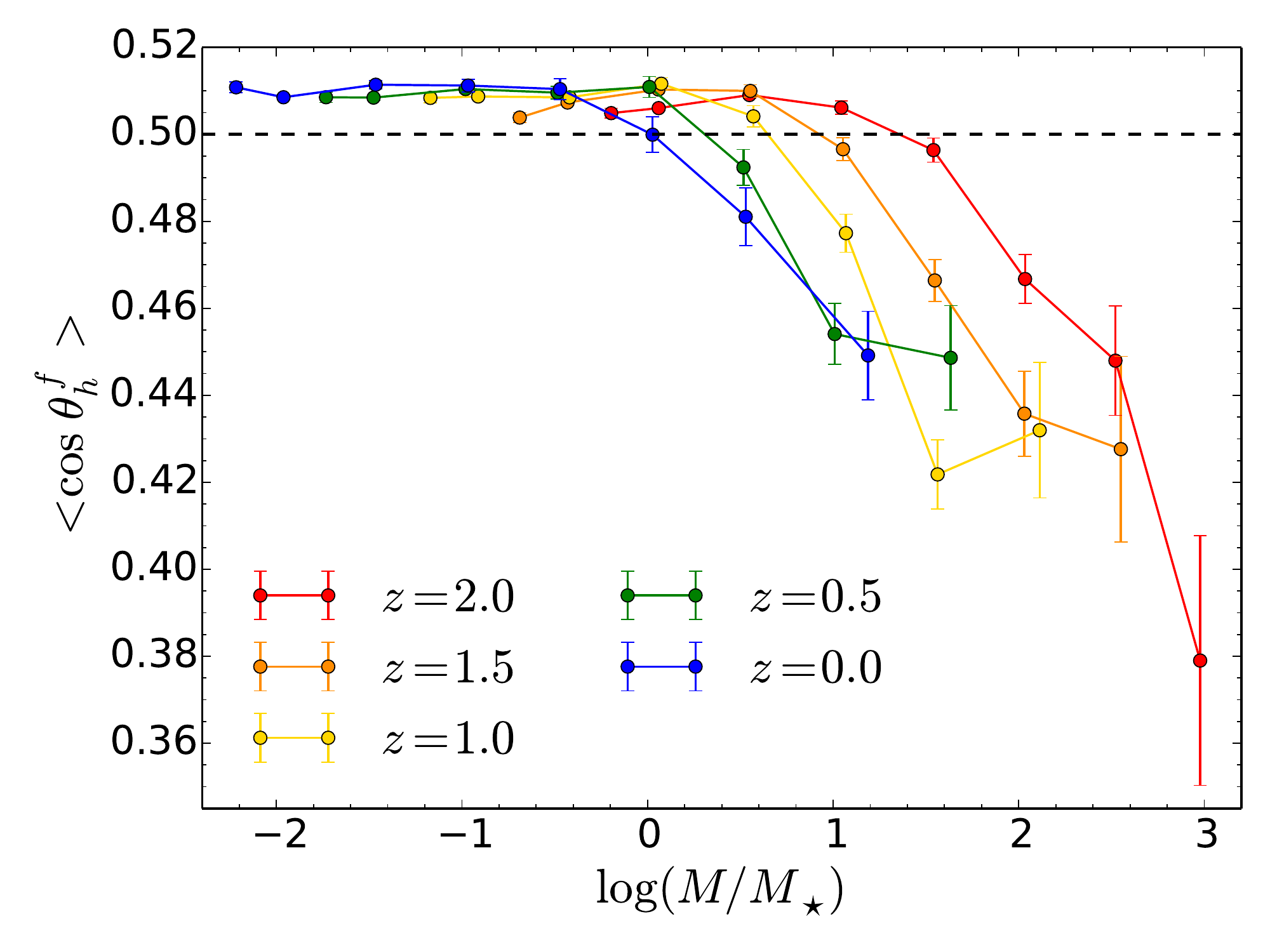}
\includegraphics[width=0.95\columnwidth]{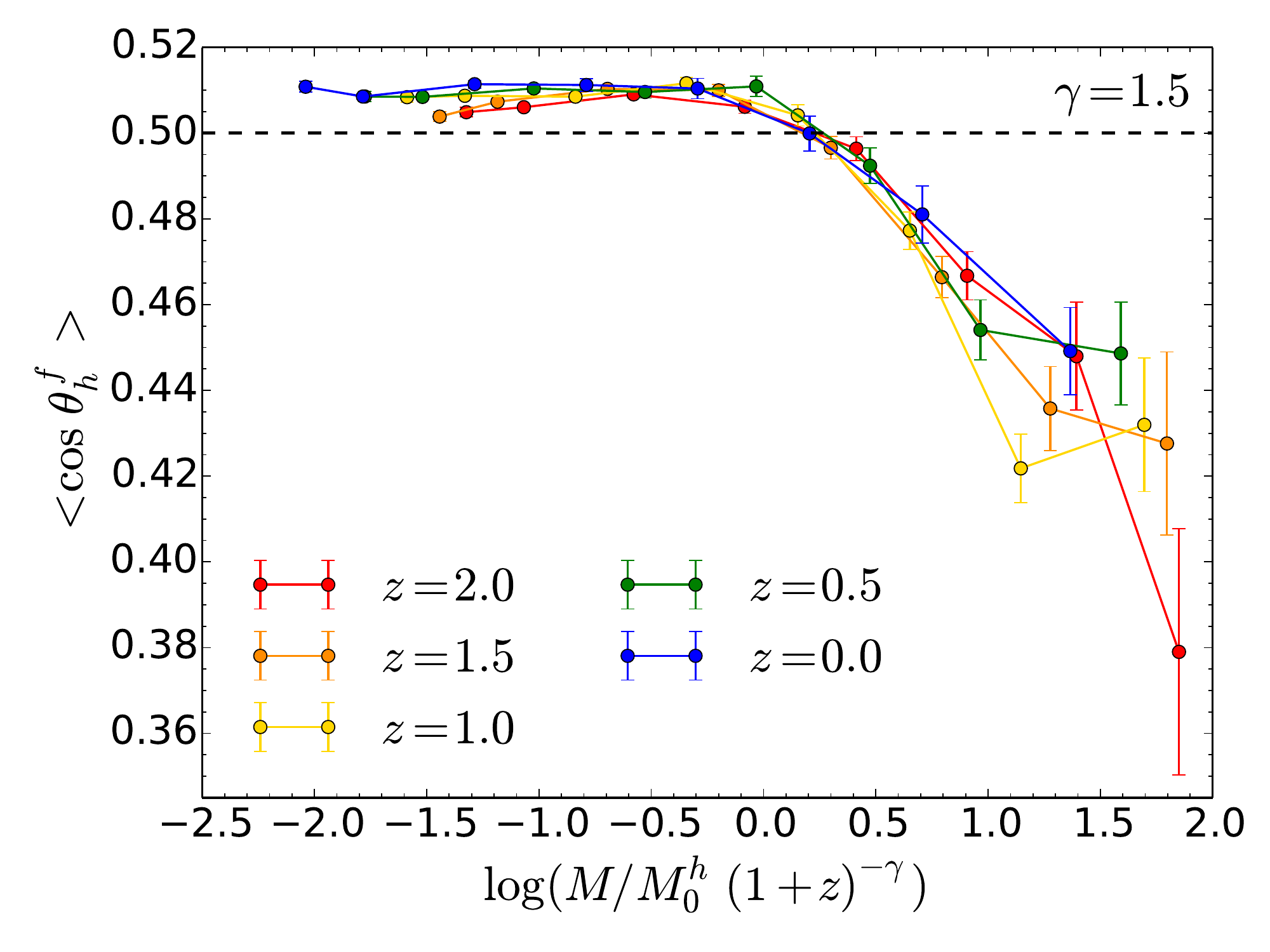}
\caption{Alignment between the spin of DM haloes and their host filament at different redshifts, similar to the top left-hand panel of  Figure~\ref{fig:CosThetaVsMass}, but when the mass is in units of the mass of non-linearity (top panel) or divided by the power law fit given in Equation~\ref{eq:fit} (bottom panel).
 }
\label{fig:Mstar}
\end{figure}

When considering the orientation of the spin of DM haloes compared to their host filament, the transition from alignment to perpendicularity was found to occur at different masses depending on the redshift.
To better assess the evolution of the transition mass across cosmic time, let us compare it to the mass of non-linearity defined as the mass enclosed in a sphere of variance 
\begin{equation}
\label{eq:sig}
\sigma(R,z)\equiv D(z) \sqrt{ \int_0^\infty \! \! P(k) W^2(k R)  {\rm d}^3 k}=1.686,
\end{equation}
where the smoothing function is a top-hat filter
\begin{equation}
W^{2}(x)= {9}\left({\sin x}/{x}-\cos x\right)^{2}/{x^{2}},
\end{equation}
$P(k)$ is the power spectrum
and $D(z)$ the growth factor.
The constraint given by Equation~\ref{eq:sig} then provides a mapping between the scale of non-linearity $R$ and the redshift from which the mass of non-linearity follows
\begin{equation}
\label{eq:mass}
 M_{\star} (z)\equiv \frac 4 3 \pi  {\bar\rho} R(z)^3\,,
 \end{equation}
 with ${\bar \rho}$ the present-day average density of matter in the Universe.
 
 When masses are expressed in units of the mass of non-linearity, the mean alignment between halo spin and filaments as a function of mass (previously top left-hand panel of Figure~\ref{fig:CosThetaVsMass}) becomes Figure~\ref{fig:Mstar}. Interestingly, at low redshift, the transition mass is very close to the non-linear mass but is increasingly larger for higher redshift. 
 
 Overall, the redshift evolution follows a power-law behaviour rather than the shape of the mass of non-linearity
  \begin{equation}
  \label{eq:fit}
 M_{\rm tr}^{h}(z)\approx M_{0}^{h}(1+z)^{-\gamma_{h}},
 \end{equation}
 with $\log M_{0}^{h}=12.2\pm0.1$ and $\gamma_{h}=1.6\pm0.3$. This fit was performed by measuring the transition mass as the first crossing $\langle \cos \theta_{h}^{f}\rangle=0.5$  at various redshifts. The estimated one-sigma error bar on this measurement was computed as the error on the mean among 8 subcubes of the simulation. The linear (in logarithmic mass and $\log 1+z$) model fitting was then performed and one-sigma confidence intervals were found for $M_{0}^{h}$ and $\gamma_{h}$.
 These findings are in excellent agreement with numerical works based on pure dark matter simulations \citep{codisetal12} once extrapolated to smaller scales.

\section{The effect of morphology}
\label{sec:vsig} 

\begin{figure*}
\includegraphics[width=0.95\columnwidth]{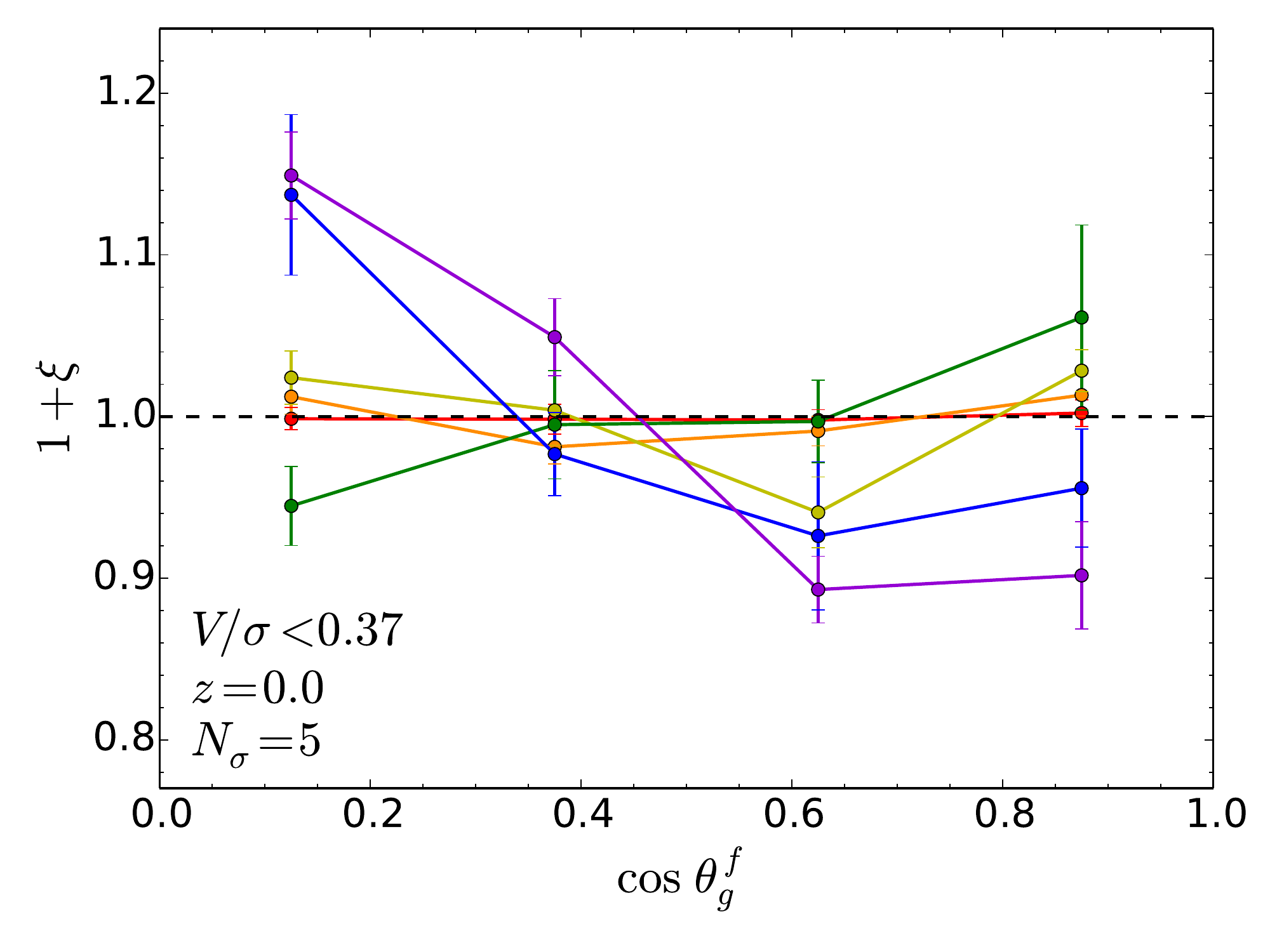}
\includegraphics[width=0.95\columnwidth]{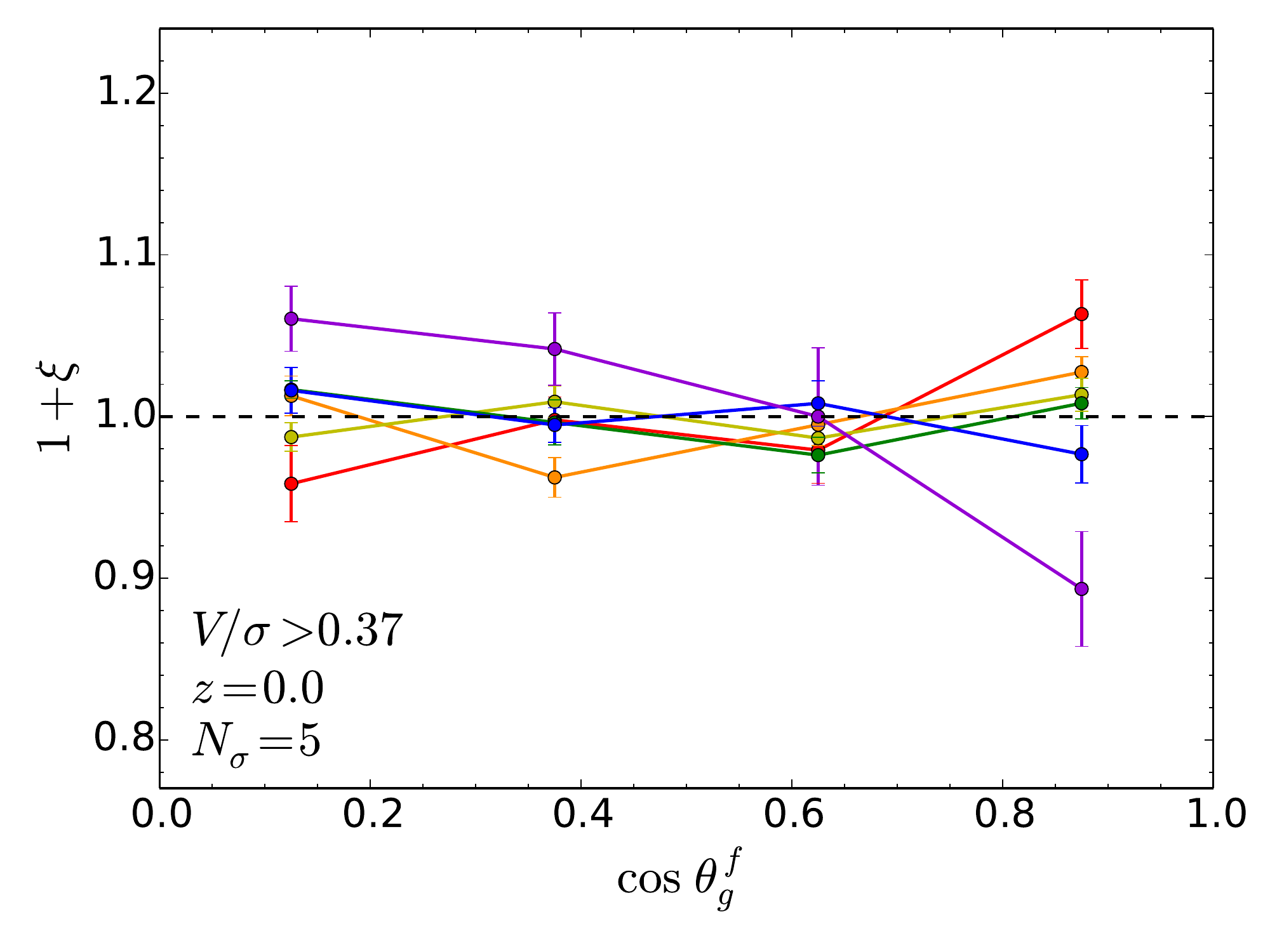}
\includegraphics[width=0.95\columnwidth]{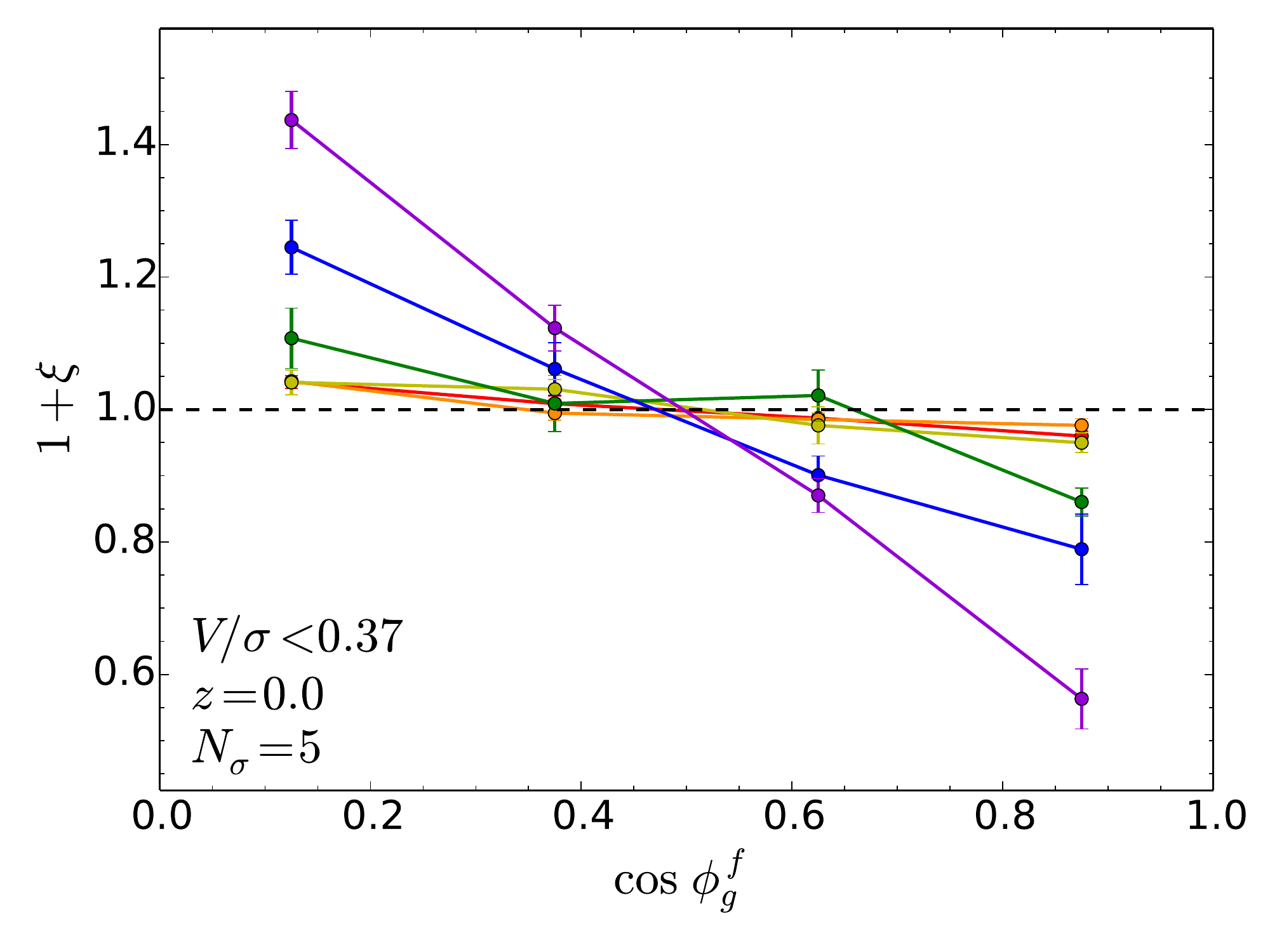}
\includegraphics[width=0.95\columnwidth]{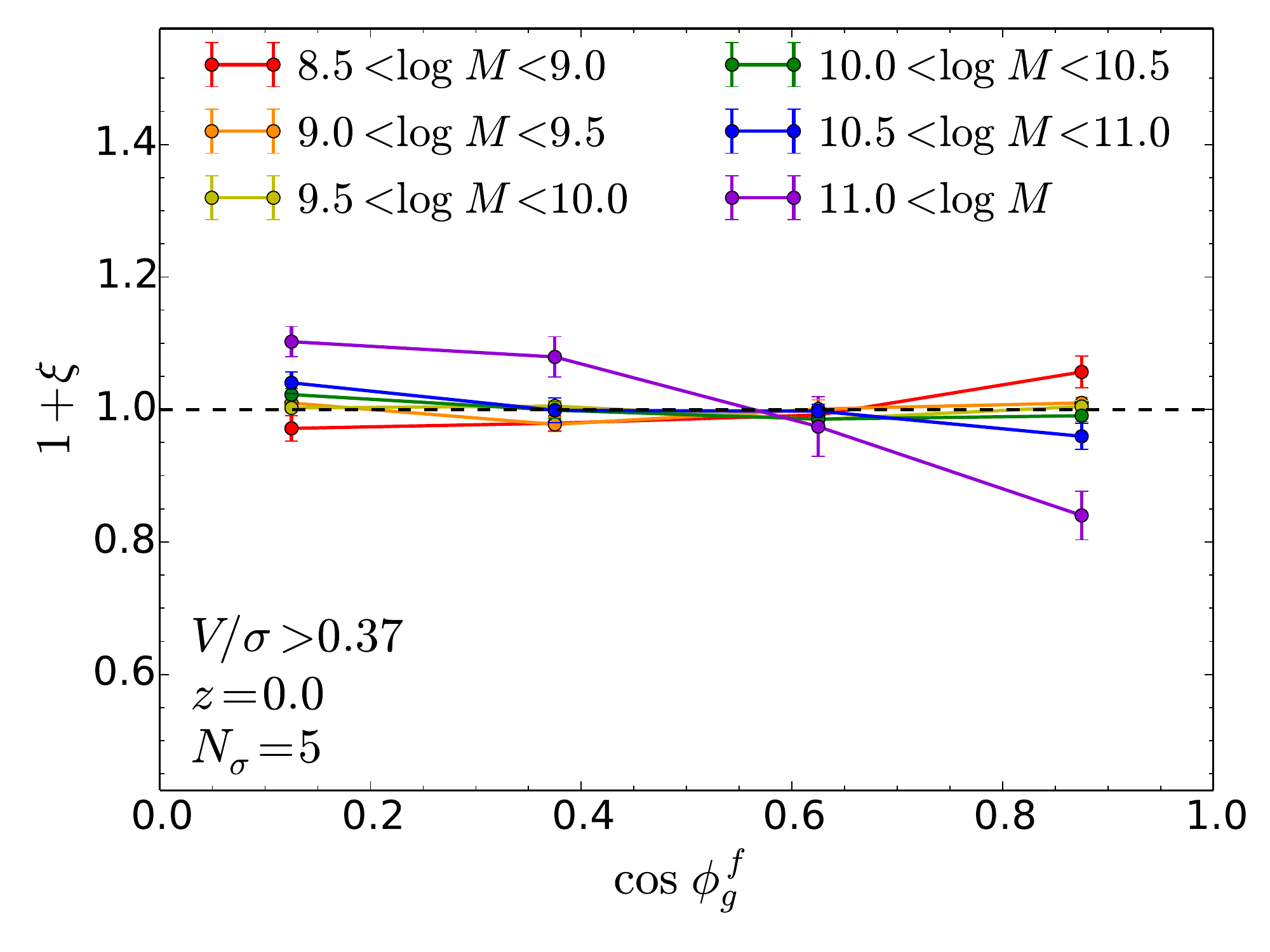}
\caption{PDF of the cosine of the angle between filaments and the spin (top panels) or minor axis (bottom panels) of galaxies at $z=0$ and for a persistence level $N_{\sigma}=5$. Two populations of galaxies are considered, one with low $V/\sigma$ (left-hand panels) corresponding to ellipticals and one with large $V/\sigma$ (right-hand panels) corresponding to spirals. Most of the shape alignments comes from elliptical galaxies for which the spin alignment is small. A smaller shape alignment is also detected for spirals with a significant transition from alignment at small mass to perpendicularity at large mass. The spin alignments for spirals is very similar to their shape alignments despite a smaller amplitude and larger error bars.
 }
\label{fig:vsig}
\end{figure*}

Let us use here the ratio $V/\sigma$ to characterize the morphology of galaxies and study whether it impacts the alignment of spins and shapes with the cosmic web described in the main text.
High $V/\sigma$ corresponds to a population of disk galaxies, supported by their angular momentum while low $V/\sigma$ corresponds to elliptical galaxies.
Let us divide our galaxy catalogue at redshift 0 in two subsamples with the same number of galaxies, corresponding to the low and high $V/\sigma$ populations.
Let us then compute the PDF of alignment of spins and shapes (minor axes) with filaments and show the result on
Figure~\ref{fig:vsig}.

 A strong tendency for elliptical galaxies with low $V/\sigma$ to be elongated along the filaments (i.e with a minor axis perpendicular to the closest segments of filaments) is found at all masses. This signal increases with mass and can be up to $\sim 40\%$ for the most massive galaxies. For this population, the spin is not a good tracer of the minor axis and the spin alignments are therefore less pronounced. In particular,  triaxiality should play an important role, with oblate galaxies having spin more likely aligned with their minor axes whereas prolate galaxies should tend to have a spin aligned with the major axis (and therefore perpendicular to the minor axis). Hence, those different behaviours mix the clear shape alignment signal into a more confused spin-alignment plot as displayed on the top left-hand panel of Figure~\ref{fig:vsig}.
Let us also split our elliptical population into bins of different prolateness
\begin{equation}
T=\frac{1-q^{2}}{1-s^{2}}\,,
\end{equation}
where $q=b/a$ and $s=c/a$ are ratios of the major ($a$), intermediate ($b$) and minor ($c$) axes of the inertia tensor's ellipsoid. $T$ therefore ranges from 0 to 1 and is split  into three bins: $T<1/3$ for oblate shapes (flattened at poles), $1/3<T<2/3$ for triaxial and $T>2/3$ for prolate (cigar-like) shapes. As expected, the alignment of the shape of elliptical galaxies with filaments is greater as $T$ increases and the galaxy is more prolate. As an illustration, the most massive bin (purple on the bottom left-hand panel of Figure~\ref{fig:vsig}) goes up to resp. 30, 40, 60 $\%$ for resp. the oblate, triaxial and prolate subsamples. 

 Spiral galaxies with high $V/\sigma$ have a different shape alignment.
 At high mass, this signal is very similar to the case of ellipticals but with a significantly smaller amplitude (by a factor of 4). At lower mass,
 some hints for a parallel alignment is found of the minor axis with the filaments, meaning that small mass spirals are oriented perpendicular to their host filament. This is very similar to the spin alignment of this population of galaxies (top right-hand panel of Figure~\ref{fig:vsig}) but with a slightly larger amplitude and smaller variance, meaning that spins are a fairly good tracer of shape alignment for spiral galaxies. 

\section{Theory of spin near walls} 
\label{app:ATTT}
In the linear stage of structure formation, 
spin acquisition is well described by linear tidal torquing  \citep{hoyle49,peebles69,doroshkevich70,white84,Cri++01,schaefer09} which predicts a spin, $s_i$, proportional to the misalignment between the inertia tensor of the proto-object denoted $I_{ij}$ and the tidal tensor $T_{ij}$ according to $s_i=\sum_{j,k,l}a^{2}(t)\dot D_{+}(t)\epsilon_{ijk} I_{jl}T_{lk}$,
with 
$\epsilon$ the rank 3 Levi-Civita tensor, $a(t)$ the scale factor and $D_{+}$ the growth factor. Following closely \cite{ATTT}, we will assume that a protohalo is well described by a peak in the initial condition and its inertia tensor by the shape of that peak for which the density Hessian matrix is a good proxy. This approximation  allows for  an observable which only depends on local variables by means of the gravitational potential and its successive derivatives.

Imposing the presence of a sheet-like structure can be implemented in this formalism using
 the theory of constrained Gaussian random field in order to compute the expectation of any function of the density field and its derivatives subject to a wall-type saddle point constraint.
This critical point appears when the gradient of the density field is zero and is characterised by its height $\nu$ defined as the density contrast divided by its rms $\sigma_0=\sqrt{\left\langle \delta^2\right\rangle}$  and its curvature by means of the three eigenvalues of the Hessian matrix of the density contrast rescaled again by their rms  $\sigma_2=\sqrt{\left\langle (\Delta\delta)^2\right\rangle}$. For a wall-type saddle point, $\lambda_{1}\geq\lambda_{2}\geq0\geq\lambda_{3}$ and we also impose that the curvature is larger along the normal than inside the sheet.

Peak theory \citep{Kac1943,Rice1945} can then be used to predict all statistical properties of critical points as soon as the (supposedly Gaussian here) probability density function (PDF) of the field $\nu=\delta/\sigma_0$, its first $\nu_i=\delta_{,i}/\sigma_1$ and second derivatives $\nu_{ij}=\delta_{,ij}/\sigma_2$ is known. The constraint for a wall-type saddle point can be written

\begin{equation}
{\cal C}_{\rm wall}=\frac{1}{R_\star^3}\lambda_1 \lambda_2 \lambda_3 \Theta_{\rm H}(\lambda_2)\Theta_{\rm H}(-\lambda_3)\delta_{\rm D}(\nu_i)\,,
\end{equation}
where the Dirac delta function imposes the gradient to be zero, the Heaviside Theta functions ensure the right sign for the eigenvalues of the density Hessian, the Jacobian $\lambda_1 \lambda_2 \lambda_3=\det \nu_{ij}$ measures the volume associated with a critical point and we define $R_{\star}={\sigma_2}/{\sigma_1}$.

The mean density map around a wall-type saddle point can then be computed from the joint statistics of $(\nu,\nu_i,\nu_{ij})$ at the location of the saddle point together with the density field $\nu'$ at a distance $\bf r$ from the saddle point where the mean map is computed.
These 11 fields are gathered in a vector $\mathbf{X}=\{\nu',\nu,\nu_1,\nu_2,\nu_3,\nu_{11},\nu_{22},\nu_{33},\nu_{12},\nu_{13},\nu_{23}\}$ whose PDF reads

\begin{equation}
{\cal P}(\mathbf{X})=\! \frac{1}{\sqrt{{\rm det}| 2\pi \mathbf{C}|}  }
\exp\left(\!-\frac{1}{2}X^{\rm T}
\cdot \mathbf{C}^{-1} \cdot X\right)\,,
\end{equation}
where the covariance matrix $C= \langle  \mathbf{X}\cdot \mathbf{X}^{\rm T} \rangle$ depends on the separation vector $\bf r$ and the linear power spectrum
$P_k(k) $, which can include a filter function on a given scale. Here, a power-law power spectrum with spectral index $n_s=2$ is used  with a Gaussian filter defined in Fourier space by
\begin{equation}
    W_{\rm G}(\mathbf{k},L)=
    \frac{1}{(2\pi)^{3/2}}\exp\left(\frac{-k^{2}L^{2}}{2}\right)
    \,,
    \end{equation}
where $L$ is the smoothing length.
By definition, $C_{11}$ is $1$ and the block corresponding to the saddle position, ${\mathbf C}_0=(C_{ij})_{i,j>1}$, reads
\setlength{\arraycolsep}{1.5pt}
{\footnotesize
 \begin{equation}
\mathbf{C}_{0}\!=\!
\!\! \left(\!\!\!
\begin{array}{ccccccccccc}
1\!&\!0\!&\!0\!&\!0\!&\!-\!\gamma/3\!&\!-\!\gamma/3\!&\!-\!\gamma/3\!&\!0\!&\!0\!&\!0\\
0\!&\!1/3&0&0&0&0&0&0&0&0\!\\
0\!&\!0&1/3&0&0&0&0&0&0&0\!\\
0\!&\!0&0&1/3&0&0&0&0&0&0\!\\
-\gamma/3\!&\!0&0&0&1/5&1/15&1/15&0&0&0\!\\
-\gamma/3\!&\!0&0&0&1/15&1/5&1/15&0&0&0\!\\
-\gamma/3\!&\!0&0&0&1/15&1/15&1/5&0&0&0\!\\
0\!&\!0&0&0&0&0&0&1/15&0&0\!\\
0\!&\!0&0&0&0&0&0&0&1/15&0\!\\
0\!&\!0&0&0&0&0&0&0&0&1/15\\
\end{array}
\!\!\!\right)\!. \nonumber
 \end{equation}
 }
The cross correlations between $\nu'$ and the fields at the position of the saddle depend on the separation vector and can be analytically computed. They read for $j$ between 2 and 11
 \begin{equation}
\hskip -0.1cm   \left\langle \nu' X_j\right\rangle =\frac{\displaystyle\int {\rm d}^3 {\bf k} P_k (k) \prod_{i=1}^3 (-\imath k_i)^{\alpha_{ij}} \exp \left(\imath {\bf k}\cdot {\bf r}\right)}{\displaystyle\sqrt{ \int {\rm d}^3 {\bf k} P_k (k)\times \int {\rm d}^3 {\bf k} P_k (k) \prod_{i=1}^3 k_i^{2\alpha_{ij}}}}\,,
 \end{equation}
where $\alpha_{ij}$ counts the number of derivatives with respect to index $i$ of $X_j$.

The mean density map around a wall-type saddle point of fixed height and curvatures is analytical \citep{ATTT}
 \begin{multline}
\left\langle \nu'|{\cal S}\right\rangle=
\frac{(\lambda_{1}\!+\!\lambda_2\!+\!\lambda_3) (\left\langle\nu' {\rm tr}\,\nu_{ij}\right\rangle+\gamma \left\langle\nu'\nu\right\rangle )}{1-\gamma ^2}\\
+
\frac{\nu  (\left\langle\nu'\nu\right\rangle+\gamma \left\langle\nu' {\rm tr}\,\nu_{ij}\right\rangle)}{1-\gamma ^2}
+\frac {45}{4}\!\left(\mathbf{\hat r}^{\rm T} \cdot \overline{\mathbf{ H}}\cdot\mathbf{\hat r}\right)\! \left\langle \nu' \!\left(\mathbf{\hat r}^{\rm T} \cdot \overline{\mathbf{ H}}\cdot\mathbf{\hat r}\right)\!\right\rangle,
\nonumber\end{multline}
where $\overline{\mathbf{ H}}$ is the {\sl detraced} Hessian of the density and
$\mathbf{\hat r}={\mathbf{r} }/{r}$.

Similarly, one can get an analytical prediction for the mean spin map in the vicinity of the wall-type saddle point.
Because, we only care about the spin direction here and not its amplitude, let us first define the spin vector as
\begin{equation}
{s}_i = \sum_{j,k,l}\epsilon_{ijk}  \nu_{kl}\phi_{lj} \,, \label{eq:defL3D}
\end{equation}
where $\phi_{ij}$ is the tidal tensor rescaled by $\sigma_0$ whose trace is nothing but ${\rm tr}\phi_{ij}= \nu$.
In the vicinity  of a wall-type saddle point, the spin vector is found to be orthogonal to the separation and reads
 \begin{equation}
 \mathbf{s}(\mathbf{r}| {\rm crit},I_{1},\kappa_{1},\kappa_{2},\nu)=-15
(\mathbf s^{(1)}
+\mathbf s^{(2)})\cdot(\hat {\mathbf r}^{\rm T}\!\!\cdot \epsilon\cdot \overline {\mathbf{ H}}\cdot \hat {\mathbf{r} })
\,, \label{eq:defL3Dsol}
\end{equation}
where 
  $\mathbf s^{(1)}$ can be written as a function of the wall height $\nu$ and trace of the Hessian $I_{1}$ 
     \begin{align}	
 \mathbf{ s}^{(1)}=&\left(  \frac{\nu}{1-\gamma^{2}}  \left[
  (\xi_{\phi \phi}^{\Delta +}+\gamma \xi_{\phi x}^{\Delta +})\xi_{x x}^{\times\times}
  -(\xi_{\phi x}^{\Delta +}+\gamma \xi_{x x}^{\Delta +})\xi_{\phi x}^{\times\times}
  \right]\right.\nonumber
  \\
 & \hskip -1cm+\!\!\! \left.
 \frac{I_{1}}{1-\gamma^{2}}
  \left[
 (\xi_{\phi x}^{\Delta +}+\gamma \xi_{\phi \phi}^{\Delta +})\xi_{x x}^{\times\times}
-
  (\xi_{x x}^{\Delta +}+\gamma \xi_{\phi x}^{\Delta +})\xi_{\phi x}^{\times\times}
    \right]\,\right) {I}_{3}\,,\nonumber
     \end{align}
     and $\mathrm s^{(2)}$ depends on the traceless part of the Hessian
          \begin{multline}	
\mathbf{s}^{(2)}=
  -\frac{5}{8}
   \left[
 2((\xi_{\phi x}^{\Delta +}-\xi_{\phi x}^{\Delta \Delta})\xi_{x x}^{\times\times}
  -( \xi_{x x}^{\Delta +}-\xi_{x x}^{\Delta \Delta})\xi_{\phi x}^{\times\times}) 
  \overline{\mathbf{ H}}
 \right.\nonumber
  \\
+  \left.
\Big( (7\xi_{x x}^{\Delta \Delta}+5 \xi_{x x}^{\Delta +})\xi_{\phi x}^{\times\times}-(7\xi_{\phi x}^{\Delta \Delta}+5\xi_{\phi x}^{\Delta +})\xi_{x x}^{\times\times}
  \Big)
(\hat {\mathbf r}^{\rm T}\cdot  \overline{\mathbf{ H}}\cdot \hat {\mathbf{r} }){I}_{3}
    \right]\,
   \nonumber
     \end{multline}
with    ${I}_{3}$ the identity matrix and
     \begin{align}
     \hat {\mathbf r}^{\rm T}\!\!\cdot \epsilon\cdot \overline{\mathbf{ H}}\cdot \hat {\mathbf{r} }=   \sum_{ikl} \hat {\mathbf r}^{}_{i}\epsilon_{ijk}\overline{\mathbf{ H}}_{kl}\hat {\mathbf{r} }_{l}\,.
     \end{align}

As an illustration, we pick the mean wall-type saddle which can be shown to be described by $\bar \nu=0.55 \gamma$, $\bar \lambda_1=0.59$, $\bar \lambda_2=0.27$ and $\bar \lambda_3=-0.83$ and show the mean spin map around it in Figure~\ref{fig:spinwallATTT}, the $z$-axis is the normal to the wall (corresponding to the $ \lambda_3$ direction) while the $x$ and $y$-axis describes the sheet with the wall-type saddle point at the origin of the coordinate system. All the coordinates are expressed in units of the smoothing length. Linear tidal torque theory therefore predicts the spins to be aligned with the normal to the wall inside the wall but perpendicular as we get further away. As the density field is correlated with the typical halo mass, this spatial transition implies a transition in mass: low-mass objects having a spin lying in the plane of the wall and higher mass objects with a spin perpendicular to that plane.

\section{  Resolution test}
\label{sec:res}

{ 
In order to check how the simulation resolution impacts our results, we use a degraded version of Horizon-AGN called Horizon-AGN-512 with 8 times fewer DM particles and a finest spatial resolution of 2 kpc and we measure the alignment between the spins of galaxies found in Horizon-AGN-512 and the skeleton of Horizon-AGN at redshift $z=0.055$. Figure~\ref{fig:res} shows that the results are consistent (within the error bars) therefore demonstrating that our results are not affected by simulation resolution.

\begin{figure}
\includegraphics[width=\columnwidth]{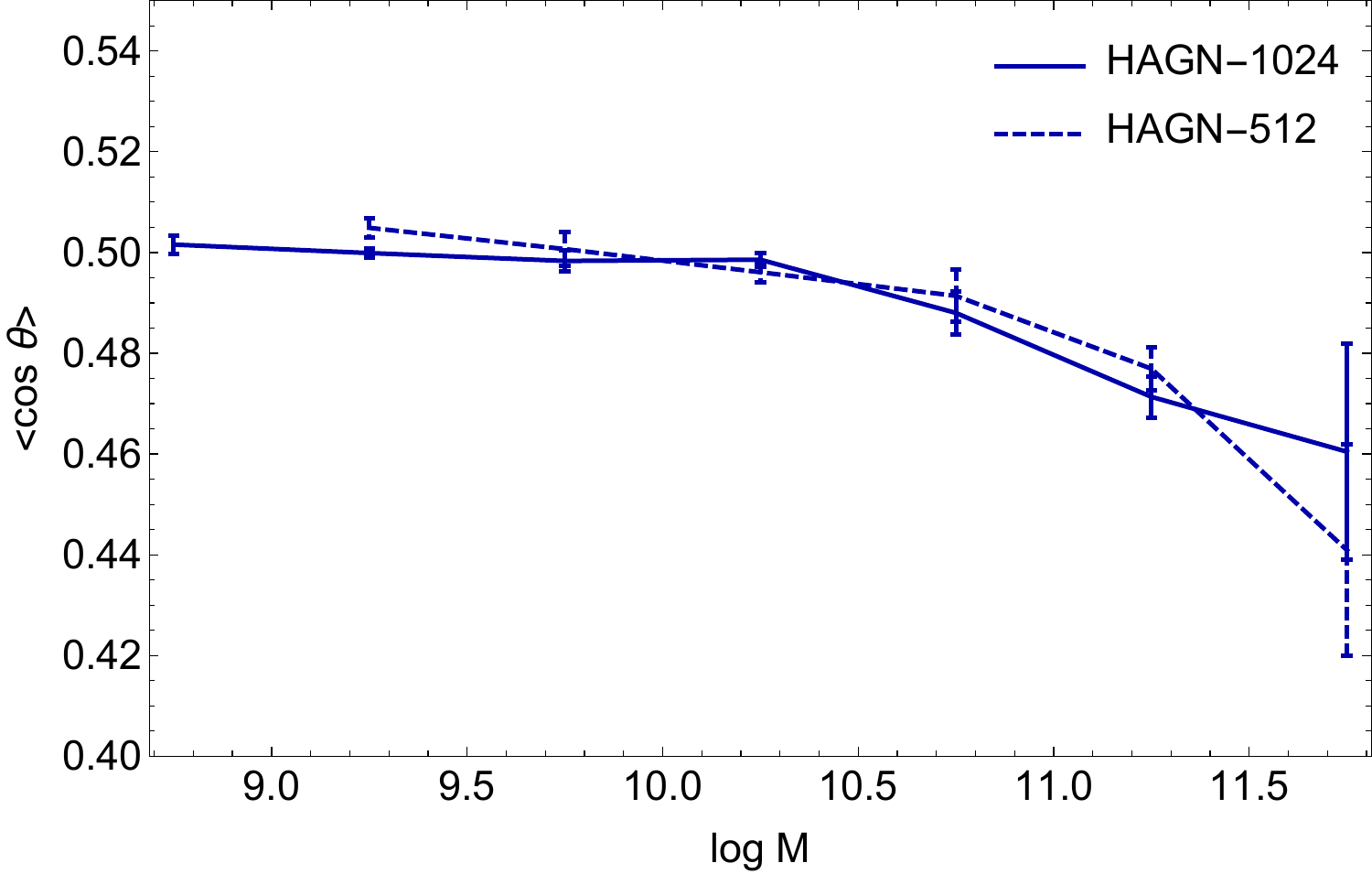}
\caption{  Mean cosine of the angle between the filaments and the spin of galaxies of different mass found in Horizon-AGN (solid line) and its companion run with less resolution Horizon-AGN-512 (dashed).
 }
\label{fig:res}
\end{figure}

}

\end{document}